\documentclass[11pt]{article}
\usepackage{axodraw}
\usepackage{epsfig}
\newcommand{\la}[1]{\label{#1}}
\newcommand{\be}{\begin{equation}}
\newcommand{\ee}{\end{equation}}
\newcommand{\ba}{\begin{eqnarray}}
\newcommand{\ea}{\end{eqnarray}}
\newcommand{\rmi}[1]{{\mbox{\scriptsize #1}}}
\newcommand{\fig}{Fig.~}
\newcommand{\eq}{Eq.~}
\newcommand{\se}{Sec.~}

\newcommand{\nr}[1]{(\ref{#1})}
\newcommand{\tr}{{\rm Tr\,}}

\renewcommand{\vec}[1]{{\bf #1}}

\def\lsi{\raise0.3ex\hbox{$<$\kern-0.75em\raise-1.1ex\hbox{$\sim$}}}
\def\gsi{\raise0.3ex\hbox{$>$\kern-0.75em\raise-1.1ex\hbox{$\sim$}}}
\newcommand{\lsim}{\mathop{\;\lsi\;}}
\newcommand{\gsim}{\mathop{\;\gsi\;}}

\newcommand{\nF}[1]{n_\rmi{F{#1}}}

\newcommand{\im}{\mathop{\mbox{Im}}}
\newcommand{\bsl}[1]{\,\slash\!\!\!\!{#1}\,}

\newcommand{\Tint}[1]{{\hbox{$\sum$}\!\!\!\!\!\!\int}_{\!\!\!\!\raise-0.9ex\hbox{$\scriptstyle{#1}$}}}

\newcommand{\piccc}[1]{\;\parbox[c]{90pt}{\begin{picture}(90,30)(0,0)
\SetWidth{1.0}\SetScale{1.0} #1 \end{picture}}\;}

\def\Lwidth{1}

\def\Agl(#1,#2)(#3,#4,#5){\PhotonArc(#1,#2)(#3,#4,#5){\Lwidth}
{6.283 #3 mul 360 div #4 #5 sub #4 #5 sub mul sqrt mul Ldensity mul}}
\def\Lgl(#1,#2)(#3,#4){\Photon(#1,#2)(#3,#4){\Lwidth}
{#1 #3 sub #1 #3 sub mul #2 #4 sub #2 #4 sub mul add sqrt Ldensity mul}}
\def\Agh(#1,#2)(#3,#4,#5){\DashArrowArc(#1,#2)(#3,#4,#5){1}}
\def\Aagh(#1,#2)(#3,#4,#5){\DashArrowArcn(#1,#2)(#3,#5,#4){1}}
\def\Lgh(#1,#2)(#3,#4){\DashArrowLine(#1,#2)(#3,#4){1}}
\def\Lagh(#1,#2)(#3,#4){\DashArrowLine(#3,#4)(#1,#2){1}}
\def\Ahh(#1,#2)(#3,#4,#5){\DashCArc(#1,#2)(#3,#4,#5){1}}
\def\Lhh(#1,#2)(#3,#4){\DashLine(#1,#2)(#3,#4){1}}
\def\Aqu(#1,#2)(#3,#4,#5){\ArrowArc(#1,#2)(#3,#4,#5)}
\def\Aaqu(#1,#2)(#3,#4,#5){\ArrowArcn(#1,#2)(#3,#5,#4)}
\def\Lqu(#1,#2)(#3,#4){\ArrowLine(#1,#2)(#3,#4)}
\def\Laqu(#1,#2)(#3,#4){\ArrowLine(#3,#4)(#1,#2)}
\def\Aqq(#1,#2)(#3,#4,#5){\CArc(#1,#2)(#3,#4,#5)}
\def\Lqq(#1,#2)(#3,#4){\Line(#1,#2)(#3,#4)}
\def\Asc(#1,#2)(#3,#4,#5){\CArc(#1,#2)(#3,#4,#5)}
\def\Lsc(#1,#2)(#3,#4){\Line(#1,#2)(#3,#4)}

\def\Textt(#1,#2,#3){\Text(#1,#2)[t]{{$\scriptstyle #3$}}}
\def\Textb(#1,#2,#3){\Text(#1,#2)[b]{{$\scriptstyle #3$}}}
\def\Textl(#1,#2,#3){\Text(#1,#2)[l]{{$\scriptstyle #3$}}}
\def\Textr(#1,#2,#3){\Text(#1,#2)[r]{{$\scriptstyle #3$}}}
\def\Texttl(#1,#2,#3){\Text(#1,#2)[tl]{{$\scriptstyle #3$}}}
\def\Textbl(#1,#2,#3){\Text(#1,#2)[bl]{{$\scriptstyle #3$}}}
\def\Texttr(#1,#2,#3){\Text(#1,#2)[tr]{{$\scriptstyle #3$}}}
\def\Textbr(#1,#2,#3){\Text(#1,#2)[br]{{$\scriptstyle #3$}}}
\def\Textst(#1,#2,#3){\Text(#1,#2)[t]{{$\scriptscriptstyle #3$}}}
\def\Textsb(#1,#2,#3){\Text(#1,#2)[b]{{$\scriptscriptstyle #3$}}}
\def\Textsl(#1,#2,#3){\Text(#1,#2)[l]{{$\scriptscriptstyle #3$}}}
\def\Textsr(#1,#2,#3){\Text(#1,#2)[r]{{$\scriptscriptstyle #3$}}}
\def\Textstl(#1,#2,#3){\Text(#1,#2)[tl]{{$\scriptscriptstyle #3$}}}
\def\Textsbl(#1,#2,#3){\Text(#1,#2)[bl]{{$\scriptscriptstyle #3$}}}
\def\Textstr(#1,#2,#3){\Text(#1,#2)[tr]{{$\scriptscriptstyle #3$}}}
\def\Textsbr(#1,#2,#3){\Text(#1,#2)[br]{{$\scriptscriptstyle #3$}}}

\def\Lwidth{1}

\def\TopoSBnew(#1,#2,#3){\piccc{#1(0,15)(15,15) #2(30,15)(15,0,180)%
 #3(30,15)(15,180,360) #1(45,15)(60,15)%
 \Textl(45,30,\tilde R)\Textl(45,0,\tilde Q+\tilde R)\Textb(3,18,\tilde Q)}}
\def\Bqu(#1,#2)(#3,#4,#5){\SetWidth{2.0}\ArrowArc(#1,#2)(#3,#4,#5)%
\SetWidth{1.0}}

\makeatletter \@addtoreset{equation}{section} \makeatother
\renewcommand{\theequation}{\arabic{section}.\arabic{equation}}
\makeatletter
\renewcommand\section{\@startsection {section}{1}{\z@}%
                                   {-5.5ex \@plus -1ex \@minus -.2ex}
                                   {2.3ex \@plus.2ex}%
                                   {\normalfont\large\bfseries}}
\renewcommand\subsection{\@startsection{subsection}{2}{\z@}%
                                     {-3.25ex\@plus -1ex \@minus -.2ex}%
                                     {1.5ex \@plus .2ex}%
                                     {\normalfont\normalsize\bfseries}}
\renewcommand\thesection {\@arabic\c@section}
\renewcommand\thesubsection   {\thesection.\@arabic\c@subsection}
\renewcommand{\@seccntformat}[1]{%
\csname the#1\endcsname.\hspace{1.0em}}
\makeatother

\begin{document}

\begin{titlepage}
\begin{flushright}
\date{today}
\end{flushright}
\begin{centering}
\vfill

{\Large{\bf The $\nu$MSM, leptonic asymmetries, and \\[2mm] properties of
singlet fermions}}

\vspace{0.8cm}

Mikhail~Shaposhnikov

{\em Institut de Th\'eorie des Ph\'enom\`enes Physiques, EPFL, 
CH-1015 Lausanne, Switzerland}

\vspace{0.3cm}
 
\mbox{\bf Abstract}

\end{centering}

\vspace*{0.3cm}
 
\noindent  
We study in detail the mechanism of baryon and lepton asymmetry 
generation in the framework of the $\nu$MSM (an extension of the
Standard Model by three singlet fermions with masses smaller than the
electroweak scale). We elucidate the issue of CP-violation in the
model and define the phase relevant for baryogenesis. We clarify the
question of quantum-mechanical coherence, essential for the lepton
asymmetry generation in singlet fermion oscillations and compute the
relevant damping rates. The range of masses and couplings of singlet
leptons which can lead to successful baryogenesis is determined. The
conditions which ensure survival of primordial (existing above the
electroweak temperatures) asymmetries in different leptonic numbers
are analysed. We address the question whether CP-violating reactions
with lepton number non-conservation can produce leptonic asymmetry
{\em below} the sphaleron freeze-out temperature. This asymmetry, if
created, leads to resonant production of dark matter sterile
neutrinos.  We show that the requirement that a significant lepton
asymmetry be produced puts stringent constraints on the properties of
a pair of nearly degenerate singlet fermions, which can be tested in
accelerator experiments. In this region of parameters the $\nu$MSM
provides a common mechanism for production of baryonic matter and dark
matter in the universe. We analyse different fine-tunings of the model
and discuss possible symmetries of the $\nu$MSM Lagrangian that can
lead to them.
 


\vspace*{1cm}
 
\noindent

\date{today}

\vfill

\end{titlepage}

%
\section{Introduction}
\la{se:intro}
This paper is a continuation of the works \cite{Asaka:2005an} - 
\cite{Bezrukov:2007ep} addressing the cosmological and
phenomenological consequences of the
$\nu$MSM. The $\nu$MSM (neutrino Minimal Standard Model) is a
renormalizable  extension of the  Standard Model (SM) by three light
singlet fermions -- right-handed,  or sterile neutrinos. Amazingly
enough, this simple theory allows to solve in a unified way four
observational problems of the SM \cite{Asaka:2005pn,Bezrukov:2007ep}.
It leads to neutrino masses and thus gives rise to neutrino
oscillations, absent in the SM. It provides a candidate for dark
matter particle in the form of a long-lived sterile
neutrino\footnote{A number of interesting astrophysical applications
of keV scale sterile neutrinos can be found in \cite{astro}.},
discussed already in  \cite{Dodelson:1993je,Shi:1998km,Dolgov:2000ew}.
It allows for baryon asymmetry generation due to coherent oscillation
of the other two singlet fermions \cite{Asaka:2005pn,Akhmedov:1998qx}
and electroweak anomalous fermion number non-conservation at high
temperatures \cite{Kuzmin:1985mm}, associated with sphalerons
\cite{Klinkhamer:1984di}. A non-minimal coupling of the Higgs field to
gravity would lead to inflation consistent with cosmological
observations \cite{Bezrukov:2007ep}.  

Let us recall the essential features of the $\nu$MSM. The lightest
sterile neutrino $N_1$ plays the role of the dark matter particle.  It
should have a mass above $0.3$ keV
\cite{Tremaine:1979we,Lin:1983vq,Dalcanton:2000hn} (the most
conservative Tremaine-Gunn bound), following from observations of
rotational curves of dwarf galaxies. It is practically decoupled from
other fields of the Standard Model (its Yukawa coupling to the Higgs
and active neutrino must be smaller than $10^{-12}$, as follows from
the requirement that the mass density of the sterile neutrinos
produced in the early universe does not exceed the dark matter
abundance \cite{Asaka:2005an,Asaka:2006nq} and from astrophysical
X-ray constraints \cite{Boyarsky:2006jm}). Because of the very weak
coupling, it does not contribute to the mass matrix for active
neutrinos \cite{Asaka:2005an,Asaka:2005pn,Shaposhnikov:2006nn} and
does not play a role in baryogenesis. The two other singlet fermions,
$N_2$ and $N_3$, have masses above $140$ MeV (the constraint is coming
from accelerator experiments combined with BBN bounds
\cite{Gorbunov:2007ak}, see also \cite{Kusenko:2004qc}). These
particles must be nearly degenerate in mass, to ensure coherent
CP-violating oscillations leading to baryon asymmetry of the universe
\cite{Asaka:2005pn,Akhmedov:1998qx}. 

In this paper we are going to study several related issues. The first
one is an elaboration of the mechanism of baryogenesis in singlet
fermion oscillations. Though the master equations for leptogenesis in
the $\nu$MSM have already been written in \cite{Asaka:2005pn}, they
were only analysed in the symmetric phase of the electroweak theory
and in the regime where all singlet fermions are out of thermal
equilibrium. We would like to answer the following questions. What is
the CP-violating phase driving the baryogenesis? What happens with
lepton asymmetry when these particles thermalize? When is the
coherence of their quantum-mechanical oscillations, crucial for the
resonant lepton asymmetry production lost? Are the effects of
electroweak symmetry breaking essential for leptogenesis? What are the
rates of non-conservation of different leptonic flavours in the
$\nu$MSM? What is the time evolution of deviations from thermal
equilibrium due to singlet fermions, essential for leptogenesis? Can
the primordial lepton asymmetries be protected from erasure in the
$\nu$MSM?

The second problem is a phenomenological one. The parameter space of
the $\nu$MSM which can lead to the observed baryon asymmetry has never
been explored in detail. Therefore, we would like to improve the
existing cosmological constraints on masses and couplings of singlet
fermions, which could be helpful for their experimental search.

The third question we address in this paper is: Can we have large
lepton asymmetries {\em well below} the electroweak scale? 
The motivation for this consideration is the following. 
In \cite{Asaka:2006nq} we computed the abundance of dark matter
sterile neutrino in the Dodelson-Widrow (DW) scenario
\cite{Dodelson:1993je} (for earlier works see 
\cite{Dolgov:2000ew,Abazajian:2001nj,Abazajian:2002yz,Abazajian:2005gj}).
This scenario assumes that \\
(i) no sterile neutrinos existed at temperatures above $1$ GeV;\\ 
(ii) the only interactions the sterile neutrinos have are those with
the ordinary neutrinos; \\ 
(iii) the universe was (leptonic) charge symmetric at temperatures
below $1$ GeV.

\noindent
The result was compared with two types of astrophysical bounds.
The first one deals with X-ray observations of diffuse X-ray
background of  our and distant galaxies and Milky-Way satellites
\cite{Dolgov:2000ew}, \cite{Abazajian:2001vt} -
\cite{Boyarsky:2007ge} and gives an upper limit on the mixing angle
of dark matter sterile neutrino as a function of its mass. The second
bound limits the free streaming length of the dark matter particle
from observation of Lyman-$\alpha$ clouds \cite{Hansen:2001zv} -
\cite{Viel:2007mv}. The prediction, even with the largest
uncertainties resulting from poor knowledge of QCD dynamics at the
epoch of the quark-hadron crossover, is in conflict with astrophysical
bounds. This rules out the DW mechanism as a source of sterile
neutrino production, if one takes for granted that the results of
\cite{Seljak:2006qw,Viel:2006kd} are robust. If the weaker, but more
conservative Tremaine-Gunn bound is applied, then the DW mechanism
can account for sterile neutrino dark matter in the universe,
provided the mass of sterile neutrino is below 3.5 keV  (the most
conservative bound is 6 keV,  see  \cite{Asaka:2006nq}).

Since there are three essential assumptions involved, these
considerations force to challenge one or more of them. As was found
in  \cite{Shaposhnikov:2006xi}, (i) and (ii) are not valid in an
extension of the $\nu$MSM by a light scalar singlet\footnote{In ref.
\cite{Shaposhnikov:2006xi} this scalar boson was playing the role of
the inflaton. The same mechanism was used in a similar model for
another choice of parameters in \cite{Kusenko:2006rh,Petraki:2007gq}.
}, interacting with the dark matter sterile neutrino.  The decays of
this scalar field  provide an efficient mechanism for the production
of dark matter particles, which is in perfect agreement with all
astrophysical constraints. 

In \cite{Shi:1998km} it was shown that the assumption (iii) is also
crucial. Namely, Shi and Fuller (SF) demonstrated that large lepton
asymmetries can boost the transitions between active and sterile
neutrinos leading to a possibility of resonant creation of dark
matter sterile neutrinos, satisfying both the Lyman-$\alpha$ and
X-ray constraints even if (i) and (ii) are correct. Qualitatively,
the presence of a lepton asymmetry changes the dispersion relation
for active neutrinos in a way that it intersects with the dispersion
relation for the sterile neutrino at some particular momentum. The
level crossing leads to a transfer of the leptonic excess in active
neutrinos to the sterile ones, so that the dark matter abundance is
roughly proportional to the lepton asymmetry.

However, for the mechanism to work, the required lepton asymmetry
must exist at temperatures ${\cal O}(1)$ GeV and must be much larger
than the baryon asymmetry, $\frac{\Delta L}{\Delta B}\gsi 3\times
10^5$, \cite{Shi:1998km,Laine:2008pg}. At the same time, in the majority
of the models with baryon and lepton number violation, proposed so
far, the lepton asymmetry is of the same order of magnitude as the
baryon asymmetry. The reason is that the source of lepton number
violation is associated with an energy scale which is of the order or
greater than the electroweak scale $M_W$. For example, in Grand
Unified Theories the baryon and lepton numbers are broken at the
scale of the order of $M_{\rm GUT} \sim 10^{15}$ GeV, in see-saw
models the masses of singlet Majorana fermions are, as a rule,
greater than $10^9$ GeV. In the SM (or in its supersymmetric
extension), the breaking of lepton and baryon numbers is related to
anomaly, without any other violation terms. In these models
baryogenesis takes place at temperatures $T > M_W$, and  the
equilibrium character of sphaleron processes ensures the relation
$B=\sigma L$ \cite{Kuzmin:1987wn} -
\cite{Laine:1999wv}, where $\sigma$ is a coefficient of the order of
one, depending on the particle content of the Standard Model or its
extensions\footnote{A breakdown of the relation $L\simeq B$ may happen
in Affleck-Dine baryogenesis \cite{Affleck:1984fy}, if it takes place
below the electroweak scale and if because of some reason the decay
of squark-slepton condensate produces considerably more leptons than
quarks.}.  A lepton asymmetry so small is irrelevant for the Big Bang
Nucleosynthesis (BBN)  and for dark matter sterile neutrino
production. In conclusion, the existence of large lepton asymmetries
seems to be very unlikely, if not impossible. 

In this paper we will show that this is not necessarily the case in
the $\nu$MSM. Indeed, the $\nu$MSM is very different from the models
mentioned above.  In particular, the energy scale of the breaking of
lepton number $L$, existing due to Majorana neutrino masses of singlet
fermions, is small (below the electroweak scale), whereas the only
source for baryon number ($B$) violation is the electroweak chiral
anomaly. We will see that these facts change the situation so that the
generation of (large) leptonic asymmetries becomes possible.
Basically, the baryon asymmetry of the universe is related to the
lepton asymmetry at the temperature of the sphaleron freeze-out
$T_{EW}$, and the lepton asymmetry generation {\em below} $T_{EW}$
leaves no trace on baryon asymmetry. The requirement that large enough
lepton asymmetry is generated below the electroweak scale puts a
number of stringent constraints on the properties of the singlet
fermions, which can be tested in a number of accelerator experiments,
discussed in  \cite{Gorbunov:2007ak}.

Motivated by the fact that large low temperature lepton asymmetries
can be a consequence of the $\nu$MSM, in an accompanying paper
\cite{Laine:2008pg} we reanalyze the SF mechanism for production of dark
matter sterile neutrinos in a charge asymmetric medium. This can be
rigorously done  with the use of the formalism of \cite{Asaka:2006rw}
that allows the computation of the abundance of dark matter neutrinos
from first principles of statistical mechanics and quantum field
theory. In particular, we find the spectra of dark matter neutrinos
which can be used in warm dark matter simulations, in the subsequent
Lyman-$\alpha$ analysis and for the study of core profiles of dwarf
spheroidal galaxies. In \cite{Laine:2008pg} we also establish a lower
bound on the leptonic asymmetry $\Delta \equiv \Delta L/L \equiv
(n_L-{\bar n}_L)/(n_L+{\bar n}_L) \gsim 2 \times 10^{-3}$ which is
needed to make the SF mechanism for sterile neutrino production 
consistent with X-ray and Lyman-$\alpha$ observations (here $n_L$ and
${\bar n}_L$ are the number densities of leptons and anti-leptons
correspondingly;   other conventions to characterize the presence of a
non-zero lepton asymmetry are described in Appendix A of
\cite{Laine:2008pg}).  Only this result from \cite{Laine:2008pg} will be
used in the present paper. 

Our findings concerning the parameter-space of the $\nu$MSM, leading
to correct baryon asymmetry and dark matter abundance, leads us to the
fourth question we address in this paper. Namely, we will identify 
different fine-tunings between the parameters of the model, required
for its phenomenological success, and discuss their possible origins.

Throughout the paper we will {\em assume} the validity of the standard
Big Bang theory below temperatures of order $1$~TeV and that the only
relevant degrees of freedom are those of the $\nu$MSM (i.e. of the
Standard Model plus three singlet fermions). After all, one of the
strong motivations for considering the $\nu$MSM as a theory providing
the physics beyond the SM is the possibility to explain neutrino
oscillations, dark matter, inflation and baryon asymmetry in the
framework of a minimal model, and the creation of a baryon asymmetry
requires the presence of temperatures above the electroweak scale.  We
also {\em assume} that at temperatures well above the electroweak
scale  the concentrations of {\em all} singlet fermions were zero and
that the universe was lepton and baryon charge symmetric at this time.
This type of initial conditions may arise in the $\nu$MSM where
inflaton is associated with the SM Higgs field
\cite{Bezrukov:2007ep}.  Note also that a number of calculations in
this work are on the level of approximate estimates which are
valid within a factor of a few and thus may be refined. However, since
even this analysis happened to be rather involved, we prefer to
postpone the detailed study until the $\nu$MSM gains some direct
experimental support. 

The paper is organised as follows. In \se\ref{se:lagrangian} we review
the Lagrangian of the $\nu$MSM and fix the notation. We also discuss
different contributions to the mass difference of singlet fermions,
essential for baryogenesis and formulate several possible scenarios
for its value. In \se\ref{se:CP} we analyse the structure of
CP-violation in the model and identify the the CP-violating phase that
drives baryogenesis. In \se\ref{se:asymmetry} we set up the master
equations for analysis of kinetics of leptogenesis. In
\se\ref{se:CPeven} we analyse CP-odd deviations from thermal
equilibrium and in  \se\ref{se:CPodd} CP-even perturbations. In
\se\ref{se:generation} we  examine in the mechanism of leptogenesis
via singlet fermion oscillations. We derive constraints on the masses
and couplings of neutral leptons from the requirement that the
produced baryon asymmetry has the observed value and analyse the
question whether large lepton asymmetries, which can boost the dark
matter production, can be generated below the electroweak scale. We
also determine the parameters of the model which allow for the
survival of primordial lepton asymmetries to low temperatures and are
consistent with the observed baryon asymmetry. In \se\ref{se:symmetry}
we discuss the fine-tunings and possible symmetries of the $\nu$MSM
and speculate on the origin of the $\nu$MSM Lagrangian.
\se\ref{se:conclusion} is conclusions, where we summarize the results.

%
\section{The $\nu$MSM and constraints on its parameters}
\la{se:lagrangian}
For our aim it is convenient to use the Lagrangian of the $\nu$MSM in
the parametrisation of Ref. \cite{Shaposhnikov:2006nn}:
\begin{eqnarray}
  {\cal L}_{\nu\rm{MSM}}&=& {\cal L}_0 +  \Delta {\cal L}~,
    \label{lagr}\\
  {\cal L}_0&=&{\cal L}_{\rm{SM}} + 
  \bar{N_I} i \partial_\mu \gamma^\mu N_I
  - (h_{\alpha 2} \,  \bar L_\alpha N_2 \tilde{\Phi}
  + M \bar {N_2}^c N_3 + \rm{h.c.}),\nonumber\\
\Delta {\cal L}&=& 
-  h_{\alpha 3} \,  \bar L_\alpha N_3 \tilde{\Phi}
-  h_{\alpha 1} \,  \bar L_\alpha  N_1 \tilde{\Phi}
  -  \frac{\Delta M_{IJ}}{2} \; \bar N_I{}^c N_J + \rm{h.c.} \,,
  \nonumber 
\end{eqnarray}  
where $N_I$ are the right-handed singlet leptons, $\Phi$ and
$L_\alpha$ ($\alpha=e,\mu,\tau$) are  the Higgs and lepton doublets
respectively, $h$ is a matrix of Yukawa coupling constants, $M$ is
the common mass of two heavy neutral fermions, different elements of 
$\Delta M_{IJ}\ll M$ provide a mass to the lightest sterile neutrino
$N_1$,  responsible for dark matter ($M_1\simeq\Delta M_{11}$) and
produce the small splitting of the masses of $N_2$ and $N_3$, $\Delta
M_{23}=0,~~\tilde\Phi_i=\epsilon_{ij}\Phi^*_j$, and $M$ is taken to be
real.

Another parametrization of the same Lagrangian is related to the mass
basis of Majorana neutrinos. We write
\be
{\cal M} = U \lambda U^T~,
\label{calM}
\ee
where ${\cal M}_{IJ}= M \delta_{I2}\delta_{J3} + M
\delta_{I3}\delta_{J2} + \Delta M_{IJ}$, $\lambda$ is a diagonal
matrix with positive values $\lambda_I$, and $U$ is a unitary matrix.
The values  $\lambda_I^2$ are nothing but eigenvalues of the
hermitean matrix ${\cal M}{\cal M}^\dagger$ which can be diagonalized
with the help of $U$, $\lambda_1\approx M_1$.  

Yet another possibility is to use the basis in which the matrix of
Yukawa couplings $h_{\alpha I}$ is diagonal, 
\be
h={\tilde K}_L f_d  {\tilde K}_R^\dagger
\ee
with $f_d={\rm diag}(f_1,f_2,f_3)$. Definitions of the matrices
${\tilde K}_L$ and ${\tilde K}_R$ can be found in 
\cite{Asaka:2005pn}.

The Yukawa coupling constants of the dark  matter neutrino $N_1$ are
strongly bounded by cosmological considerations \cite{Asaka:2005an}
and by X-ray observations \cite{Boyarsky:2006jm,Boyarsky:2006fg}:  
$\sum_{\alpha,I}|h_{\alpha I}U_{I1}|^2 \lsim 10^{-24}$.  As has been
demonstrated in \cite{Asaka:2005an}, the contribution of the dark
matter sterile neutrino to the masses of active neutrinos via the
see-saw formula
\be
 [\delta M_\nu]_{\alpha\beta} = -
 \frac{{M_D}_{\alpha 1} {M_D}_{\beta 1}}{M_1} \; 
 \label{dmcontr}
\ee
is much smaller than the solar neutrino mass difference,  and can
thus safely be neglected (here ${M_D}_{\alpha 1}= h_{\alpha1} v$ and
$v=174$ GeV is the vacuum expectation value of the Higgs field).
Therefore, we set  $h_{\alpha 1} = 0$ in the following and omit $N_1$
from the Lagrangian.  Note that eq. (\ref{dmcontr}) implies that the
mass of one of the active neutrinos is much smaller
\cite{Asaka:2005an} than the solar mass difference $\sim 0.01$ eV and
can be  put to zero in what follows.

The values of the Yukawa coupling constants $h_{\alpha 2}$ and
$h_{\alpha 3}$ are further constrained by the requirement that the
$\nu$MSM must describe the observed pattern of neutrino masses and
mixings. The following relation must hold:
\be
[M_\nu]_{\alpha\beta} = - h_{\alpha I}h_{\beta J}
\left[\frac{v^2}{{\cal M}_N}\right]_{IJ}~,
\ee
where $M_\nu$ is the mass matrix of active neutrinos, and  we denoted
by  ${\cal M}_N$ the $2\times2$ mass matrix of the second and third
singlet fermion, $[{\cal M}_N]_{IJ}={\cal M}_{IJ}$ for $I,J=2,3$. This
formula can be simplified further by  noting that $N_2,N_3$ must be
highly degenerate in mass in order to  ensure successful
baryogenesis~\cite{Asaka:2005pn,Akhmedov:1998qx}. In fact, any
non-zero mass difference that still remains (it will be discussed
later) is inessential for discussion of masses and mixings of active
neutrinos and can be ignored \cite{Shaposhnikov:2006nn}. We have,
therefore,
\be
[M_\nu]_{\alpha\beta} = -\frac{v^2}{M}
\left(h_{\alpha 2}h_{\beta 3}+h_{\alpha 3}h_{\beta 2}\right)~.
\label{mass}
\ee
As was shown in ref.~\cite{Shaposhnikov:2006nn}, this simplified
situation allows to determine Yukawa coupling constants from the mass
matrix of active neutrinos up to rescaling  $h_{\alpha 2} \rightarrow
h_{\alpha 2}/z, ~~h_{\alpha 3} \rightarrow z h_{\alpha 3}$,
where $z$ is an arbitrary complex number. In addition, one can 
solve for the active neutrino masses explicitly:  
\be
 m \in \{ 0,v^2 [F_2 F_3
\pm |h^\dagger h|_{23}] / M \}~,
\label{mrel}
\ee
where $F_i^2 \equiv [h^\dagger h]_{ii}$. This leads to two
qualitatively different cases,  namely the ``normal hierarchy'',  $
m_1 =0,~m_2 = m_\rmi{sol},~m_3 = m_\rmi{atm}$,  and the ``inverted
hierarchy'',  $ m_1 \approx m_2 \approx m_\rmi{atm},~ m_3 = 0$.  Here
$m_\rmi{sol} \equiv \sqrt{\Delta m_{sol}^2},~~ m_\rmi{atm} \equiv
\sqrt{\Delta m_{atm}^2}$, and  $\Delta m_{sol}^2 \simeq 8.0\times
10^{-5}$ eV$^2$,    $\Delta m_{atm}^2 \simeq 2.5\times 10^{-3}$
eV$^2$ \cite{Strumia:2006db}. Normal hierarchy corresponds to the
case  $ |h^\dagger h|_{23} \approx F_2 F_3$,  and the inverted
hierarchy to the case $ |h^\dagger h|_{23} \ll F_2 F_3$. From here it
follows that 
\be
 \nonumber
 2 F_2 F_3 v^2 /M \simeq \kappa m_{\rmi{atm}}~,
 \label{Ffix}
\ee
where $\kappa=1~(2)$ for normal (inverted) hierarchy, $m_\rmi{atm}
\approx 0.05$~eV. If $F_3$ is taken to be very small,  $F\equiv F_2$
is required to be large to keep the atmospheric mass difference in
the right place. The ratio of Yukawa couplings $F_2$ and $F_3$ will
play an important role in what follows and is denoted by $\epsilon$,
\be
\epsilon = \frac{F_3}{F_2}~.
\label{epsilon}
\ee
Then 
\be
F^2 =\frac{\kappa M m_{atm}}{2 \epsilon v^2}. 
\ee
In the limit $\epsilon\rightarrow 0,~\Delta M_{IJ}\rightarrow 0$ the
Lagrangian acquires the global leptonic U(1) symmetry
\cite{Shaposhnikov:2006nn} which guarantees the degeneracy of the
$N_2$ and $N_3$ of singlet fermions (necessary for baryogenesis),
absence of mass for $N_1$ and absence of interactions of $N_1$ with
active fermions (providing thus an approximate description of the
required  parameters of the dark matter sterile neutrino). This
symmetry can be made explicit by introducing the 4-component 
Dirac spinor $\Psi = N_2 +N^c_3$ unifying a pair of two degenerate
Majorana fermions.

For the discussion of the baryon and lepton asymmetries of the
universe an essential parameter is the mass difference $\delta M$
between the mass eigenstates of the two heaviest neutrinos
\cite{Asaka:2005pn,Akhmedov:1998qx}. Indeed,  successful baryogenesis
can take place provided  the  mass difference is small enough. 
In the theory defined by the Lagrangian (\ref{lagr}) there are two
sources for the mass difference: the first one is related to the 
Majorana mass matrix, and the second one is due to the Higgs vacuum
expectation value and Yukawa couplings to active fermionic flavours.
The mass square difference of the physical states at the leading order
of perturbation theory with respect to Yukawa couplings and $\Delta
M_{IJ}$  is given by
\be
\delta M = \frac{|m^2|}{M}~,
\label{zeroTmd}
\ee
where
\be
m^2 \equiv 2(h^\dagger h)_{23} v^2 + M(\Delta M_{22}^*+\Delta M_{33})~.
\label{mdef}
\ee
The one-loop corrections to this result are of the order of
$\frac{\Delta m_\nu}{16\pi^2}\frac{M^2}{v^2}$. 

One can distinguish three essentially different situations, depending
on the relative importance of the mass difference induced by the
Higgs field and the difference associated with Majorana masses.  

In the first one $\Delta\lambda=\lambda_3-\lambda_2$ is negligibly
small so that the mass difference is entirely due to the Higgs
condensate. One can easily find in this case from
(\ref{mrel},\ref{mdef}) that the mass difference of heavy neutrinos is
the same as that of active neutrinos,
\be
\delta M = \Delta m_\nu~,
\label{deltam}
\ee
and is given by $\delta M = m_{\rmi{atm}} - m_{\rmi{sol}}\simeq 0.04$
eV for the case of normal hierarchy or  $\delta M \simeq 
\Delta m_{\rmi{sol}}^2/2m_{\rmi{atm}}\simeq 8\times10^{-4}$ eV for the case of
inverted. Quite amazingly, these mass differences are roughly those
at which the production of baryon asymmetry is extremized, $M \delta
M \simeq \frac{M_W^3}{M_{Pl}}$, where $M_{Pl}$ is the Planck mass,
$M_{Pl}=1.22\times10^{19}$ GeV (see below for a more detailed
discussion). We will refer to this  situation as {\bf Scenario I} for
singlet fermion mass difference.

The second option is when the mass differences coming from two
different sources are of the same order of magnitude. We will call
this choice of parameters {\bf Scenario II}. An extreme case is an
exact compensation of the two leading contributions, which would
allow to have a (low temperature) mass difference be much smaller
than the active neutrino mass difference. To realize this fine
tuning, the following condition is required to hold:
\be
\delta M \ll \Delta m_\nu ~.
\label{finetun}
\ee
Though this possibility may be considered  to be bizarre (as it 
requires that contributions of seemingly different nature are exactly
the same in magnitude and different in sign) it must not be discarded
since the origin of different terms in the $\nu$MSM is unknown, so
that eq.  (\ref{finetun}) could be a consequence of some underlying
(Planck scale?) physics. We will call this option {\bf Scenario IIa}
and discuss this fine-tuning in more detail in 
\se\ref{se:conclusion}.

Yet another possibility is when $\Delta\lambda \gg \Delta m_\nu$ so
that the mass difference is entirely due to the Majorana masses. We
will refer to this possibility as {\bf Scenario III}.

For all three cases the mass eigenstates are related to the fields 
$N_{2,3}$ by a rotation with the maximal angle $\pi/4$ (up to complex
phases and corrections $\sim{\cal O}(M_D/M)$).

%
\section{The structure of CP violation in the $\nu$MSM}
\la{se:CP}
The generic Lagrangian  (\ref{lagr}) contains a number of new physical
parameters (18) in comparison with the Standard Model. They can be
counted as follows: 3 Majorana masses of singlet fermions, 3 Dirac
masses, 6 mixing angles and 6 CP-violating phases. As we have already
mentioned,  one of the singlet fermions, playing the role of dark
matter, has very small Yukawa couplings, and thus is irrelevant for
baryogenesis and for active neutrino mixing matrix.  Thus, the number
of parameters of the $\nu$MSM, responsible for physics of heavier
singlet leptons is smaller. The aim of this Section is to identify
these parameters and to find the CP-violating phase, relevant for
baryogenesis.

After dropping the dark matter sterile neutrino $N_1$  from the 
Lagrangian (\ref{lagr}) the theory contains 11 new parameters in
comparison with the SM. These are 2 Majorana masses, 2 Dirac masses, 4
mixing angles and 3 CP-violating phases. In principle, all these 11
physical parameters of the $\nu$MSM can be determined experimentally
by the detailed study of decays of the singlet fermions. At the same
time, the mass matrix of active neutrinos in this case (note that one
of neutrinos is massless in this approximation) depends on 7
parameters\footnote{We use the notations and  parametrisation of ref. 
\cite{Strumia:2006db}.}: 3 mixing angles $\theta_{12},~\theta_{23}$
and $\theta_{13}$, one Dirac phase $\phi$, one Majorana phase
($\alpha$ in the normal hierarchy case and the combination
$\zeta=(\alpha-\beta)/2$ in the inverted hierarchy) and two masses
($m_2,~m_3$ for normal hierarchy, and $m_1,~m_2$ for  inverted). So,
the 11 new parameters of the $\nu$MSM can be mapped to 7 parameters of
the active neutrino mass matrix plus 4 extra ones. It is convenient to
select these 4 parameters as follows. The first one is the average
Majorana mass of singlet fermions $N_2$ and $N_3$,
\be
M = \frac{\lambda_3 + \lambda_2}{2}~.
\ee
The second is related to the diagonal elements of the Majorana mass
matrix of singlet fermions, which after phase transformations without
loss of generality can be written as
\be
\Delta M_M =\frac{\lambda_3 - \lambda_2}{2}=\Delta M_{22}=\Delta M_{33}~.
\ee
The third one is the parameter $\epsilon$ defined in (\ref{epsilon}),
and the fourth is an extra CP-violating phase $\eta$, associated with
it (see below).

With these notations, the relevant part of the $\nu$MSM Lagrangian is:
\ba
\nonumber
L_{\rm singlet}&=&\left(\frac{\kappa M m_{atm}}{2 v^2}\right)^{\frac{1}{2}}
\left[\frac{1}{\sqrt{\epsilon e^{i\eta}}}\bar L_2 N_2 +
\sqrt{\epsilon e^{i\eta}}\bar L_3 N_3\right] \tilde{\Phi}\\
&-& M \bar {N_2}^c N_3 - 
\frac{1}{2}\Delta M_M (\bar {N_2}^c N_2 + \bar {N_3}^c N_3)+ \rm{h.c.} \,,
\label{pmm}
\ea
where $L_2$ and $L_3$ are the combinations of $L_e,~L_\mu$ and
$L_\tau$
\be
L_{2}=\frac{\sum_\alpha h_{\alpha 2}^*L_\alpha}{F_2}~,~~~~
L_{3}=\frac{\sum_\alpha h_{\alpha 3}^*L_\alpha}{F_3}~.
\label{L23def}
\ee
We stress that the relation between $L_{2,3}$ and
$L_{e,\mu,\tau}$ is not unitary, in general.  In fact, there are 4
different relations between $L_{2,3}$ and leptonic flavours for each
type of hierarchy \cite{Shaposhnikov:2006nn} (some of them may be
equivalent to each other after phase redefinition of leptonic
flavours), leading to  one and the same active neutrino mixing matrix.
We will present and analyse below only one of them for each hierarchy,
others can be treated in a similar way.

To simplify the analysis, we consider the case (suggested by
experiments) when the angle $\theta_{13}$ and deviation of
$\theta_{23}$ from its maximal value, $\delta\theta_{23}=\theta_{23}
-\frac{\pi}{4}$ are small. Let us introduce the notations
\ba
\nonumber
D_1&=&
\delta\theta_{23}\cos\theta_{12}+\theta_{13}\sin\theta_{12}e^{i\phi}~,~~~~
D_2=
\delta\theta_{23}\cos\theta_{12}-\theta_{13}\sin\theta_{12}e^{i\phi}~,\\
D_3&=&
\delta\theta_{23}\sin\theta_{12}+\theta_{13}\cos\theta_{12}e^{i\phi}~,~~~~
D_4=
\delta\theta_{23}\sin\theta_{12}-\theta_{13}\cos\theta_{12}e^{i\phi}~.
\ea

Then, for the normal hierarchy case we can write:
\ba
\nonumber
L_2&=& 
+a_1(1+z_1+z_2)\frac{L_\mu-L\tau}{\sqrt{2}}
+a_2(1+z_2-z_1)L_e 
+a_3\frac{L_\mu+L\tau}{\sqrt{2}}~,\\
L_3&=& 
-a_1(1-z_1-z_2)\frac{L_\mu-L\tau}{\sqrt{2}} 
-a_2(1-z_2+z_1)L_e
+a_3\frac{L_\mu+L\tau}{\sqrt{2}}~,
\label{L23norm}
\ea
where the main terms are
\ba
\nonumber
a_1&=&i e^{-i(\alpha+\phi)} \sin\rho\cos\theta_{12}~,\\
\nonumber
a_2&=&i e^{-i\alpha}        \sin\rho\sin\theta_{12}~,\\
\label{normain}
a_3&=&\cos\rho~,\\
\nonumber
\tan\rho&=& \sqrt{\frac{m_2}{m_3}}\simeq 
\left(\frac{\Delta m^2_{\rm sol}}{\Delta m^2_{\rm
atm}}\right)^{\frac{1}{4}}\simeq 0.4~.
\ea
The corrections are given by
\ba
\label{norcor}
z_1&=&-i D_4^* e^{i(\alpha+\phi)}\frac{\cot\rho}{\sin 2\theta_{12}}~,\\
\nonumber
z_2&=&+i D_3^* e^{i(\alpha+\phi)}\frac{\cot\rho}{\sin 2\theta_{12}}-
e^{-i(\alpha+\phi)}D_1 \tan\rho ~.
\ea

For the inverted hierarchy the corresponding equations are:
\ba
\nonumber
L_2&=& 
+i e^{-i\phi} b_1(1+t_1^*-t_2^*)\frac{L_\mu-L\tau}{\sqrt{2}}
+             b_2(1+t_1  +t_2  )L_e 
+(z_3-t_3) \frac{L_\mu+L\tau}{\sqrt{2}}~,\\
L_3&=&
-i e^{-i\phi}b_2^*(1-t_1^*-t_2^*)\frac{L_\mu-L\tau}{\sqrt{2}} 
+            b_1^*(1-t_1  +t_2  )L_e
+(z_3+t_3)\frac{L_\mu+L\tau}{\sqrt{2}}~,
\label{L23inv}
\ea
where the main terms are
\ba
\nonumber
b_1&=&\frac{1}{\sqrt{2}}
\left[\cos\theta_{12}e^{-i\zeta}+i\sin\theta_{12}e^{+i\zeta}\right]~~~,\\
\label{invmain}
b_2&=&\frac{1}{\sqrt{2}}
\left[\cos\theta_{12}e^{+i\zeta}+i\sin\theta_{12}e^{-i\zeta}\right]~,
\ea
and the corrections are given by
\ba
\nonumber
t_1=\delta_{\rm inv}
\frac{2i\cos2\zeta\sin2\theta_{12}}{3+\cos4\zeta+2\sin^2
2\zeta\cos4\theta_{12}}~,\\
\label{invcor}
t_2=\delta_{\rm inv}
\left[\frac{1}{2}-\frac{1}{1+e^{-4i\zeta}\tan^2\theta_{12}}\right]~,\\
\nonumber
z_3= D_4 e^{+i(\zeta-\phi)},~~~t_3=i  D_1
e^{-i(\zeta+\phi)}~,
\ea
where
\be
\delta_{\rm inv} = \left(\frac{m_2-m_1}{m_2+m_1}\right)\simeq
\frac{\Delta m^2_{\rm sol}}{4\Delta m^2_{\rm atm}}\simeq
8\times10^{-3}~.
\label{deltainv}
\ee 

In general, baryon asymmetry of the universe in the $\nu$MSM may
depend on all three CP-violating phases described above. However, in a
specific limit, when all charged lepton Yukawa couplings are the same,
the baryogenesis is driven by a single phase, which we identify below.
This can be done on general grounds and does not require complicated
computation. 

If the Yukawas in the charged sector are the same, one can choose
a basis for leptonic doublets, in which interactions of singlet
fermions have a simple form
\be
(f_2 \bar l_2 N_2 +f_3 \bar l_3 N_3 +f_{23}\bar l_2 N_3)\tilde{\Phi}~,
\label{Yusi}
\ee
where $l_{1,2,3}$ are related to $L_{e,\mu,\tau}$ by some {\em
unitary} transformation. Now, by the phase redefinition of $l_2$ and
$l_3$ the constants $f_2$ and $f_3$ can be made real. Then, in this
parametrization, the CP-violation effects must be proportional to the
complex phase of the coupling $f_{23}$, they must vanish in the limit 
$f_{23}\to 0$. It is easy to see that 
\be
f_{23}=\frac{[h^\dagger h]_{23}}{F}~.
\ee

Now, with the use of eqns. (\ref{normain},\ref{norcor}) we get for the
normal hierarchy :
\be
f_{23} = \epsilon F e^{i\eta}\cos2\rho\left[1-2i\tan\rho
\left(\delta\theta_{23}\cos(\alpha+\phi)
\cos\theta_{12} +
\theta_{13}\cos\alpha\sin\theta_{12}\right)\right]~.
\label{CPnorm}
\ee
The similar expression for the inverted hierarchy is obtained with the
use of (\ref{invmain},\ref{invcor}):
\be
f_{23} =- \epsilon F e^{i\eta}
\left[\delta_{\rm inv}+\frac{1}{2}
\left(t_3^*-z_3^*\right)\left(t_3+z_3\right)\right]~.
\label{CPinv}
\ee

Yet another parameter in (\ref{pmm}) which is important for the issue
of CP-violation is the mass splitting $\Delta M_M$. Indeed, in the
limit   $\Delta M_M=0$ (the extreme case of {\bf Scenario I}), the
Lagrangian (\ref{Yusi}) acquires the global U(1) symmetry, so that
the  phases of $N_{2,3}$ cannot be fixed anymore by the mass terms,
and the CP phase of $f_{23}$ in the interaction (\ref{Yusi}) can be
rotated away. This means that the measure of CP-violation, relevant
for baryogenesis for the case when all charged lepton Yukawa couplings
are the same, can be conveniently parametrised as
\be
\delta_{CP}^0 = \epsilon \sin(\arg f_{23})\sin \theta~,
\label{deltaCP0}
\ee
where
\be
\tan \theta = \frac{\Delta M_M}{\Delta m_\nu}~.
\label{thM}
\ee

In reality the charged lepton Yukawa couplings are different, and the
structure of CP-breaking relevant for baryogenesis is more involved.
It has been found in  \cite{Asaka:2005pn} for the case when all
reactions of singlet fermions are out of thermal equilibrium, see eq.
(29) of that paper\footnote{The fact that this result does not depend
on Yukawa couplings of charged fermions does not mean that they can be
neglected. Indeed,  the computation of baryon asymmetry takes into
account that {\em all} reactions of the particles of the SM
equilibrate, and the fact that charged Yukawas are non-zero and
different is essential for kinetic description of the system with the
use of eqs. (\ref{kineq1},\ref{kineq2},\ref{kineq3}).}. For
convenience, we present it here in notations of our work, factoring
out the largest Yukawa coupling $F$:
\be
\delta_{CP}=\frac{1}{F^6}\left[{\rm Im}[h^\dagger h]_{23}
\sum_\alpha \left(|h_{\alpha 2}|^4-|h_{\alpha 3}|^4\right)
-\left(F_2^2-F_3^2\right)
\sum_\alpha \left(|h_{\alpha 2}|^2+|h_{\alpha
3}|^2\right){\rm Im}[h^*_{\alpha 2} h_{\alpha 3}]\right]
\label{deltaCP}~.
\ee
With the use of relations (\ref{L23def},\ref{L23norm},\ref{L23inv}),
the expression (\ref{deltaCP}) can be rewritten through the parameters
of the neutrino mixing matrix, the value of $\epsilon$ and phase
$\eta$. The corresponding relations are not very illuminating, so that
we just summarize the main qualitative features of (\ref{deltaCP}).\\
(i) The sign of baryon
asymmetry cannot be found even if the active neutrino mixing matrix is
completely known.\\ 
(ii) For small $\epsilon$,  $\delta_{CP}\propto \epsilon$, similar to
eq. (\ref{deltaCP0}).\\
(iii) Baryon asymmetry is non-zero even if $\theta_{13}=0$ and 
$\delta\theta_{23}=0$. In other words, the details of the active
neutrino mass matrix have little influence on baryogenesis, if these
angles are small.\\
(iv) Baryon asymmetry does not vanish even if $\tan \theta=0$, i.e.
in the {\bf Scenario I} for singlet fermion mass differences. In this
case the result is determined by the parameters of the neutrino mixing
matrix only.\\
(v) In general, $\delta_{CP}\neq 0$ for $\epsilon = 1$.\\
(vi) For inverted hierarchy case, $\delta_{CP}\neq 0$ even if one takes
a limit $m_1=m_2,~\theta_{13}=0,~\delta\theta_{23}=0$.

The computation, leading to eq. (\ref{deltaCP}), cannot be applied to
the low temperature leptogenesis that occurs at temperatures in the GeV
range (see \se\ref{se:generation}), because the mass of $\tau$-lepton
is comparable with a relevant temperature and with the mass of singlet
leptons $N_{2,3}$. In this case the charged lepton masses certainly
cannot be neglected, and the structure of CP-violation is even richer
than that above the electroweak temperature. Still, CP-violating
effects are proportional to $\epsilon$, since in the limit $\epsilon
\to 0$ all couplings in (\ref{pmm}) can be made real.

For numerical estimates of CP-violating effects in the paper we will
assume that the relevant CP-violating phase is of the order of one, so
that the effects are suppresses by $\epsilon$, writing explicitly this
factor in the formulas.
%
\section{Lepton asymmetry generation: review of theoretical framework}
\la{se:asymmetry}
The detailed description of the system of singlet leptons and active
fermions in the early universe is necessarily quite complicated. The
number of relevant zero-temperature degrees of freedom (3 active and
3 sterile neutrinos and their antiparticles) is large\footnote{In
fact, the number of types of particle excitations in high temperature
plasma is even higher, but we will assume in this paper that only
zero-temperature degrees of freedom are relevant.}, and the
time-scales of different processes can vary by many orders of
magnitude. Moreover, due to the smallness of the sterile-active
Yukawa couplings the processes with singlet fermions have in general
a coherent character, making the approach based on Boltzmann equation
for particle concentrations useless. 

Probably, the simplest way to deal with coherent effects is to use
the equation for the density matrix \cite{Dolgov:1980cq,Sigl:1992fn,
Akhmedov:1998qx,Asaka:2005pn}. In our case this is a $12 \times 12$
matrix ($12=3\times 2\times 2$ degrees of freedom for all active and
sterile neutrino states), satisfying the kinetic equation  (11) of
\cite{Asaka:2005pn}:
\be
i\frac{d\rho}{dt} =
[H,\rho]-\frac{i}{2}\{\Gamma,\rho\}+\frac{i}{2}\{\Gamma^p,1-\rho\}~,
\label{cineq}
\ee
where $H=p(t)+H_0+H_{int}$ is the Hermitean effective Hamiltonian
incorporating the medium effects on neutrino propagation, $p(t)$ is
the neutrino momentum, with $\langle p(t)\rangle \sim 3T$ (we will
assume that all the neutral fermion masses are much smaller than the
temperature), $H_0 = \frac{M^2}{2p(t)}$ (we include $\Delta M_{IJ}$ to
$H_{int}$), $\Gamma$ and $\Gamma^p$ are the Hermitean matrices
associated with destruction and production rates correspondingly, and
$[~,~]$ ($\{~,~\}$) corresponds to the commutator  (anti-commutator)
\footnote{We stress that eq.(\ref{cineq}), though it looks identical
to eq. (1) of the earlier work \cite{Akhmedov:1998qx}, is in fact very
different. In ref. \cite{Akhmedov:1998qx} $\rho$ is a $3\times 3$
matrix, associated with singlet fermions only, whereas
eq.(\ref{cineq}) accounts for all leptonic degrees of freedom.}.
Following refs.  \cite{Akhmedov:1998qx,Asaka:2005pn} we will use the
Boltzmann statistics for estimates and replace the last term in
(\ref{cineq}) by $i \Gamma^p$. Also, following \cite{Asaka:2005pn} we
will replace  $\Gamma^p$ by $\frac{1}{2}\{\Gamma,\rho^{eq}\}$ with
$\rho^{eq}=\exp{(-p/T)}$ being an equilibrium diagonal density matrix,
ensuring the correct approach to thermal equilibrium. After these
substitutions the kinetic equation takes a simple form 
\be
i \frac{d\rho}{dt}= [H, \rho]
-\frac{i}{2}\{\Gamma, \rho - \rho^{eq}\}~.
\label{NN}
\ee
This is a relaxation time approximation for the density matrix,
fairly standard one in non-equilibrium statistical physics. 

This equation can be simplified even further (for details see
\cite{Asaka:2005pn}) accounting for the following facts: \\
(i) The rates of interactions between active neutrinos are much
higher that the rate of the universe expansion. Therefore, coherent
effects for active neutrinos are not essential and the part of the
general density matrix $\rho$ related to active leptonic flavours can
be replaced by equilibrium concentrations characterised by 3
dimensionless chemical potentials $\mu_\alpha$ (the ordinary chemical
potential divided by the temperature) giving the leptonic asymmetry
in each flavour.\\
(ii) Active neutrinos get temperature dependent masses that are quite
different from those of singlet fermions. Therefore, all non-diagonal
elements of the density matrix involving simultaneously the  active
and sterile states can be put to zero.\\
(iii) The coupling of the dark matter neutrino is so weak that it
decouples from the system.  

This leaves us with the $2\times2$ density matrix $\rho_N$ for singlet
fermions $N_2$ and $N_3$, charge conjugated density matrix
$\bar{\rho}_N$ for corresponding antiparticles (or, to be more
precise, opposite chirality states), and 3 chemical potentials
$\mu_\alpha$. The corresponding equations can be written as
\cite{Asaka:2005pn}:
\ba
\label{kineq1}
i \frac{d\rho_N}{dt}&=& [H, \rho_N]
-\frac{i}{2}\{\Gamma_N, \rho_N - \rho^{eq}\} +
i \mu_\alpha{\tilde\Gamma^\alpha}_N~,\\
i \frac{d\bar\rho_N}{dt}&=& [H^*, \bar\rho_N]
-\frac{i}{2}\{\Gamma^*_N, \bar\rho_N - \rho^{eq}\} -i \mu_\alpha{
\tilde\Gamma^{\alpha *}}_N~,
\label{kineq2}\\
i \frac{d\mu_\alpha}{dt}&=&-i\Gamma^\alpha_L\mu_\alpha +
i {\rm Tr}\left[{\tilde \Gamma^\alpha}_L(\rho_N -\rho^{eq})\right] -
i {\rm Tr}\left[{\tilde \Gamma^{\alpha*}}_L(\bar\rho_N -\rho^{eq})\right]
~.
\label{kineq3}
\ea 
In the equation for $\mu_\alpha$ there is no summation over $\alpha$
and $\Gamma^\alpha_L$ are real. The explicit expressions for the
matrices describing different equilibration rates
$(\Gamma_N,~{\tilde\Gamma^\alpha}_N,~  \Gamma^\alpha_L,~{\tilde
\Gamma^\alpha}_L)$ via Yukawa coupling constants can be found in
\cite{Asaka:2005pn} for the case when the temperature is higher than
the electroweak scale. They are all related to the absorptive parts of
the two point functions for active or sterile neutrino states and
contain a square of Yukawa couplings $h_{\alpha I}$.  The real parts
of the corresponding graphs together with mass squared difference
between $N_2$ and $N_3$ determine the effective Hamiltonian $H$. For
high temperatures $T\gsim T_{EW}$  the equilibration processes are
associated with Higgs, $W$ and $Z$ decays  to singlet and active
fermions, to corresponding inverse processes, and to $t\bar t\to
N\bar\nu$ scattering ($t$ is the top-quark). At smaller temperatures
$T\lsim T_{EW}$ the rates are associated with $W$ and $Z$ exchange and
singlet-active mixing through the Higgs vev, see Fig. \ref{fig:prod}.

\begin{figure}[tb]
\centerline{%
\epsfysize=8.0cm\epsfbox{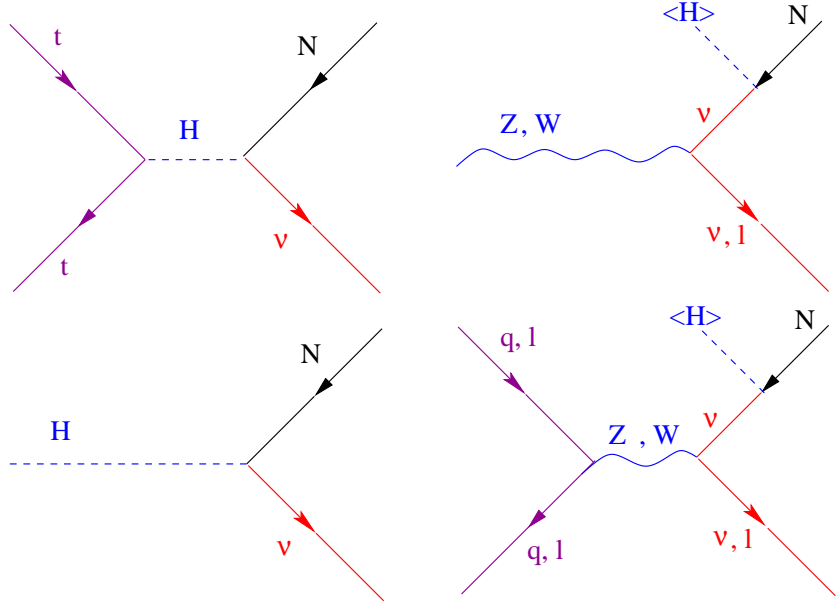}%
}
\caption[a]{\small Diagrams for the processes which contribute to
equilibration rates.} 
\la{fig:prod}
\end{figure}

In the earlier work \cite{Akhmedov:1998qx} the computations and
qualitative discussion were based on incomplete kinetic equations,
which did not include the last terms in eqns. (\ref{kineq1}) and
(\ref{kineq2}), as well as eq. (\ref{kineq3}).  As was shown in
\cite{Asaka:2005pn}, these terms are absolutely essential for the
analysis of lepton asymmetry generation. Therefore, all the {\em
quantitative} results for baryonic excess of 
\cite{Asaka:2005pn,Shaposhnikov:2006nn} and of the present paper are
different from those of \cite{Akhmedov:1998qx}. If some of the {\em
qualitative} conclusions happen to be the same,  we cite both papers
\cite{Akhmedov:1998qx} and \cite{Asaka:2005pn} simultaneously.

The eqs. (\ref{kineq1},\ref{kineq2},\ref{kineq3}), supplemented by an
initial condition $\rho_N= \bar\rho_N = \mu_\alpha =0$ which may be 
fixed by inflation, provides a basis for the analysis of the lepton
asymmetry generation \cite{Asaka:2005pn}. In that paper an analytic
perturbative  expression for lepton asymmetry has been derived. The
expression (33) of \cite{Asaka:2005pn} is valid provided the following
requirements are met:\\ 
(i) All reactions, in which the singlet fermions participate, are out
of thermal equilibrium above the sphaleron freezing temperature. In
this case a straightforward perturbation theory on Yukawa couplings of
singlet fermions can be used.  If this requirement is not satisfied, a
perturbative expansion contains the so-called secular terms, which
diverge with time and require resummations. It is one of the aims of
the present paper to find out what happens if the singlet fermions
equilibrate before the sphalerons decouple.\\ 
(ii) The mass difference between singlet fermions is sufficiently
large so that the Higgs field contribution to it can be neglected.  In
other words, only {\bf Scenario III} was considered. Naturally, we
would like to extend the analysis of baryogenesis to {\bf Scenarios I,
II}, and {\bf IIa}.\\ 
(iii) The number of oscillations  of singlet fermions, related to
their mass difference (see exact definition in eq. (\ref{x})) is much
greater than one at the time of the electroweak cross-over. In this
case the baryogenesis occurs in the symmetric high temperature phase
of the SM and the Higgs vacuum expectation value does not play any
role. We would like to understand what happens if this assumption is
not satisfied, which is the case, in particular, in {\bf Scenarios I,
II}, and {\bf IIa}.

To address all these questions we make a number of helpful
transformations of kinetic equations
(\ref{kineq1},\ref{kineq2},\ref{kineq3}). In particular, a further
simplification of the system (\ref{kineq1},\ref{kineq2},\ref{kineq3})
can be made under assumption that the CP-violating effects are small.
Let us introduce the CP-odd ($\rho_-$) and CP-even deviations
($\rho_+$) from thermal equilibrium by writing
\be
\rho_N-\rho^{eq}=\delta\rho_+ + \frac{\delta\rho_-}{2}~,~~~
\bar \rho_N-\rho^{eq}=\delta\rho_+ - \frac{\delta\rho_-}{2}~.
\ee 
and neglect in (\ref{kineq1},\ref{kineq2},\ref{kineq3}) all terms that
are of the second order in CP-odd quantities (such as
$(\Gamma_N-\Gamma_N^*)\delta\rho_-$) etc.  In this approximation  one
can decouple the equations for the  CP-even deviations $\rho_+$ and
get
\be
i \frac{d\delta\rho_+}{dt}= [{\rm Re}~H, \delta\rho_+]
-\frac{i}{2}\{{\rm Re}~\Gamma_N, \delta\rho_+\}~,
\label{even}
\ee
with the initial condition $\delta\rho_+ =-\rho^{eq}$. The equations
for the CP-odd part in this approximation have the form:
\ba
i \frac{d\delta\rho_-}{dt}&=& [{\rm Re}~H, \delta\rho_-]
-\frac{i}{2}\{{\rm Re}~\Gamma_N, \delta\rho_-\} +i\mu_\alpha{{\rm Re}~\tilde
\Gamma}^\alpha_N~+S,
\nonumber\\
i \frac{d\mu_\alpha}{dt}&=&-i \Gamma^\alpha_L\mu_\alpha +
i {\rm Tr}\left[{\rm Re}~{\tilde\Gamma}^\alpha_L\delta\rho_-\right]
+S_\alpha~,
\label{odd}
\ea
with zero initial conditions for $\delta\rho_-$ and leptonic chemical
potentials. Here the source terms $S$ and $S_\mu$ are proportional to
CP breaking parameters and given by:
\ba
S&=&2 i [{\rm Im}~H,\delta\rho_+]+ \{{\rm Im}~\Gamma_N,
\delta\rho_+\}~,\\
S_\alpha&=&-2{\rm Tr}\left[{\rm Im}~{\tilde\Gamma}^\alpha_L 
\delta\rho_+\right]~.
\label{source} 
\ea
They are only non-zero when CP-even deviations from thermal
equilibrium exist, which is a key issue for baryogenesis and
leptogenesis  \cite{Sakharov:1967dj}. At the same time, if different
damping rates in (\ref{odd}) are all  larger than the rate of the
universe expansion after leptogenesis, the created asymmetry
disappears. Therefore, to find whether baryogenesis is possible at
all, one can study first the rates of different processes that
equilibrate  CP-odd and CP-even deviations from thermal equilibrium.
This can be only skipped if {\em all} reactions with singlet fermions
are out of thermal equilibrium, which was the case considered in
\cite{Asaka:2005pn}. If the necessary conditions for baryogenesis are
found to be satisfied, an analysis of the CP-violating effects must
follow.

So, we will consider first the system (\ref{even},\ref{odd})
neglecting all CP-violating effects. To simplify the notations, we
will take away the symbol Re of the real part from the equations.

\section{CP-even deviations from thermal equilibrium}
\label{se:CPeven}
\subsection{High temperature singlet fermion masses and mass eigenstates}
\label{ssec:mass}
The behaviour of the CP-even perturbations is determined by $H$ and
$\Gamma_N$, see eq. (\ref{even}). Let us start from a discussion of
the Hamiltonian $H_{int}$, describing the oscillations.

The Hamiltonian $H_{int}$ has the form
\be
H_{int}=\frac{\Delta M^2(T)}{2p}~,
\label{hint}
\ee
where $\Delta M^2(T)$ is the temperature dependent (non-diagonal) 
matrix of mass differences between singlet fermions. It is
determined by the zero-temperature mass difference and by real parts
of propagator-type graphs for sterile fermions, see Fig.
\ref{fig:deltam}. 
\begin{figure}[tb]
\centerline{%
\epsfysize=6.0cm\epsfbox{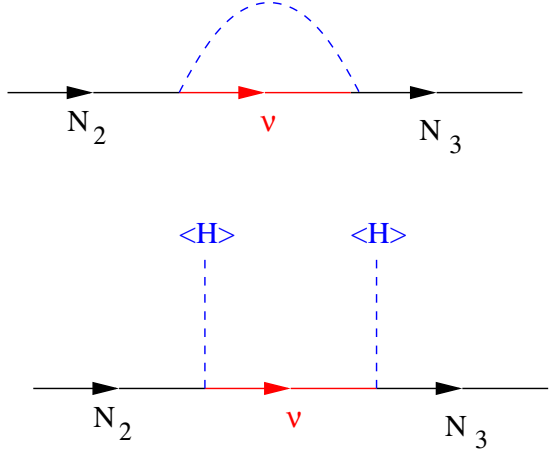}%
}
\caption[a]{\small ``Soft'' contribution to the mass difference of
singlet fermions coming from electroweak spontaneous symmetry
breaking (lower panel) and from radiative correction (upper panel).
Non-zero temperature neutrino propagator has to be used.} 
\la{fig:deltam}
\end{figure}

There are two different temperature-dependent contributions to the
mass difference. The bottom one is proportional to the square of the
temperature dependent vacuum expectation value of the Higgs field
$v(T)$ whereas the top one is proportional to $T^2$, coming from the
Higgs exchange. At temperatures around and below the sphaleron
freezing, interesting to us, the contribution related to the Higgs
vev dominates because of the usual loop suppression and since $v(T)
\gsim T$. We get for the high temperature case $M^2 \ll T^2$:
\be
\Delta M^2(T)_{IJ} \simeq   
\left(
\begin{array}{c c}
0 & m^2(T)\\
{m^*}^2(T) & 0
\end{array}
\right)
- v^2(T)h_{\alpha I}h_{\alpha J}^* \frac{2 b p}{M^2 + 2 b p}~,
\label{massdifft}
\ee
where $m^2(T)$ is determined by eq. (\ref{mdef}) with the replacement $v
\to v(T)$. The function  $b$ is defined by the active
neutrino propagator $1/({\bsl p}+{\bsl\Sigma}_\alpha)$: $\bsl \Sigma
= a \bsl p + b \bsl u$, where $u$ is 4-vector of the medium.
The function $b$ in different limits is given by
\cite{Weldon:1982bn,Notzold:1987ik}:
\be
b =
\left\{\begin{array}{c c}
-\frac{\pi\alpha_W T^2}{8p}
\left(2+\frac{1}{\cos^2\theta_W}\right)~,&~~~ T \gg M_W\\
\frac{16 G_F^2}{\pi\alpha_W}\left(2+\cos^2\theta_W\right)
\frac{7 \pi^2T^4p}{360}~,&~~~ T \ll M_W~.
\end{array}\right.~
\label{bdef}
\ee
This is the so-called potential contribution to active neutrino
dispersion in the medium.

From (\ref{massdifft}) we get for the temperature-dependent mass
difference $\delta M(T)$
\be
\delta M(T)\simeq
\frac{v^2(T)}{2M}\sqrt{\left(F_2^2-F_3^2\right)^2+
4\left|(h^\dagger h)_{23}-\frac{m^2(T)}{v^2(T)}\right|^2}~.
\label{mdiffhigh}
\ee
For small $\epsilon$ and high temperatures this gives in {\bf
Scenarios I, II}:
\be
\delta M \simeq \frac{\kappa m_{atm}}{4\epsilon}\frac{v^2(T)}{v^2}~.
\label{htdm}
\ee

The temperature-dependent contribution to the mass difference is
suppressed in comparison with the zero-temperature one at $2 bp \ll
M^2$. Taking for an estimate the typical momentum of a particle in
high temperature plasma $p \simeq 3T$ one finds that this inequality
is satisfied at 
\be
T \lsim T_{pot} = 13\left(\frac{M}{\rm GeV}\right)^{1/3}~{\rm GeV}~,
\label{mix}
\ee
and that for these temperatures and  $\epsilon \ll 1$ the mass
difference is 
\be
\delta M(T) \simeq\sqrt{\left(\frac{\kappa m_{atm}}{2\epsilon}\frac{b
p}{M^2}\right)^2 
+ \delta M^2(0)}~.
\label{mssdeff}
\ee

Note that depending on parameters $\delta M(T)$ can go through zero at
some particular temperature, leading to level crossing and to the
resonant production of lepton asymmetry, see \se\ref{se:generation}.
As follows from eq. (\ref{mdiffhigh}), this may only happen at $T >
T_{pot}$ if $\epsilon =1$. 

Let us discuss the high temperature mass eigenstates for three
different scenarios of the singlet fermion mass differences (see the
end of \se \ref{se:lagrangian} for definition).

{\bf Scenario I}. In this case $\Delta\lambda=0$ and the mass
difference comes entirely from the interaction with the Higgs field. 

For $\epsilon \ll 1$ the high temperature mass eigenstates $N_2^T$ and
$N_3^T$ are close to $N_2$ and $N_3$,
\ba
N_2^T &\simeq& \cos\beta_0 N_2 + \sin\beta_0 N_3~,
\nonumber\\
N_3^T &\simeq& \cos\beta_0 N_3 - \sin\beta_0 N_2~,
\ea
where 
\be
\beta_0 \simeq
\frac{(h^\dagger h)_{23}}{F^2}~, 
\label{hightmix}
\ee
whereas for $\epsilon \sim 1$ they represent the mixing of $N_2$ and
$N_3$ with the angle of the order of 1. With the use of eq. 
(\ref{Ffix})
the ratio of Yukawa couplings that appear in (\ref{hightmix}) can be
written as
\be
\beta_0\simeq \frac{\epsilon \Delta
m_\nu}{\kappa m_{atm}}\simeq \epsilon
\left\{\begin{array}{c c}1&
{\rm Normal~hierarchy}\\
8\times10^{-3}&{\rm Inverted~hierarchy}~.\end{array}\right.
\ee

It is not difficult to see that for temperatures below $T_{pot}$ the
mixing angle $\beta$ gets modified,
\be
\beta_0 \to \beta\simeq \beta_0\left(1+\frac{M^2}{2b p}\right)~,
\label{betachange}
\ee
leading to (\ref{hightmix}) for $b p \gg M^2$ (this expression is
valid provided $\beta \lsim 1$). Therefore, the temperature
$T_{\beta}$ at which the mixing $\beta$ is of the order of one is
given by
\be
T_{\beta} \simeq \beta_0^{\frac{1}{6}}T_{pot}
\simeq 16~{\rm GeV} \left(\frac{\epsilon\delta
M(0)}{\kappa m_{atm}}\right)^{\frac{1}{6}}
\left(\frac{M}{\rm GeV}\right)^{\frac{1}{3}}~.
\label{tbeta}
\ee
This is derived with the use of eq. (\ref{bdef}) for $T\ll M_W$ and
$\epsilon \ll 1$.

{\bf Scenario II}. In this case both terms in eq. (\ref{massdifft})
have the same order of magnitude and the mixing angle $\beta$ is in
general of the order of one. It goes to the zero-temperature value
$\pi/4$ at temperatures below $T_{\beta}$, see eq. (\ref{tbeta}). For
the {\bf Scenario IIa} at $T>T_\beta$ the mass difference is of the
order of $m_{atm}/\epsilon$ (see eq. (\ref{htdm})) and is much
smaller than $\Delta m_\nu$ at lower temperatures, see eq.
(\ref{mssdeff}). Note that at $T \gsim T_{EW}$ $v(T)\neq v$ and,
therefore, $m(T)\neq 0$ even if $m(0) = 0$.

{\bf Scenario III}. In this case the mass difference comes entirely from
the tree Majorana mass, and the high temperature mixing angle is
always close to $\pi/4$.

\subsection{Damping of CP-even perturbations}
\label{ssec:dampeven}

Let us turn now to the part of equation (\ref{even}) describing
creation and destruction of singlet fermions. To find $\Gamma_N$ we
note that for initial condition $\rho=0,~\mu=0$ one gets for
sufficiently small times from eq. (\ref{even}) that
\be
\frac{d\rho}{dt}\simeq\Gamma_N(\vec{q})~.
\label{gamma}
\ee
At the same time, for the time scales smaller than  $1/\Gamma_N$ but 
much larger than microscopic time scales such as $1/T$, the
derivative $\frac{d\rho}{dt}$ can be found from first principles of
statistical mechanics and quantum field theory as described in
\cite{Asaka:2006rw,Asaka:2006nq}. Therefore, we can use the methods
of these papers to define the rate $\Gamma_N$.

First we note that for the computation of $\Gamma$ the term 
$\frac{\Delta M_{IJ}}{2}\; \bar N_I{}^c N_J$ can be neglected (this
term, however, must be kept in $H_{int}$, as has been done above). In
this case the fields $N_2$ and $N_3$ can be unified in one Dirac
spinor as $\Psi=N_2 + N_3^c$. As usual, $\Psi$ can be decomposed in
creation and annihilation operators as 
\be
 \Psi(x)  = 
 \int\! \frac{{\rm d}^3 \vec{p}}{\sqrt{\raise-0.1ex\hbox{$(2\pi)^3 2 p^0$}}}
 \sum_{s=\pm}
 \Bigl[ 
  \hat{a}^{\mbox{ }}_{\vec{p},s} 
   u(\vec{p},s) e^{-iP\cdot x} + 
  \hat{b}^{\dagger}_{\vec{p},s} 
   v(\vec{p},s) e^{iP\cdot x}   
 \Bigr]~,
 \ee
where the spinors $u,v$ satisfy the completeness relations 
\be 
\sum_s u(\vec{p},s) \bar u(\vec{p},s) = \slash \!\!\!{p}\,\,\, + M,~~ 
\sum_s v(\vec{p},s) \bar v(\vec{p},s) = \slash \!\!\!{p}\,\,\, - M~.
\ee 
The operators $ \hat a^\dagger_{\vec{q},s}$ and  $ \hat
b^\dagger_{\vec{q},s}$ are the creation operators of a singlet
fermion $N_2$ and an anti-fermion $N_3$ with momentum $\vec{q}$, and
helicity state $s$. These operators are normalized as 
\be
 \{ \hat a^{\mbox{ }}_{\vec{p},s} , \hat a^\dagger_{\vec{q},t} \}
 = \delta^{(3)}(\vec{p}-\vec{q})\delta_{st}
 \;, \la{norm}
\ee
and $V$ is the volume of the system. The density matrix
$\rho_N$ is associated with operators
\be
\hat \rho_N=\frac{1}{V}
\left(
\begin{array}{c c}
\hat a^\dagger_{\vec{q},+} \hat a^{\mbox{}}_{\vec{q},+}&
\hat a^\dagger_{\vec{q},+} \hat b^{\mbox{}}_{\vec{q},+}\\
\hat b^\dagger_{\vec{q},+} \hat a^{\mbox{}}_{\vec{q},+}&
\hat b^\dagger_{\vec{q},+} \hat b^{\mbox{}}_{\vec{q},+}
\end{array}
\right)~.
\label{operator}
\ee
Now, repeating literally the discussion of the Section 2 of ref.
\cite{Asaka:2006rw} we arrive at
\be
\Gamma_N^{IJ}(\vec{q})=
  \frac{2 \nF{}(q^0)}{(2\pi)^3 2 q^0}
   \sum_{\alpha = 1}^{3}  
  \tr\Bigl\{ 
  \Pi_{IJ}^\alpha
  a_L 
   \Bigl[
    \rho_{\alpha\alpha}(-Q) + \rho_{\alpha\alpha}(Q) 
   \Bigr]
  a_R
 \Bigr\} 
 \;, \la{master}
\ee
where $\rho$ is the spectral function defined in Appendix B of ref.
\cite{Asaka:2006rw}, and matrices $\Pi_{IJ}$ are given by
\ba
\nonumber
\Pi_{22}^\alpha&=&v^2(T)\left[|h_{\alpha2}|^2 P_u(\vec{p})+
|h_{\alpha3}|^2 P_u^c(-\vec{p})\right]~,\\
\nonumber
\Pi_{23}^\alpha&=& h_{\alpha2}h_{\alpha3}^*v^2(T)
P_u(\vec{p})~,\\
\nonumber
\Pi_{32}^\alpha&=& h_{\alpha2}^*h_{\alpha3} v^2(T)
P_u^\dagger(\vec{p})~,\\
\label{PIJ}
\Pi_{33}^\alpha&=&v^2(T)\left[|h_{\alpha2}|^2 P_v(\vec{p})+
|h_{\alpha3}|^2 P_v^c(-\vec{p})\right]~.
\ea
The spin operators $P_u=u(\vec{p},+) \bar u(\vec{p},+)$ and 
$P_v =v(\vec{p},+) \bar v(\vec{p},+)$ are 
\ba
\nonumber
P_{u,v}&=&\frac{1}{2}(p^0+p)
\left(\gamma^0-\frac{\vec{\gamma}\vec{p}}{p}\right)a_{L,R} +
\frac{1}{2}(p^0-p)
\left(\gamma^0 +\frac{\vec{\gamma}\vec{p}}{p}\right)a_{R,L}~,\\
P_u^c&=&\gamma^2 P_u \gamma^2,~~~P_v^c=\gamma^2P_v\gamma^2,~u^c=v,~
v^c=u,~~p=|\vec{p}|~.
\ea

Let us discuss the structure of the matrix $\Gamma_N^{IJ}$ in more
detail. We will be interested in a total  (integrated over momenta)
rate appearing in eqs. (\ref{kineq1},\ref{kineq2},\ref{kineq3}):
\be
\Gamma_N(T,M) = \frac{1}{T^3}\int d^3q \Gamma_N^{IJ}(\vec{q})~.
\ee
Then the structure of $\Gamma_N$ is:
\be
\Gamma_N =
\frac{F^2}{F_0^2}
\left(
\begin{array}{c c}
R(T,M) + \epsilon^2 R_M(T,M)& \frac{(h^\dagger h)_{23}}{F^2} R(T,M)\\
\frac{(h^\dagger h)_{32}}{F^2} R(T,M)&
\epsilon^2 R(T,M) + R_M(T,M)
\end{array}
\right)~,
\label{gammaNtot}
\ee
where $F_0=2\times 10^{-9}$ is a convenient normalisation constant,
and
\be
R(T,M) = \frac{F_0^2}{F^2 T^3} \int d^3q 
\Gamma_N^{11}(\vec{q}){\vert}_{\epsilon=0}
\ee
can be called the rate of the singlet fermion production at $F=F_0$.
The quantity 
\be
R_M(T,M) = \frac{F_0^2}{F^2 T^3} \int d^3q \frac{q^0-q}{q^0+q}
\Gamma_N^{11}(\vec{q}){\vert}_{\epsilon=0}
\label{eq:RM}
\ee
vanishes in the limit $M \to 0$ and represents the rate of the
processes with violation of total lepton number (to be defined exactly
below).

Computation of   $R_M(T,M)$ and $R(T,M)$ is quite involved and is
discussed in detail in Appendix A. A large number of processes, such
as $W,Z$ and Higgs decays, together with $ 2\to 2$ reactions
incorporating quark and lepton initial and final states must be taken
into account. The result of the computation is presented in Fig.
\ref{fig:Y}. The vertical axis is the temperature $T$,
and the horizontal axis is  the temperature derivative of the yield
parameter, defined in eq.(4.8) of \cite{Asaka:2006nq} :
\be
T \frac{dY}{dT}= - \kappa(T) R(T,M),~~~
T \frac{dY_M}{dT}= - \kappa(T) R_M(T,M),~~~
\kappa(T)=\frac{30 M_0(T)}{4 \pi^2 c_s^2(T) h_{eff}(T) T^2}~,
\label{mT1}
\ee
where $c_s$ is a speed of sound, the temperature-time relation is
given by $t=\frac{M_0}{2T^2},~~ M_0\simeq M_{Pl}/1.66\sqrt{g_{eff}}$,
and the temperature dependence of the numbers of degrees of freedom
$g_{eff}$ and $h_{eff}$ can be taken from \cite{Asaka:2006nq}. The
combination 
\be
\frac{1}{Y_{eq}} T \frac{dY}{dT}
\ee
 is nothing but the ratio of the singlet fermion production rate to
the Hubble constant.
\begin{figure}[tb]

\centerline{%
\epsfysize=8.0cm\epsfbox{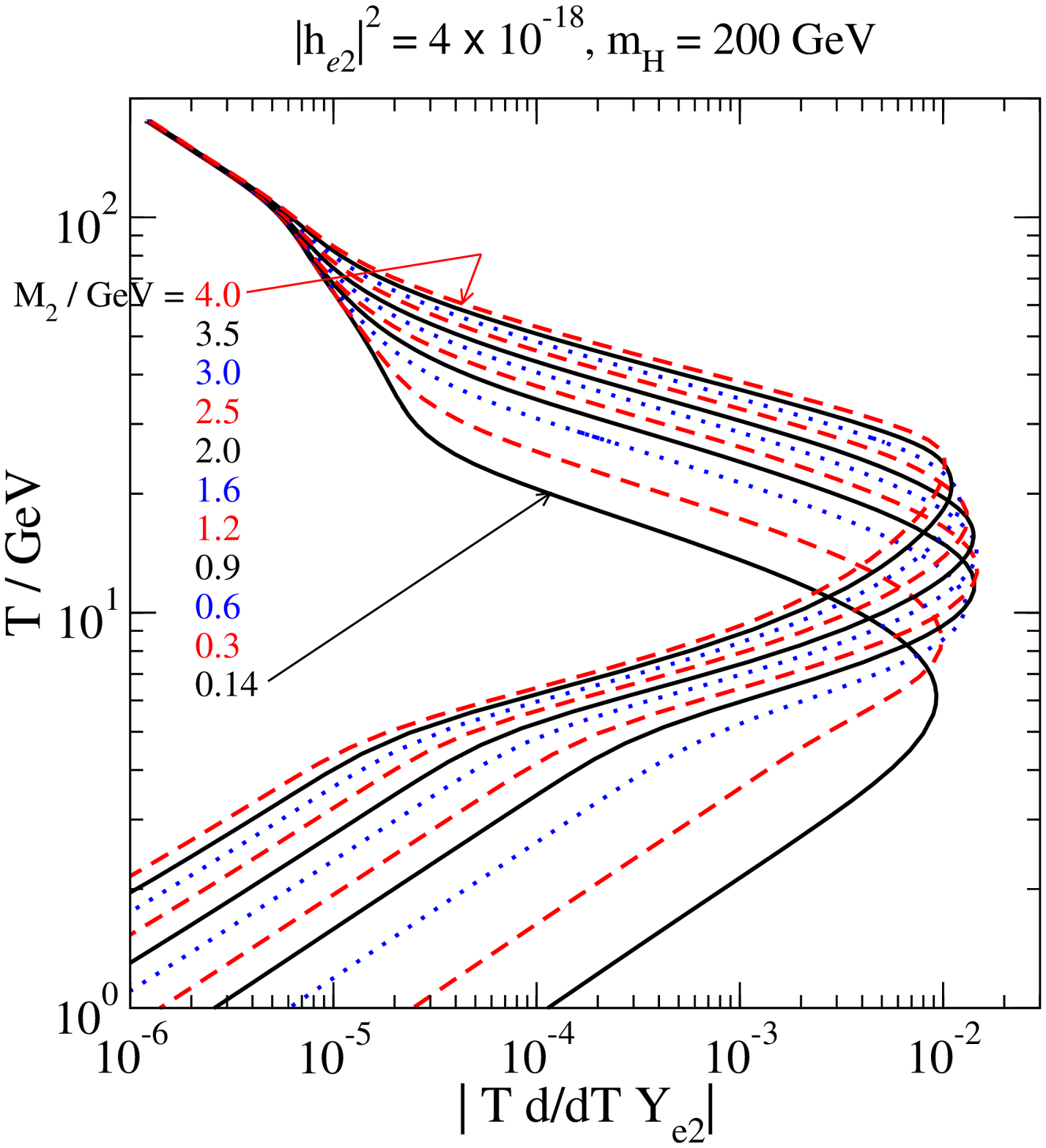}%
~~~\epsfysize=8.0cm\epsfbox{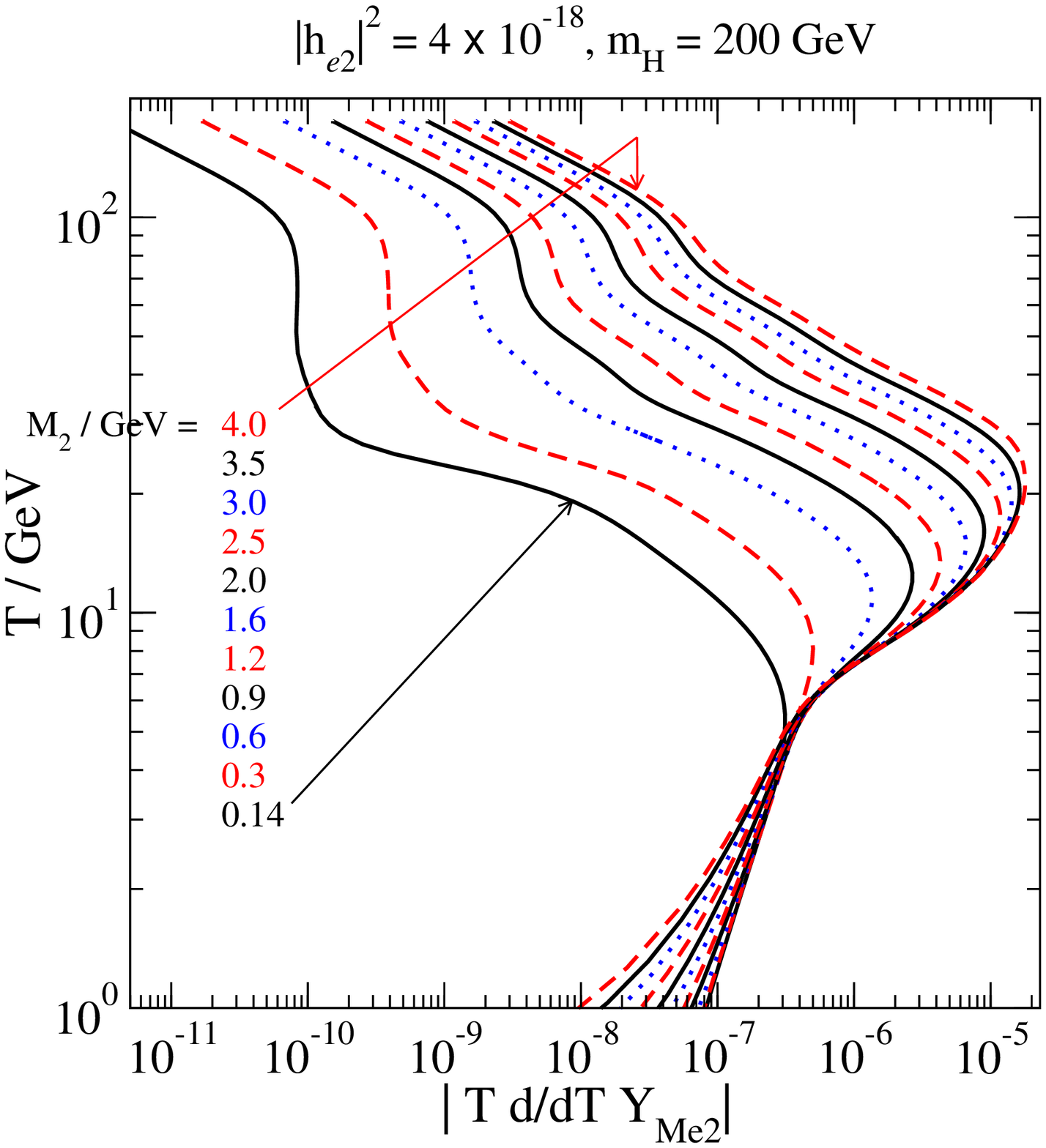}%
}
\caption[a]{\small The temperature derivative of the yield parameter
related to the rate $R(T,M)$ for the Higgs mass $m_H=200$ GeV, $F=F_0$
and different values of the singlet fermion mass (left panel). Right
panel: the same for $R_M(T,M)$.
} 
\la{fig:Y}
\end{figure}

For temperatures smaller than the peak temperature the
rate $R(T,M)$ can be reasonably approximated by
\be
\frac{F^2}{F_0^2} R(T,M) \simeq B G_F^2 T^5 \theta_0^2~,
\label{apph}
\ee
where $B\simeq 5$ is a numerical constant found by fitting of the
numerical result, $\theta_0^2= \frac{F^2 v^2}{M^2}$ is the
zero-temperature mixing angle between the singlet fermion and active
neutrinos. At the temperatures above and around  the peak the
suppression of the transitions due to the medium effects
\cite{Dodelson:1993je,Shi:1998km,Dolgov:2000ew} becomes important.
Also, the  decays  of the vector bosons and of the Higgs must be taken
into account.  At temperatures in the region  $100 - 200$ GeV the rate
scales as $R(T,M) \propto 1/T$, while at temperatures above the peak
roughly as $R(T,M) \propto 1/T^4$. In the symmetric phase of the
electroweak theory, $T\gsim 250$ GeV, studied previously for
baryogenesis via singlet fermion oscillations in
\cite{Asaka:2005pn,Akhmedov:1998qx},  the rate scales like $R(T,M)
\propto T$.

\subsection{Time evolution of CP-even perturbations}
\label{ssec:time}
Having defined the mass matrix of singlet fermions and the matrix of
the damping rates we are ready to consider the behaviour of CP-even
deviations from thermal equilibrium. Let us choose the basis in which
$\Delta M^2(T)_{IJ}$ is diagonal:
\be
\Delta E=\left(\begin{array}{c c}E_2 &0\\0&E_3\end{array}\right)
\label{Ediag}
\ee
and 
\be
\Gamma_N = \left(\begin{array}{c c}
\Gamma_{22}&\Gamma_{23}\\\Gamma_{32}&\Gamma_{33}\end{array}\right)~.
\ee
Since for practically all temperatures $\Delta E \gg \Gamma_{tot}^N$,
one easily finds four different exponentials describing the time
behaviour of the density matrix:
\be
\exp[-((\Gamma_{22}+\Gamma_{33})/2 \pm i (E_2-E_3)) t]~,~~
\exp(-\Gamma_{22} t)~~ {\rm and}~~ \exp(-\Gamma_{33} t)~. 
\label{exponents}
\ee
The first one corresponds to the behaviour of the off-diagonal
elements of $\Delta\rho_+$ and thus to the damping of
quantum-mechanical coherence in the oscillations of singlet fermions.
Two others represent the approach to thermal equilibrium of the
diagonal elements of the density matrix. In general, if  $\Gamma_{22}
t \gsim 1$ and $\Gamma_{33} t \gsim 1$ the system equilibrates
completely.

Let us consider again different scenarios for the singlet fermion
mass matrix.

In the {\bf Scenario I}  the matrices $H_{int}$ and $\Gamma_{tot}^N$
can be simultaneously diagonalised for  $T > T_{pot}$ and $\epsilon
\ll 1$ (up to the mass corrections $M^2/T^2$). Then, for $\epsilon
\ll 1$ we have two very different relaxation rates  for the diagonal
elements of the density matrix,
\be
\Gamma_{22} \simeq \frac{F^2}{F_0^2} R(T,M)~,~~
\Gamma_{33} \simeq r_\epsilon R(T,M) +\frac{F^2}{F_0^2}R_M(T,M),~~
r_\epsilon =\frac{F^2}{F_0^2}
\left(\epsilon^2-\frac{|(h^\dagger h)_{23}|^2}{F^4}\right) ~,
\label{gg}
\ee
whereas the rate of coherence loss is related to $\Gamma_{22}$. With
the use of relation (\ref{mrel}) the combination of Yukawa couplings
which appears in (\ref{gg}) can be represented as
\be
r_\epsilon
\simeq \frac{\epsilon M m_{atm}}{v^2 F_0^2}
\left\{\begin{array}{c c}0.36&
{\rm Normal~hierarchy}\\1&{\rm Inverted~hierarchy}
\end{array}\right.~,
\label{reps}
\ee
leading to  $\Gamma_{33}/\Gamma_{22} \propto \epsilon^2$. When the
temperature falls down from $T_{pot}$ to $T_{\beta}$ the mixing angle
$\beta$ changes from small values $\sim \epsilon$ to $\beta \sim 1$,
which modifies the (smaller) rate $\Gamma_{33}$ as
\be
\Gamma_{33}\to \Gamma_{33} + \sin^2\!\beta \, \Gamma_{22}~.
\label{rrmix}
\ee
As a result, at $T \lsim T_\beta$ both rates are of the same order
of magnitude and are related to the largest one $\Gamma_{22}$. Of course,
for $\epsilon \sim 1$ all damping rates have the same order of
magnitude for all temperatures.

In the {\bf Scenario II} the matrix $H_{int}$ can be diagonalised
with the help of orthogonal transformation $O$, $O^T H_{int} O =
{\rm diag}$, characterised by the angle $\beta\sim 1$. In general, the rates
$\Gamma_{22}$ and $\Gamma_{33}$ are of the same order. The same is
also true for  the {\bf Scenario III} with $\Delta \lambda \gsim \Delta
m_\nu/\epsilon$, leading to the mixing angle $\beta \simeq \pi/4$.
For  $\epsilon \ll 1$  all the damping rates are nearly the same and
equal to $\frac{F^2}{2F_0^2} R(T,M)$.  Qualitatively, if the rate of
oscillations between strongly coupled singlet fermion (rate
$[\Gamma_N^{tot}]_{22}$) and weakly interacting fermion (rate
$[\Gamma_N^{tot}]_{33}$) is large, the approach to thermal
equilibrium is determined by the largest rate since the system spends
half of the time in the strongly interacting state. For $\Delta
\lambda \lsim \Delta m_\nu/\epsilon$  the mixing angle is between
$\pi/4$ and zero, $\beta \sim \epsilon \frac{\Delta\lambda}{\Delta
m_\nu}$. Varying $\Delta\lambda$ one goes smoothly from one regime to
another.

We define the temperature $T_+$ at which the singlet fermion {\em
enters} in thermal equilibrium from the equation
\be
S_+(T_+)\equiv \frac{1}{Y_{eq}(T_+)}
\int_{T_+}^\infty \left(T \frac{dY}{dT}\right)\frac{dT}{T} = 1~,
\label{Rplus}
\ee
which tells that at $T=T_+$ the number of created particles  is equal
to the equilibrium one $Y_{eq}$. If $S_+(T) \geq 1$, the initial
deviations from thermal equilibrium are damped as $\exp(-S_+(T))$.
The behaviour of the integrated rate $S_+(T)$ as a function of
temperature is shown in Fig. \ref{intY}. 
\begin{figure}[tb]

\centerline{%
\epsfysize=4.5cm\epsfbox{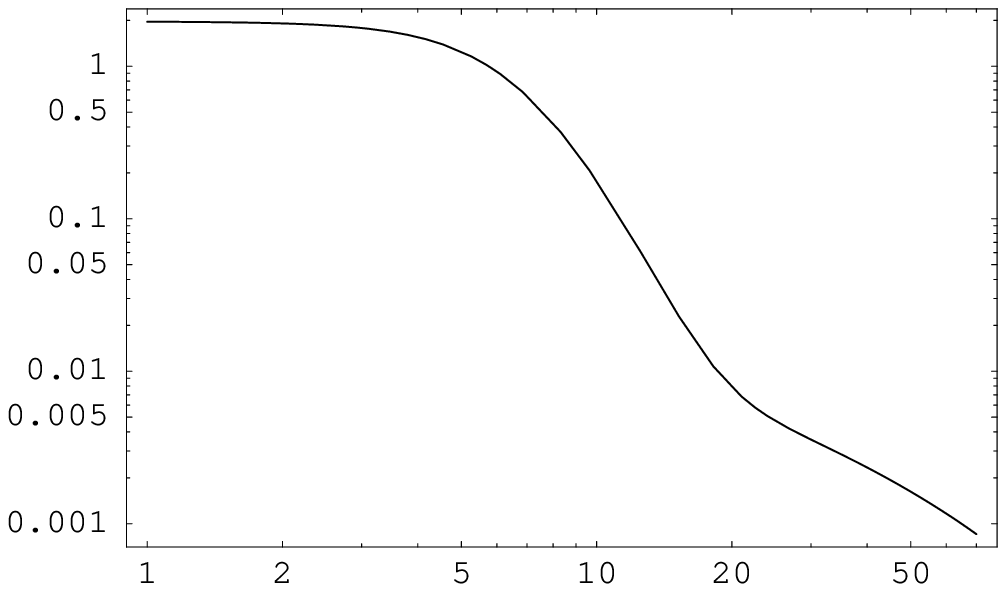}%
~~~\epsfysize=4.5cm\epsfbox{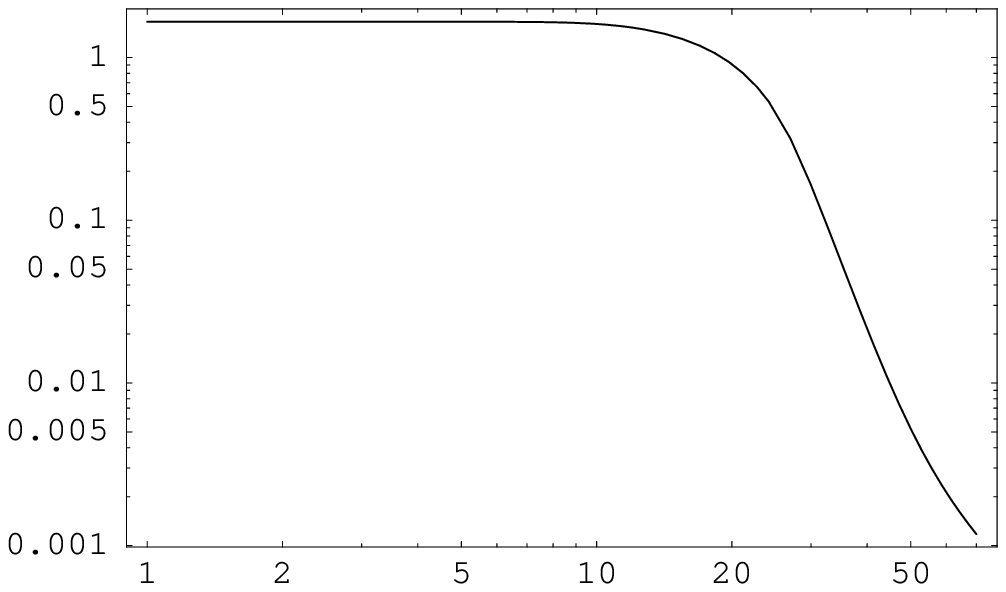}%
}
\caption[a]{\small The ratio of the integrated rate to the equilibrium
concentration of the singlet fermions for $F=F_0$ as a function of
temperature (in GeV). The system enters in thermal equilibrium when
this ratio is equal to one. Left panel: $M=0.14$ GeV, right panel: 
$M=4$ GeV. 
} 
\la{intY}
\end{figure}

In full analogy, the temperature $T_-$ at which the singlet fermions
{\em go out} of thermal equilibrium is determined by
\be
S_-(T_-)\equiv \frac{1}{Y_{eq}(T_-)}
\int_0^{T_-} \left(T \frac{dY}{dT}\right)\frac{dT}{T} = 1~.
\label{Rminus}
\ee
We show the temperatures $T_+$, $T_-$ and the
temperature at which the rate is maximal in Figs. \ref{Tplusminus}. 
The temperature $T_+$ (given roughly by  $T_+\simeq T_{EW}(0.02
\kappa M/\epsilon)^{\frac{1}{3}}$ at $100$ GeV $<T<300$ GeV)  is
below the sphaleron freeze-out temperature $T_{EW} \simeq 175$ GeV
(we take $M_H = 200$ GeV) for
\be
\epsilon \gsim 0.02\kappa \frac{M}{\rm GeV}~.
\label{tpew}
\ee

\begin{figure}[tb]

\centerline{%
\epsfysize=4.0cm\epsfbox{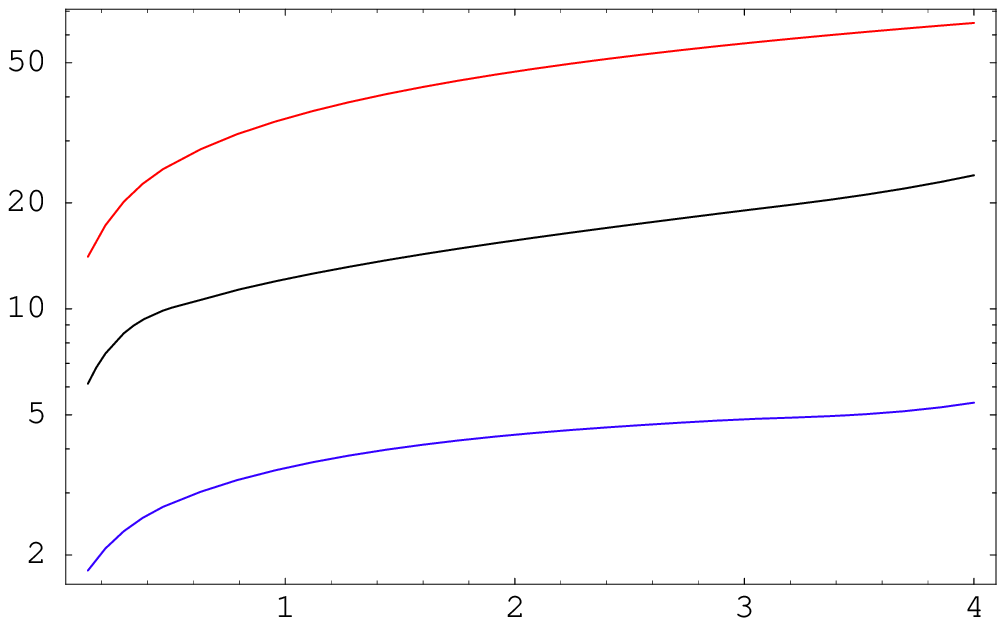}%
~~~\epsfysize=4.0cm\epsfbox{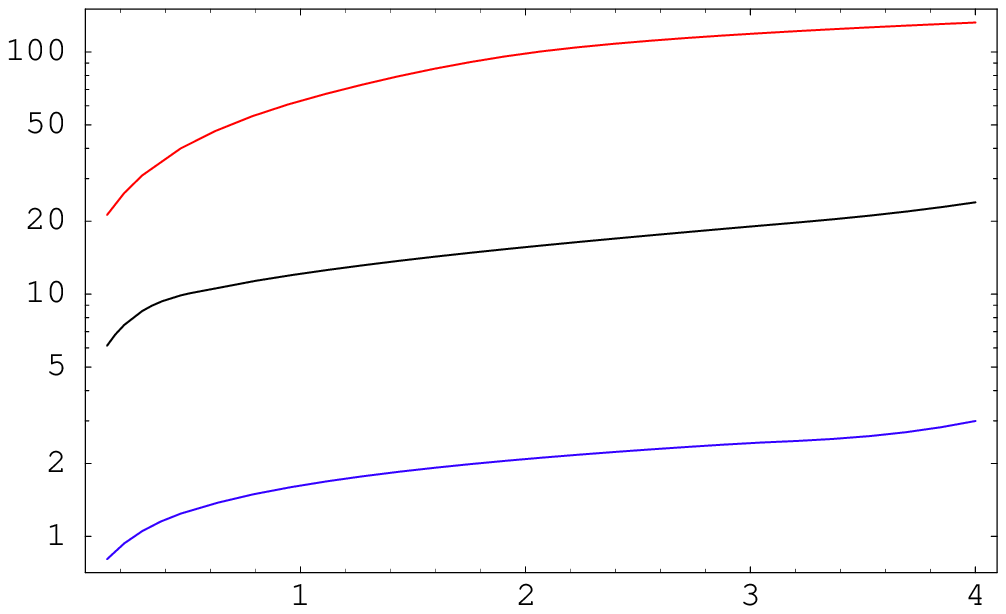}%
}
\centerline{%
\epsfysize=4.0cm\epsfbox{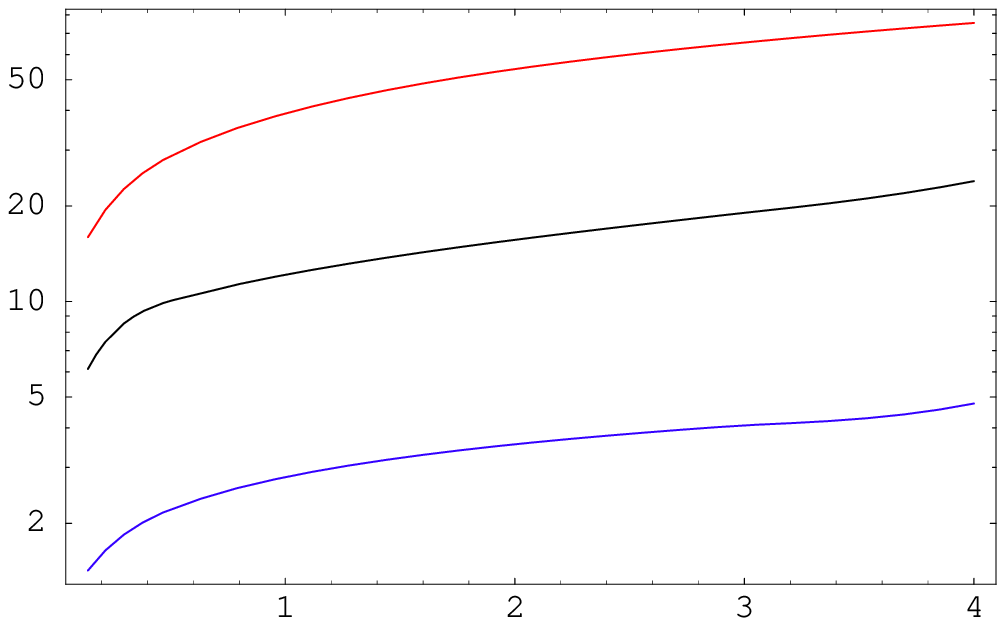}%
~~~\epsfysize=4.0cm\epsfbox{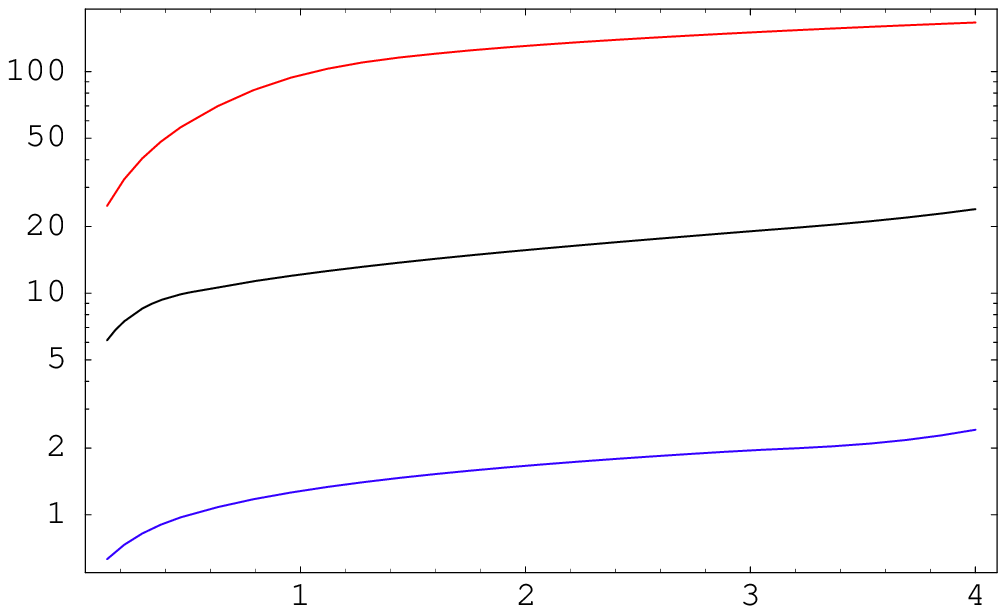}%
}
\caption[a]{\small The temperatures (in GeV) $T_+$ (upper curves), $T_-$ (lower
curves) and the peak rate temperature (central curves) as a function
of singlet fermion mass (in GeV). Upper panels: normal hierarchy, lower panels:
inverted hierarchy. Left panels: $\epsilon=1$, right panels:
$\epsilon=0.1$.
} 
\la{Tplusminus}
\end{figure}

Looking at Figs. \ref{fig:Y}, \ref{intY} and \ref{Tplusminus} one can
see that thermal equilibrium exists for the range of temperatures $T_-
< T < T_+$. As a numerical example let us take the minimal possible
mass $M = m_\pi$ and minimal value of Yukawa coupling,  $F^2 \simeq 
10^{-16}$ (see Appendix B). It corresponds to the choice $\epsilon =
1$ and leads to $\Gamma_{22} = \Gamma_{33}$. Then the solutions to
eqns. (\ref{Rplus},\ref{Rminus}) are $T_+ \simeq 15$ GeV and $T_-
\simeq 2$ GeV telling that the system is in thermal equilibrium for
temperatures $2$ GeV $ < T < 15$ GeV. Asymptotically, the integrated
rate approaches $S_+(T_-) \simeq 58$.

Since the Yukawa coupling chosen for this example is the minimal
possible one, we reach an important conclusion that the reactions
associated with the Yukawa coupling $h_{\alpha2}$ were certainly in
thermal equilibrium during some stage of the universe expansion.
Moreover, in the second and third scenarios for the mass difference
between singlet fermions with $\Delta \lambda \gsim \Delta
m_\nu/\epsilon$ the same conclusion is valid for all elements of the
density matrix $\delta\rho_+$ due to rapid oscillations between $N_2$
and $N_3$ states.

The case when $\epsilon \ll 1$  is somewhat more delicate. At first
sight one may choose $\epsilon$ in such a way that the rate
$\Gamma_{33}$ found in (\ref{gg}) is always smaller than the rate of
the universe expansion. And, indeed, the part of it, proportional to
Yukawa coupling $h_{\alpha3}$ (see (\ref{gg})) is smaller than the
Hubble rate $H$ for all temperatures if
\be
\epsilon
\lsim 3.4 \times10^{-3} \frac{\rm GeV}{M}
\left\{\begin{array}{c c}1 &
{\rm Normal~hierarchy}\\0.36& {\rm
Inverted~hierarchy}\end{array}\right.~.
\label{epsconst}
\ee
Since the mass of singlet fermion is bounded from below by the pion
mass, we get that this can only happen at $\epsilon < 2.4\times
10^{-2}$. At the same time, the mixing angle $\beta$ gets large at $T \sim
T_\beta$. So, if $T_\beta > T_-$, the system equilibrates even if
(\ref{epsconst}) is satisfied. This does not happen only if the zero
temperature mass difference of singlet fermions is very small, as in
{\bf Scenario IIa},
\be
\frac{\delta M}{m_{atm}} < 8\times 10^{-5} \frac{\epsilon}{\kappa}~.
\label{emix}
\ee  

To summarise,  for {\em any} values of parameters, consistent with
observed pattern of neutrino oscillations, with the exception of the
{\bf Scenario IIa}  and for $M \ll M_W$, the CP-even deviations from
thermal equilibrium are damped in some temperature interval
$[T_+,T_-]$ below the electroweak scale. The ratio of the peak rate
for equilibration of any element of $\delta\rho_+$ to the Hubble rate
is at least $58$. If the {\bf Scenario IIa} is realized, and relations
(\ref{epsconst},\ref{emix}) are satisfied, deviations from thermal
equilibrium in CP-even perturbations are substantial for all
temperatures. Moreover, in {\em any} scenario for singlet fermion mass
difference, the coherence in $N_2\leftrightarrow N_3$ oscillations is
lost in the temperature interval $[T_+,T_-]$. Thus, the lepton
asymmetry generation may occur either above $T_+$ or below $T_-$ (see
\se\ref{se:generation} for details). 

As we discussed, the CP-even deviations are important for {\em
generation} of the lepton asymmetry. The produced asymmetry must not
be diluted by reactions that can change it. Thus, we consider the
CP-odd deviations from thermal equilibrium in the next subsection in
order to understand whether the asymmetry that was generated before $T
\simeq T_+$ or below $T_-$ can survive the subsequent evolution. 

\section{CP-odd deviations from thermal equilibrium}
\label{se:CPodd}
The CP-odd deviations from thermal equilibrium are described by eq. 
(\ref{odd}). Having found the matrices $H_{int}$ and $\Gamma_N$ in the
previous subsection we still should compute six $2\times2$ matrices
$\tilde\Gamma_L^\alpha,~~\tilde\Gamma_N^\alpha$ and 3 rates
$\Gamma_L^\alpha$. They are coming from imaginary parts of the
diagrams shown in Fig. \ref{fig:gammaodd} and have the following
structure (we integrated the rates over momenta but are keeping the
same notations):

\begin{figure}[tb]

\centerline{%
\epsfysize=8.0cm\epsfbox{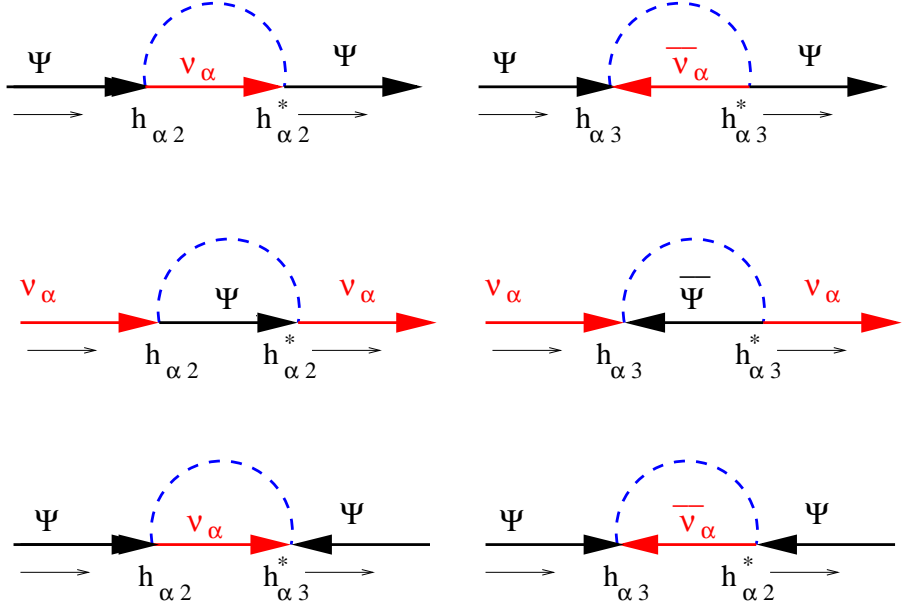}%
}
\caption[a]{\small The propagator-type diagrams for computation of
the damping rates. The Higgs line can be cut and replaced by
$v(T)^2$; the active neutrino propagator contains one-loop
corrections. The incoming  (outcoming) fermions correspond to two
arrows  entering (exiting) the vertex. Outcoming antifermion
corresponds to arrows in opposite directions, $\Psi = N_2 +N^c_3$. 
} 
\la{fig:gammaodd}
\end{figure}
\ba
\tilde \Gamma_N^\alpha&\simeq&
\frac{1}{F_0^2}
\left(
\begin{array}{c c}
|h_{\alpha2}|^2R(T,M) - |h_{\alpha3}|^2 R_M(T,M)&h_{\alpha2}^*h_{\alpha3} R(T,M)\\
h_{\alpha3}^*h_{\alpha2} R(T,M)&
|h_{\alpha3}|^2R(T,M) - |h_{\alpha2}|^2 R_M(T,M)
\end{array}
\right)~,
\nonumber\\
\nonumber\\
\tilde \Gamma_L^\alpha&\simeq&\tilde \Gamma_N^\alpha~,
\label{oddrates}\\
\nonumber\\
\Gamma_L^\alpha &\simeq& \frac{1}{F_0^2}(|h_{\alpha2}|^2 + |h_{\alpha3}|^2)
(R(T,M)+R_M(T,M))~.
\nonumber
\ea
The minus signs in eqns. (\ref{oddrates}) in front of mass
corrections come about since  the corresponding terms in
(\ref{kineq1},\ref{kineq2},\ref{kineq3}) are proportional to the chemical potentials
$\mu_\alpha$ (notice the change of direction of the fermionic line in
Fig. \ref{fig:gammaodd}).

\subsection{Approximate conservation laws and damping rates}
\label{ssec:law}
The structure of (\ref{oddrates}) is almost uniquely fixed by the
field-theoretical consideration presented below. Indeed, the CP-odd
deviations from thermal equilibrium can be considered as average
values of the densities of fermionic currents, which may be exactly
conserved for some particular choice of the parameters of the
$\nu$MSM. 

In the limit when all Yukawa couplings and Majorana masses of singlet
fermions are equal to zero the $\nu$MSM has five conserved leptonic
numbers:
\be 
{\cal L}_\alpha = \int \! {\rm d}^3 \vec{x} \, 
J_\alpha^0~,
\ee
where $\alpha = 1,...,5$.  Three of the currents are related to 
the active leptonic flavours,
\be
J_\alpha^\mu = \left[{\bar L}_\alpha \gamma^\mu  L_\alpha + 
 {\bar E}_\alpha \gamma^\mu  E_\alpha \right]~,
\ee
where $E_\alpha$ are the right charged leptons. The other two
conserved currents count the asymmetries in singlet fermions $N_2$
and $N_3$,
\be
N_2^\mu= {\bar N}_2 \gamma^\mu  N_2~,~~ N_3^\mu={\bar N}_3 \gamma^\mu 
N_3~.
\ee
When the Yukawa couplings and Majorana masses are switched on, none of
these numbers are conserved any more.

To make the discussion more transparent, consider the following
combinations of the currents introduced above:
\be
J_4^\mu= J_L^\mu =\sum_{\alpha=1}^3 J_\alpha^\mu  + N_2^\mu -N_3^\mu~.
\label{j4}
\ee
and
\be
J_5^\mu=J_F^\mu =\sum_{\alpha=1}^3 J_\alpha^\mu  + N_2^\mu + N_3^\mu~.
\label{j5}
\ee
The first current $J_L^\mu$ (total leptonic number) corresponds
precisely to the leptonic number symmetry defined in
\cite{Shaposhnikov:2006nn} which is exact in the limit $h_{\alpha3}
\rightarrow 0,~~ \Delta M_{IJ} \rightarrow 0$, whereas the second
current (it can be called total fermionic number) is conserved when
all Majorana neutrino masses are put to zero. What concerns  the 
currents $J_\alpha^\mu$ for a given $\alpha$, they are conserved in
the limit $h_{\alpha3}\to 0,~h_{\alpha2}\to 0$.

Now, if some combination of the currents introduced above is exactly
conserved, the equations (\ref{odd}) with zero source terms  must
have a time-independent solution for any choice of initial
conditions. As an example consider first the limit $M\to 0,~\epsilon
\neq 0$.  In this case the current $J_5^\mu$ is exactly conserved,
and we must have
\be
\frac{d}{dt}\left[{\rm Tr}\delta\rho_- +\sum\mu_\alpha\right]=0
\ee
for any $\delta\rho_- $ and $\mu_\alpha$. This leads to
\be
\Gamma_N= \sum_\alpha{\tilde\Gamma}^\alpha_L
~~{\rm and}~~
{\rm Tr}{\tilde\Gamma}^\alpha_N=\Gamma^\alpha_L
~~{\rm for}~M=0.
\label{param1}
\ee

In the another limit $M \neq 0,~\epsilon \to 0$ it is the current
$J_4^\mu$ which is exactly conserved and 
\be
\frac{d}{dt}\left[{\rm Tr}\tau_3\delta\rho_- +\sum_\alpha\mu_\alpha\right]=0
\ee
for any $\delta\rho_- $ and $\mu_\alpha$ (here $\tau_3$ is the Pauli
matrix). This gives
\be
\Gamma_N^{22}= \sum_\alpha{\tilde\Gamma}_L^{\alpha22},~\Gamma_N^{33}=
-\sum_\alpha{\tilde\Gamma}_L^{\alpha33},~{\tilde\Gamma}_L^{\alpha23}=\Gamma_N^{23}=0
~~{\rm and}~~
{\rm Tr}{\tau_3\tilde \Gamma}^\alpha_N=\Gamma^\alpha_L~~{\rm for}~\epsilon=0.
\label{param2}
\ee

In more general terms, the consistency condition can be formulated as
follows. Rewrite eqn. (\ref{odd}) with $S=S_\mu=0$ in the form
\be
\frac{d z}{dt} = D z~,
\ee
where $z$ is a vector with 7 components,
$z=(\delta\rho_-^{11},\delta\rho_-^{12},
\delta\rho_-^{21},\delta\rho_-^{22},\mu_\alpha)$ and $D$ is the
$7\times7$ matrix constructed from
$\tilde\Gamma_L^\alpha,~~\tilde\Gamma_N^\alpha$, $\Gamma_L^\alpha$
and $H$. The time-independent solution appears when $D$ has a zero
eigenvalue. Then, we must have ${\rm det}~D=0$ for the following
choices of parameters, corresponding to the conservation of the 5
currents introduced above: $h_{1 I}=0$ for $I=2,3$, corresponding to
conservation of the leptonic number of the first generation (and
similar relations for the second and third generation), $h_{\alpha
3}=0$, corresponding to conservation of the current $J_4^\mu$ (and an
equivalent relation for $N_2 \leftrightarrow N_3$), and $M=0$,
leading to conservation of $J_5^\mu$. One can check that eq.
(\ref{oddrates}) indeed satisfies these requirements.

It is instructive to find the damping rates in the limit $M\to
0,~~\epsilon\to 0$. In this case the matrix $D$ has two zero
eigenvalues corresponding to the conservation of currents $J_4^\mu$
and $J_5^\mu$, 2 complex eigenvalues
\be
\frac{F^2}{2F_0^2}R(T,M) \pm i (E_2-E_3)
\label{cohodd}
\ee
corresponding, as in the case of CP-even perturbations, to the
off-diagonal elements of the density matrix $\delta\rho_-$, and three
eigenvalues related to the damping rates of three different leptonic
flavours, 
\be
\gamma_i = \frac{F^2 x_\alpha}{F_0^2}R(T,M)~,
\label{flav}
\ee
where $x_i$ are the roots of the cubic equation
\be
x^3 + 2 x^2 +\frac{3}{2}\left(1-\frac{\sum_\alpha
h_{\alpha2}^4}{F^4}\right)x
+\frac{4 h_{e2}^2 h_{\mu2}^2 h_{\tau2}^2}{F^6}=0~.
\label{qubic}
\ee
If, for example, $h_{e2}\ll h_{\mu2},~h_{e2}\ll h_{\tau2}$ then the
smallest root of eq. (\ref{qubic}) is approximately given by
${5h_{e2}^2}/{4F^2}$. From (\ref{cohodd}) we can see that the
coherence in CP-odd perturbations is lost at the same time as it is
in CP-even perturbations. As for the damping
rates of active flavours, with the use of constraints
(\ref{fnorm},\ref{finv}) (see Appendix B) one finds that the
integrated rates corresponding to $\gamma_i$
\be
S_i=\frac{1}{Y_i^{eq}}\int _{\infty}^{T_-}  
\left[\kappa(T)\gamma_i\right]\frac{dT}{T}
\ee
are at least
\be
S_1 \simeq 8.2/\epsilon,~~S_2 \simeq 50/\epsilon,~~
S_3\simeq 156/\epsilon~,~{\rm Normal~hierarchy}~,
\label{nR}
\ee
\be
S_1 \simeq 32/\epsilon,~~S_2
\simeq 22/\epsilon,~~
S_3 \simeq 122/\epsilon~,~{\rm Inverted~hierarchy}~,
\label{nL}
\ee
where the smallest number in (\ref{nR}) corresponds to the asymmetry
in the electronic flavour. Eq. (\ref{nL}) shows that if the hierarchy
is inverted,  all the rates exceed the rate of the universe expansion
by a factor of at least $22$ (corresponding to the damping of
asymmetry which existed before the equilibrium period by a factor
smaller than $\simeq e^{-22} \sim 3\times 10^{-10}$).  For the case
of the normal hierarchy eq. (\ref{nR}) shows that the damping is at
least $\simeq e^{-8.2} \sim 3\times 10^{-4}$. This leads to the
conclusion that the reactions which change leptonic numbers in each
generation were certainly in thermal equilibrium during some time
below the electroweak scale which is good enough to dilute the lepton
asymmetry below the level required for resonant production of dark
matter. At this point the $\nu$MSM is very different from the
Standard Model, where leptonic numbers are conserved (up to
electroweak anomaly). 

\subsection{Protection of lepton asymmetries}
\label{ssec:protect}
The fact that the flavour changing reactions were in thermal
equilibrium during some period of the universe expansion below the
electroweak scale would at first sight mean that no (large)
asymmetry in active leptonic flavours can exist at small
temperatures. However, this conclusion is not necessarily true since
some combination of asymmetries in active and {\em sterile} flavours
may be protected from erasure due to the existence of approximate
conservation laws of currents $J_4^\mu$ and $J_5^\mu$. The only
certain thing for the moment is that the low temperature  remnants of
high-temperature leptonic asymmetries in active neutrinos are
flavour-blind, i.e. $\mu_e\simeq\mu_\mu\simeq\mu_\tau \equiv \mu$. 
This fact allows to simplify the further analysis replacing the
system of equations (\ref{odd}) with zero sources by
\ba
i \frac{d\delta\rho_-}{dt}&=& [H, \delta\rho_-]
-\frac{i}{2}\{\Gamma_N, \delta\rho_-\} +i\mu\sum_\alpha{\tilde
\Gamma}^\alpha_N~,
\nonumber\\
i \frac{d\mu}{dt}&=&-i\mu\frac{1}{3}\sum_\alpha \Gamma^\alpha_L +
i {\rm Tr}\left[\frac{1}{3}\sum_\alpha{\tilde\Gamma}^\alpha_L\delta\rho_-\right]~.
\label{oddblind}
\ea 

To consider the possibility of protection of lepton asymmetry we
start from the {\bf Scenario I} for the mass difference of singlet
fermions. Then for $b p \gg M^2$ the Hamiltonian $H_{int}$ can be
diagonalized simultaneously with the damping rates in eq.
(\ref{oddblind}), and one finds that for small $\epsilon$ and $M$ the
rates $\gamma_4$ and $\gamma_5$ are
\ba
\gamma_4&\simeq&r_\epsilon R(T,M)~,\\
\gamma_5&\simeq& \frac{4F^2}{5F_0^2} R_M(T,M)~,
\ea
where $r_\epsilon$ is defined in (\ref{reps}). In comparison with
$\gamma_{1,2,3}$, the rate $\gamma_4$ is suppressed by $\epsilon^2$
whereas the rate $\gamma_5$  is suppressed by $M^2$. For $b p \lsim
M^2$ the mass matrix $H_{int}$ is not proportional to $\Gamma_N$,
$\sum_\alpha{\tilde \Gamma}^\alpha_N$ and 
$\sum_\alpha{\tilde\Gamma}^\alpha_L$ any longer, leading to the rate
\be
\gamma_4\to \gamma_4 + \sin^2\!\beta\, \Gamma_{22}~,
\label{odmix}
\ee
where $\Gamma_{22}$ is defined in eq. (\ref{gg}) and the angle $\beta$
in eqns. (\ref{hightmix},\ref{betachange}). At the same time, the rate
$\gamma_5$ is not changed. Now, repeating the considerations of the
previous section one finds that $J_\mu^4$ is protected from erasure
only if inequalities (\ref{epsconst},\ref{emix}) are satisfied
simultaneously, i.e. only for {\bf Scenario IIa}.

Another leptonic charge which can be protected from erasure by the processes
with lepton number non-conservation is $J_\mu^5$. If ${\rm max}
\left(\frac{\gamma_5}{H}\right) \lsim 1$ then the density matrix at low
temperatures has the form
\be
\rho_{eq} =
\exp\left(-\frac{H}{T}-\mu_5 Q_5\right)=\exp\left(-\frac{H}{T}
 -\mu_5(L+Q_2+Q_3)\right)~,
\label{eqdm}
\ee
where  $\mu_{5}$ is the chemical potential corresponding to the
effectively conserved charge $Q_5=\int d^3x J_5^0$. In this case the
previously generated asymmetry in $Q_5$ survives, and the fact that
$Q_5$ contains the currents corresponding to active flavours ensures
non-zero asymmetry in lepton number, which is essential for the
resonant production of dark matter sterile neutrinos. Independently
of the choice of parameters, the chemical potentials for $N_{2,3}$
are the same as those of the active fermions, which is the
consequence of the fact that the transitions $L_\alpha \rightarrow
N_{2,3}$ are in thermal equilibrium. Exactly the same conclusions are
valid for the second and third scenarios for the fermionic mass
difference.

The region of the parameters in which the asymmetry in $Q_5$ is
protected can be found from the condition that the peak value of
$\gamma_5/(2H)$ does not exceed 1. We plot this region in 
Fig.~\ref{fig:Range}.

\begin{figure}[tb]

\centerline{%
\epsfysize=4.5cm\epsfbox{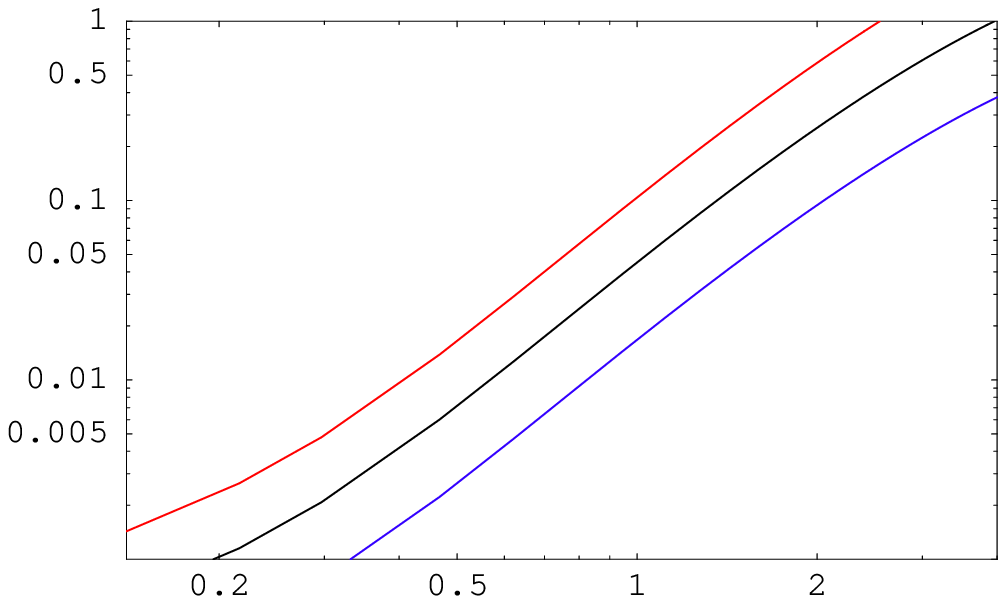}%
~~~\epsfysize=4.5cm\epsfbox{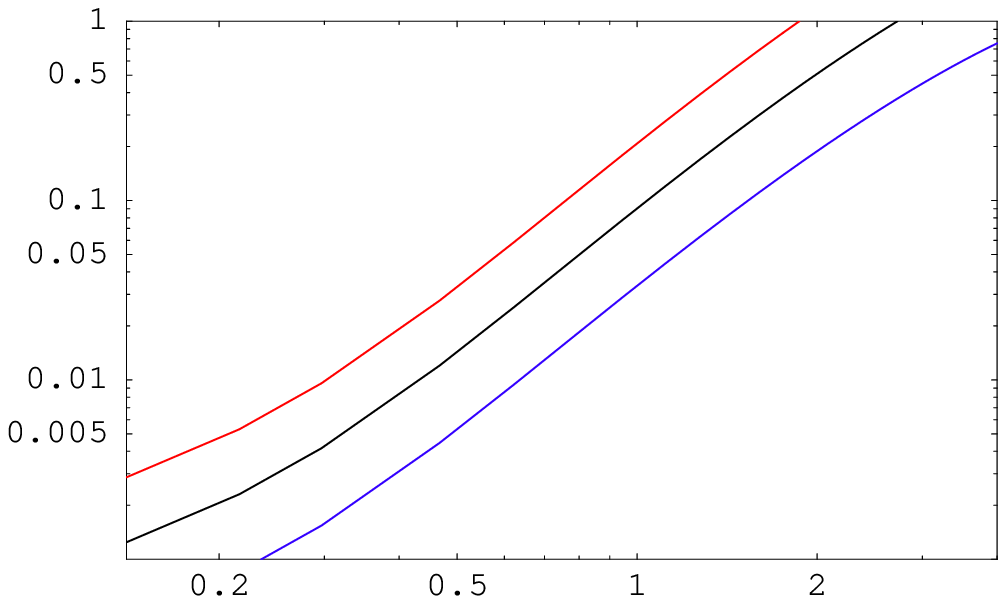}%
}
\caption[a]{\small The region of parameters in [$\epsilon$ (vertical
axis), $M/{\rm GeV}$] plane for which the low temperature lepton
asymmetry $Q_5$ is ``protected'' from erasure for normal (left panel) and
inverted (right panel) hierarchies of neutrino masses. The upper
curve corresponds to the damping factor $e^{-1}$, the lower curve to
$0.002$, and the middle one to $0.1$.
} 
\la{fig:Range}
\end{figure}

To summarize, the existence of a lepton asymmetry at small temperatures
$\sim 100$ MeV is only possible in the following situations:\\ 
(i) The asymmetry is produced  below the temperature $T_-$, when the
processes that damp the CP-even and CP-odd deviations go off thermal
equilibrium. \\ 
(ii) The asymmetry in $Q_5$ is produced above $T_+$ and the $\nu$MSM
parameters lie in the range shown in Fig.~\ref{fig:Range}
ensuring that it is not erased later on.\\
(iii) The asymmetry in $Q_4$ is produced above $T_+$ and the $\nu$MSM
parameters lie in the range (\ref{epsconst},\ref{emix}).

In the next section we will add to the analysis an input from the
dynamics of lepton asymmetry generation which will allow to choose
between these possibilities and to add further constraints.

\section{Lepton asymmetry generation and 
constraints on masses and couplings of singlet fermions}
\la{se:generation}
To find the leptonic asymmetry one should solve equations
(\ref{kineq1},\ref{kineq2},\ref{kineq3}) with zero initial conditions
for chemical potentials and for the elements of the density matrices
of singlet fermions. Due to the fact that the number of equations and 
different time scales is large (the equation count for real variables
is as follows: 4 for $\rho$, 4 for $\bar\rho$, and 3 for $\mu_\alpha$)
this cannot be done analytically. Nevertheless, the behaviour of the
system can be understood on the qualitative level with the results of 
\se\ref{se:CPeven} and \se\ref{se:CPodd} and a number of quantitative
estimates can be made.

Let us start from the small time behaviour of the system, when {\em all}
reactions involving singlet fermions are out of  thermal equilibrium,
so that the largest exponential in (\ref{exponents}), $\Gamma_{22} t$
is smaller than 1. This regime was considered in \cite{Asaka:2005pn}
for {\bf Scenario III}, assuming that the number of oscillations of
singlet fermions, defined below in eq. (\ref{x}), at the time of
electroweak cross-over, is much larger than one. We will generalize
this analysis to a more general case, accounting for the electroweak
symmetry breaking effects and considering also the time so short that
$\delta M(T) t \lsim 1$.

For these purposes it is convenient to transform the system in a
form that does not contain the term responsible for oscillations
between the two singlet fermion flavours, $[H,\rho_N]$, see
\cite{Asaka:2005pn}. This can be done by introducing $\tilde \rho_N$
related to $\rho$ in the following way:
\be
\rho_N=U(t) E(t) \tilde \rho_N 
E^\dagger(t) U^\dagger(t),~~E(t) = 
\exp\left(-i \int_0^t dt' \Delta E(t')\right)~,
\ee
where the matrix $U(t)$ converts the Hamiltonian $H_{int}$ to the diagonal
matrix $\Delta E(t)$ defined in eq. (\ref{Ediag}),
\be
H_{int}=U(t) \Delta E(t) U^\dagger(t)~. \label{DeltaEt}
\ee
Then the equation for $\tilde \rho_N$ is
\be
i \frac{d\tilde \rho_N}{dt}= [\tilde H, \tilde \rho_N]
-\frac{i}{2}\{\Gamma_{NU}, \tilde \rho_N - \rho^{eq}\} +
i \mu_\alpha{\tilde\Gamma^\alpha}_{NU}~,
\label{kinnew}
\ee
where
\ba
\tilde H &=&\frac{1}{2i}E^\dagger\left(U^\dagger \dot{U}-\dot{U}^\dagger
U\right)E~,
\nonumber\\
\Gamma_{NU}&=&E^\dagger U^\dagger\Gamma_{N}U E~,\\
{\tilde\Gamma^\alpha}_{NU}&=&E^\dagger U^\dagger
{\tilde\Gamma^\alpha}_{N}U E~.
\nonumber
\ea
Exactly the same procedure applies for the equation describing the
antiparticles.

As was explained in \cite{Asaka:2005pn}, the set of equations
(\ref{kineq1},\ref{kineq2},\ref{kineq3}) can be solved perturbatively
for the case when all damping rates (symbolically $\Gamma$) are small
enough, $\Gamma t \ll 1$. This is done in the following way: rewrite
the differential equations (\ref{kinnew}) in the integral way, e.g.
\be
\tilde \rho_N =-i \int_0^t dt'({\rm right~hand~side~of~eq.
(\ref{kinnew}))}
\ee
and then solve them iteratively. Then  asymmetries in leptonic numbers
(chemical potentials $\mu_\alpha$) are given by
\ba
\mu_\alpha &=& \int_0^t dt'\int_0^{t'} dt'' {\rm Tr}\left[
\left({\tilde\Gamma}^\alpha_L(t')V(t',t'')\Gamma_N(t'')
\right)V^\dagger(t',t'')\right]
\nonumber\\
&-&
\int_0^t dt'\int_0^{t'} dt'' {\rm Tr}\left[
\left({\tilde\Gamma}^{\alpha*}_L(t'){\bar V(t',t'')}\Gamma_N^*(t'')
\right)\bar V^\dagger(t',t'')\right]~,
\label{comp}
\ea
where 
\ba
\nonumber
V(t',t'')&=&U(t')E(t')E^\dagger(t'')U^\dagger(t'')~,\\
{\bar V(t',t'')}&=&U(t')^*E(t')E^\dagger(t'')U^T(t'')
\ea
($T$ corresponds to the transposed matrix). Equation (\ref{comp}) can
be simplified,
\ba
\mu_\alpha(t) = 4 \int_0^t dt'\int_0^{t'} dt''
&&{\rm Im}\left[(U^\dagger(t'){\tilde\Gamma}^\alpha_L(t')U(t'))_{12}
(U^\dagger(t'')\Gamma_N(t'')U(t''))_{21}\right]\times
\nonumber\\
&&{\rm Im}\left[\exp\left(i\int_{t''}^{t'}dt'''(E_2(t''')-E_3(t'''))\right)
\right]~.
\label{simpler}
\ea
As usual, the asymmetry contains a product of two imaginary parts. The
first multiplier in (\ref{simpler}) is associated with the
CP-breaking  complex phases in the Yukawa couplings, whereas the
second corresponds to the oscillations between two singlet flavours.
As was shown in \cite{Asaka:2005pn}, in the second order of
perturbation theory and neglecting mass corrections $ {\cal
O}\left(\frac{M^2}{T^2}\right)$, the total leptonic asymmetry is zero,
$\sum \mu_\alpha =0$. It appears in the third order only, leading to
an extra suppression of the order of $\Gamma t$. In this work we will
not go beyond the second order of perturbation theory, and account for
extra suppression by multiplying the results by $\Gamma t$. Note that
the resonant production of dark matter sterile neutrinos occurs even
if total lepton asymmetry is zero but individual flavour  asymmetries
are large enough \cite{Shi:1998km,Asaka:2005pn}. 

What happens for large times? For definiteness, suppose that the
smallest damping rate for CP-even deviations from thermal equilibrium
is $\Gamma_{33}$. Then, for $\Gamma_{33} t \gg 1$, the CP-even
fluctuations thermalize, $\delta\rho_+ \ll 1$, and the source terms
in (\ref{odd}) completely disappear, meaning that the production
of lepton asymmetry stops. 

In fact, an even stronger statement is true, namely that there is no
generation of lepton asymmetry for $\Gamma_{22}t \gsim 1$. Indeed, in
this regime all but one element of $\delta\rho_+$ are exponentially
damped: the oscillatory off-diagonal part of the CP-even density
matrix disappears at $(\Gamma_{22} +\Gamma_{33})t/2 \gsim 1$, and one
of the diagonal elements at    $\Gamma_{22} t \gsim 1$. So, in the
mass basis
\be
\delta\rho_+ = 
\left(
\begin{array}{c c}
0 & 0\\
0 & \delta\rho_+^{33}
\end{array}
\right)~.
\ee
For this type of deviation from thermal equilibrium the source terms
in the equation for chemical potentials $\mu_\alpha$, eq.
(\ref{source}) vanish, $S_\alpha=0$. The same is true for diagonal
elements of the density matrix $\delta\rho_-$, accounting for
asymmetries in singlet fermions. In other words, the leptogenesis
ceases to work when coherence in oscillations of singlet fermions is
lost, which happens when one of them is thermalized. The same
conclusion is reached if the damping of coherent oscillations is
inserted ``by hands'' into eq. (\ref{simpler}).

To conclude, we expect that the asymmetry is maximal at $t_{coh} \sim
2/(\Gamma_1 +\Gamma_2)$. For $t>t_{coh}$ the production of the
asymmetry is switched off, and the asymmetries in different quantum
numbers decay with the rates found in Section \ref{se:CPodd}.

Let us estimate the maximal possible asymmetry which can be created
at $t\sim t_{coh}$, corresponding to the temperature at which 
$N_2$ equilibrates, $T\simeq T_+$. 

In the {\bf Scenario III} for the singlet fermion mass differences
the matrix $U(t)$ depends on time slowly. Indeed, when the tree level
mass difference is much larger than the Higgs induced mass, the
matrix $U(t)$ corresponds to the rotation by $\pi/4$, and $\dot U
\sim \Delta m_\nu/\Delta\lambda$.  Therefore, the asymmetries at time
$t$  are of the order of
\be
\mu_\alpha(t) \simeq \delta_{CP}\frac{F^4}{F_0^2} \Phi(t),~~
\label{alt}
\ee
where
\be
\Phi(t)= \int_0^t dt'\int_0^{t'} dt''R(T',M)R(T'',M)\times
{\rm Im}\left[\exp\left(i\int_{t''}^{t'}dt'''(E_2(t''')-E_3(t'''))\right)
\right]~,
\label{Phit}
\ee
where the temperatures $T',~T''$ correspond to the times $t',~t''$,
and  $\delta_{CP}$ is defined in (\ref{deltaCP}). For $\epsilon \sim
1$, $\delta_{CP}$ can be of the order of 1. 

In the {\bf Scenario I} for $T>T_\beta$ the mass difference is
determined by the vev of the Higgs field only. Therefore,  the
temperature dependence of the  matrix $U(t)$ can also be factored out
up to mass corrections $M^2/T^2$, so that $\dot U \sim M^2/T^2$. 
However,  the asymmetry in $\mu_\alpha$ is suppressed in comparison
with eq. (\ref{alt}) by a factor (at small $\epsilon$)
\be
S_I\simeq \left(\frac{2|h^\dagger h|_{23}}{|h^\dagger
h|_{22}}\right)^2 \simeq 
\left(\frac{2\epsilon\Delta m_\nu}{\kappa m_{atm}}\right)^2~, 
\label{SI}
\ee
since in the limit $\epsilon \to 0$ the matrices $H_{int}$ and
$\Gamma_N$ can be simultaneously diagonalized (cf. eqns.
(\ref{gammaNtot}) and (\ref{massdifft})), so that off-diagonal
elements appearing in (\ref{simpler}) are suppressed either by a
factor $S_I$ or by a mass to temperature ratio  $M^2/T^2$. A similar
factor appears in the {\bf Scenario II} for $\epsilon \ll 1$.

For the generic case of {\bf Scenarios II} the phase factor cannot be
factored out and the equations are more complicated. We expect,
however, that the discussion below has a general character, at least
on the qualitative level.

It is instructive to find the behaviour of $\Phi(t)$ in different
limits. For this end we will assume that the rate $R(T,M)$ can be
approximately represented as $R(T,M) = A T^{-n}$, where $n$ is some
number. For example, for temperatures above the peak of production of
singlet fermions $n \simeq 4$, at $T> 100$ GeV $n \simeq 1$, whereas
at temperatures below the peak $n \simeq -5$.  The exponential in
(\ref{alt}) can be written as
\be
\int_{t''}^{t'}dt'''(E_2(t''')-E_3(t'''))=x(T')-x(T'')~,
\ee
where
\be
x(T)=\int_0^t dt \left\langle \frac{M\delta M(T)}{p}\right\rangle\simeq
0.15\frac{M\delta M(T) M_0}{T^3}~,
\label{x}
\ee
and $\langle ...\rangle$ is the thermal average. The physical meaning
of the parameter $x(T)$ is that $x(T)/2\pi$ gives the number of
oscillations between singlet fermions from the end of inflation till
the temperature $T$. Then one easily finds:
\ba
\Phi(t)&=& {\rm const} 
 \int_0^{x(T)}dz_1
z_1^{(n-1)/3}\int_0^{z_1}dz_2\sin(z_1-z_2)z_2^{(n-1)/3}
\nonumber\\
&=& \left(\frac{R(T)}{3H}\right)^2 F_+(x(T))~,
\label{smallx}
\ea
where $F_+(x)$ in limiting cases is given by
\be
F_+(x)=
\left\{
\begin{array}{c c}
\frac{27}{(n+2)(n+5)(2n+7)} x &,~~~ x \ll 1 \\
\frac{3}{2n+1} \frac{1}{x}&,~~~ x \gg 1 
\end{array}
\right.
\ee
which is valid for $n >-1$, true for any temperatures $T>T_{max}$,
where $T_{max}$ is the temperature at which the rate of $N$ production
is maximal. The plot of the function $F_+(x)$ for $n=4$ is shown in Fig.
\ref{fig:Fplus}.  

\begin{figure}[tb]

\centerline{%
\epsfysize=4.5cm\epsfbox{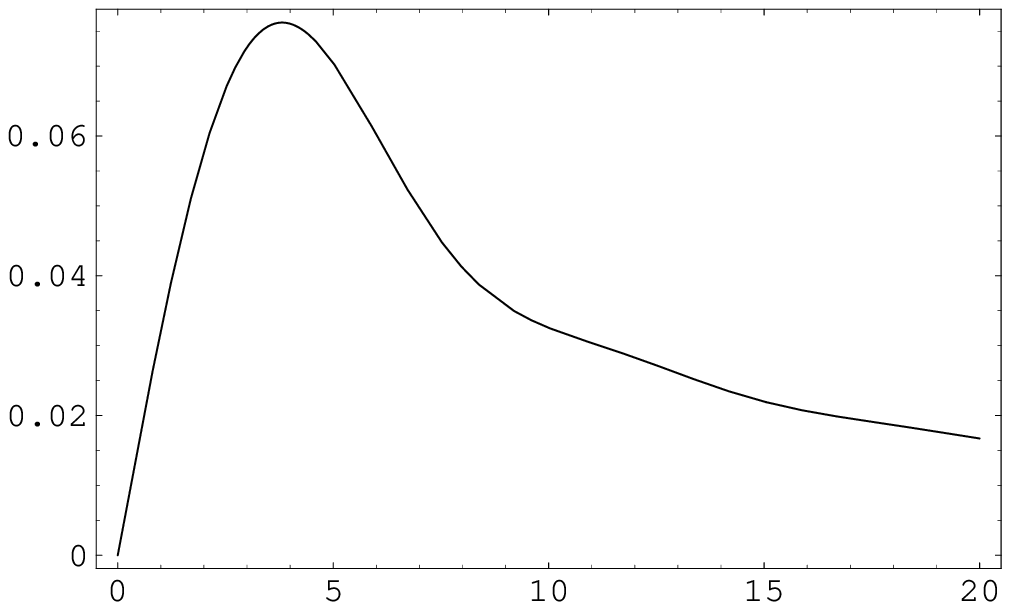}%
~~~\epsfysize=4.5cm\epsfbox{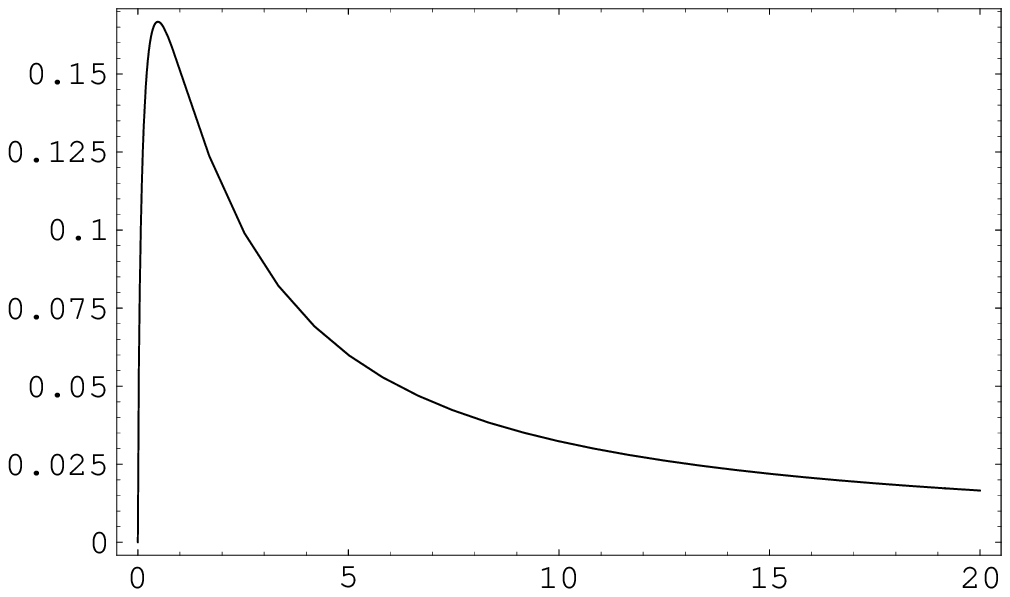}%
}
\caption[a]{\small The behaviour of functions $F_+(x)$, $n=4$  (left) and 
$F_-(x)$, $n=-5$ (right) counting the number of singlet fermion oscillations
near the temperatures $T_+$ and $T_-$.
} 
\la{fig:Fplus}
\end{figure}

Due to the very steep dependence of $\Phi(t)$ on the temperature the
baryon asymmetry, produced at $T\simeq T_{EW}$  can be much smaller
than the lepton asymmetry, created at $T\simeq T_+$.  Indeed, for
$n=4$ and for $x >1$ one gets that $\mu_\alpha \propto 1/T^{15}$, so
that a drop of the temperature by just a factor of $2$, increases the
asymmetry by a factor of $3\times 10^{4}$. Including an extra factor
$\Gamma t \simeq R(T)/3H$, accounting for the fact that baryon
asymmetry is produced in third order of perturbation theory
\cite{Asaka:2005pn} amplify the difference even further.

Let us estimate the maximal possible asymmetry which can be produced
at $T_+$. For this end suppose that the number of oscillations
maximizes the function $F_+$ ($F_+^{max}\simeq 0.076$ at $x \simeq
3.8$) and that CP-violation is maximal. Clearly, $\Delta$ cannot be
larger than $\Delta_{max}=4/(9\times2+4)=2/11$, where $4$ is the total
number of spin-states of $N_{2,3}$ and $9$ is the number of
spin-states of three leptonic generations. Thus,  
\be
\Delta \simeq \Delta_{max} \frac{ \epsilon F_+(x(T_+))}{F_+^{max}}~,
\label{LN}
\ee
where the factor $\epsilon$ accounts for the fact that CP-violation
goes away in the limit $\epsilon \to 0$.

Similar estimates apply for asymmetries in the other quantum numbers
defined in Section \ref{se:CPodd}:
\be
\delta Q_4 \simeq \Delta_{max}  \frac{\epsilon^2 F_+(x(T_+))}{F_+^{max}}~.
\label{epsN}
\ee
An extra factor $\epsilon$ appears since the rate of
creation or destruction of $Q_4$ is suppressed by $\epsilon^2$ in comparison
with the rates changing $\mu_\alpha$. As for the asymmetry in $Q_5$,
one gets
\be
\delta Q_5 \simeq 
\Delta_{max} \frac{R_M(T_+,M)}{R(T_+,M)}
\frac{\epsilon F_+(x(T_+))}{F_+^{max}}~,
\label{Q5}
\ee
where the second factor takes into account that the processes with the
change of $Q_5$ are suppressed in comparison with $L\leftrightarrow
N_2$ transitions.

The asymmetries in different quantum numbers generated at $T\sim
T_+$ are reduced later with the rates determined in Section
\ref{se:CPodd}.

\subsection{Constraints on singlet fermions from baryon asymmetry}
The estimates of the leptonic asymmetry presented above allow to find
constraints on  the masses and couplings of the singlet fermions from
the requirement that the produced lepton asymmetry is large enough to
make baryon asymmetry at the freezing point of sphaleron processes.

Consider first {\bf Scenarios I, II} for the singlet fermion mass
difference. If $T_+ > T_{EW}$,  the asymmetry generation in this case
occurs in the resonant regime as the number of oscillations at
temperature $T_+$ does not depend on parameters and is of the order of
one, 
\[
x(T_+) \simeq 12 \frac{v^2(T)}{v^2}~.
\]

We present in Fig. \ref{fig:Ncombined} the region of the parameter
space in which the baryon asymmetry (\ref{LN}), damped by a factor
$\exp(-S_+(T_{EW}))$ can exceed the observed value for normal and
inverted hierarchies. We take the sphaleron freeze-out temperature to
be $175$ GeV, corresponding to the Higgs mass $200$ GeV
\cite{Burnier:2005hp} and account for a suppression factor $S_I$
defined in eq. (\ref{SI}).   

\begin{figure}[tb]

\centerline{%
\epsfysize=4.5cm\epsfbox{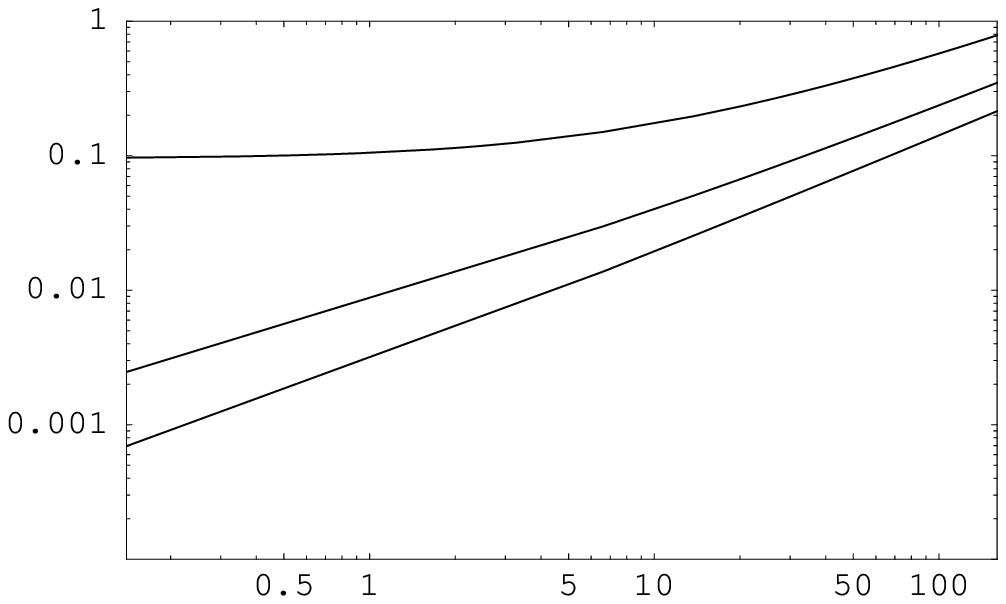}%
~~~~\epsfysize=4.5cm\epsfbox{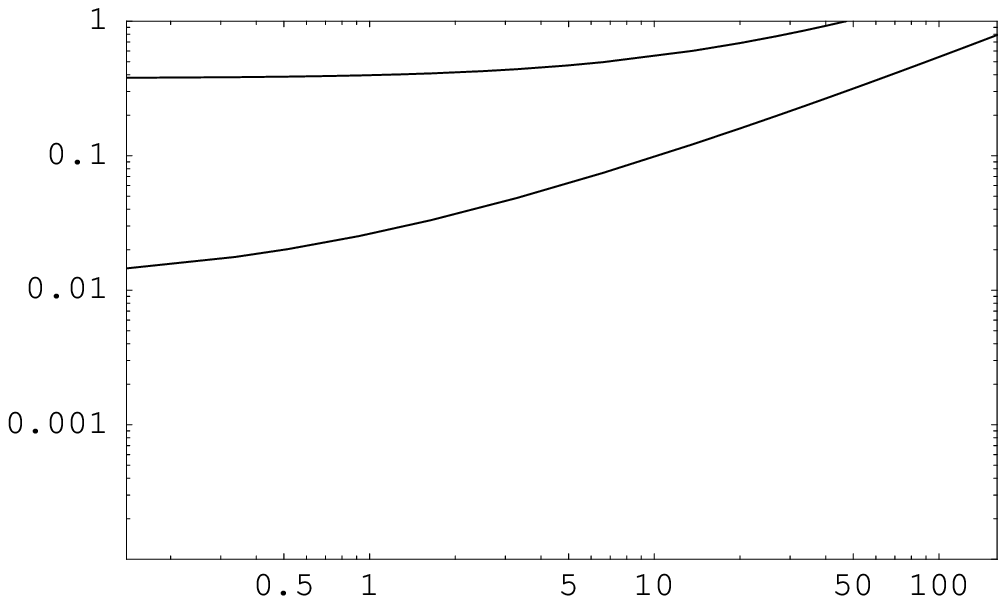}%
}
\caption[a]{\small The region of the parameter space in  [$\epsilon$
(vertical axis), $M/{\rm GeV}$] plane in which the asymmetry defined
in eq. (\ref{LN}), and then reduced due to damping,  can be
consistent with observations. The lower line corresponds to asymmetry
$\Delta = 6.6\times 10^{-9}$ (corresponding to observed baryonic
asymmetry), the middle one to $\Delta = 6.6\times 10^{-6}$ and the
upper line in left panel to $\Delta = 6.6\times 10^{-3}$.  Left panel
- normal hierarchy; right panel - inverted hierarchy.}
\la{fig:Ncombined}
\end{figure}

The asymmetry related to the charge $Q_4$, eq. (\ref{epsN}) can
exceed the observed baryon asymmetry for $\epsilon \gsim 
10^{-4}$ for normal hierarchy and for $\epsilon \gsim 
10^{-2}$ for a wide range of the singlet fermion masses, including
$M>M_W$. The fact that the baryon asymmetry generation is also
possible for masses so large was missed in \cite{Asaka:2005pn}
and is due to the fact that the charge $Q_4$ is protected from
erasure for small $\epsilon$, whatever the value of $M$ is.

In Fig. \ref{fig:AssMass} we show the region of the parameter space
where the asymmetry in $Q_5$ can exceed the observed value for the
case of the normal hierarchy. The parameter $\epsilon$ is bounded from
below by $\epsilon \simeq 7\times 10^{-5}$, and the mass from above by
$M \simeq 100$ GeV. These results refine the estimates presented in
\cite{Shaposhnikov:2006nn}. 
\begin{figure}[tb]
\centerline{%
\epsfysize=6cm\epsfbox{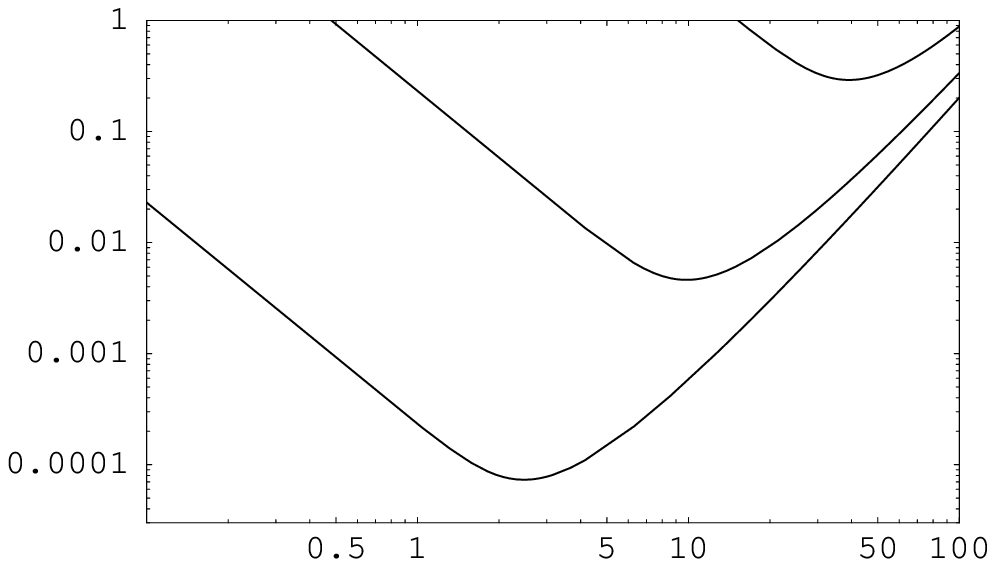}%
}
\caption[a]{\small The region of the parameter space in  [$\epsilon$
(vertical axis), $M/{\rm GeV}$] plane in which the asymmetry defined
in eq. (\ref{Q5}) and reduced later due to damping discussed in
Section \ref{se:CPodd}, can be consistent with observations. The
upper line corresponds to asymmetry $\Delta = 6.6\times 10^{-3}$, the
middle one to $\Delta = 6.6\times 10^{-6}$ and the lower line to
$\Delta = 6.6\times 10^{-9}$. We took the normal hierarchy case.}
\la{fig:AssMass}
\end{figure}

In the {\bf Scenario III} the leptogenesis goes off the resonance and
the available parameter space decreases. Comparing  eq.~(\ref{LN}) with
observed baryon asymmetry one can put an upper bound on the mass
difference difference of singlet fermions, $\delta M/M < 4\times
10^{-8}\kappa^3(M/{\rm GeV})$, valid if $T_+ < T_{EW}$, $M \lsim 50$
GeV and $\epsilon \sim 1$. If $T_+$ lies in the symmetric phase of the
electroweak theory, $T_+ \gsim 250$ GeV, a constraint from
\cite{Asaka:2005pn}, $\delta M/M < 6\times 10^{-8}(M/{\rm
GeV})^{\frac{5}{2}}$ should be used.

\subsection{Low temperature lepton asymmetry}

Let us find now the region of parameters which can lead potentially
to the generation of a large lepton asymmetry ($\Delta L/L > 2\times
10^{-3}$, as required by observational constraints, discussed in
\cite{Laine:2008pg}).  Clearly, the constraints coming from baryon asymmetry
are much weaker than those related to the large lepton asymmetry at
lower temperatures. As we have already discussed, the asymmetry can be
generated somewhat above $T_+$ or below $T_-$. 

We start from $T\simeq T_+$. Out of five different leptonic numbers
discussed in Section \ref{se:CPodd} only two can survive the subsequent
evolution. These are the asymmetry in $Q_4$ in the {\bf Scenario
IIa}, provided $\epsilon$ is small enough and in $Q_5$ which is
protected if the mass of singlet fermion is small enough, see fig.
\ref{fig:Range}. 

As we saw the number of oscillations at $T_+$ plays an essential
role in the determination of the asymmetry. So, we present in 
Fig.~\ref{fig:noscNI} the quantity $x(T_+)$ for the  {\bf
Scenario I} of singlet fermion mass difference for $\epsilon =1$ for
the case of normal and inverted hierarchies. For the generic choice
of parameters for the {\bf Scenario II} the number of oscillations is
of the same order. However, by tuning the Majorana mass difference to
the Higgs induced mass difference it can be made much smaller (for
$\epsilon = 1$, see eq. (\ref{mdiffhigh})) than the numbers appearing
in Fig.  \ref{fig:noscNI}. For the {\bf Scenario III} the number of
oscillations is much larger than that in the {\bf Scenario I} (by a
factor $\delta M/m_{atm}$ if the comparison is with normal hierarchy
case).
\begin{figure}[tb]

\centerline{%
\epsfysize=5.0cm\epsfbox{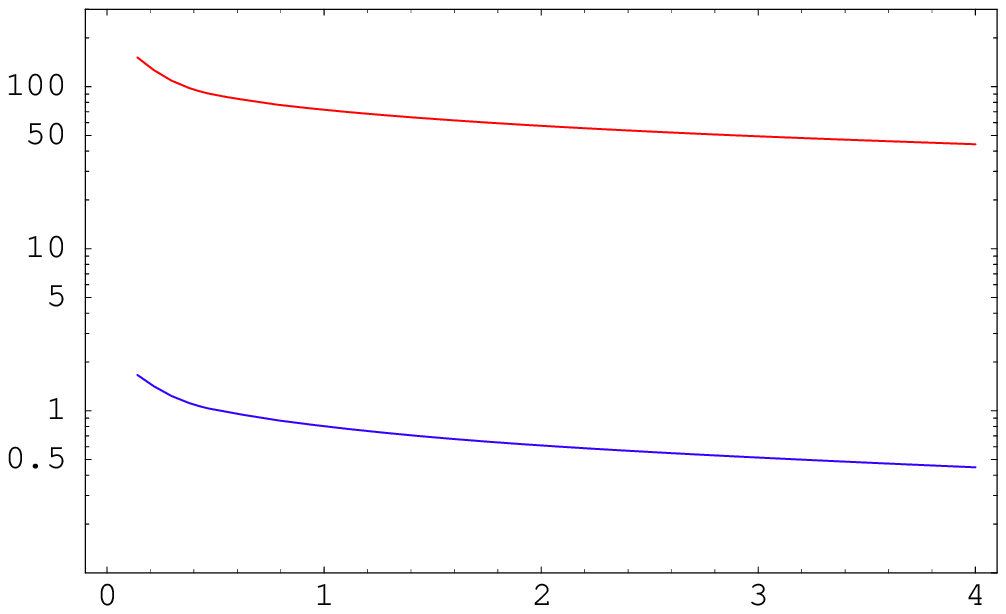}%
}
\caption[a]{\small The number of oscillations (vertical axis) of
singlet fermions  at temperature $T_+$ for the Scenario I as a
function of the fermion mass (in GeV) for normal (upper red curve)
and inverted (lower blue curve) hierarchies. We took $\epsilon =1$.}
\la{fig:noscNI}
\end{figure}
 
Consider now the lepton asymmetry in Scenarios I-III.

{\bf Scenario I.} The only possibility is to have an 
 asymmetry in $Q_5$. Inserting different rates
in (\ref{Q5}) we get for $\epsilon \simeq 1$ the asymmetries plotted
in Fig. \ref{fig:bauNI} (assuming that the number of oscillations
maximizes the asymmetry). For the normal hierarchy the asymmetry does
not exceed $2 \times 10^{-4}$ and thus is smaller than the minimal
required number ($2\times10^{-3}$) at least by a factor of $10$. For
the inverted hierarchy the maximal asymmetry is about $1\times
10^{-4}$, a factor of $20$ smaller than required. Though there
are no orders of magnitude differences between potentially produced
asymmetries and the required one, the conclusion that {\bf
Scenario I} cannot lead to necessary lepton asymmetry is robust.
Indeed, in all estimates the CP-violating affects were assumed to be
maximal, and other uncertainties were pushed in the direction which
can only increase the asymmetry (for example, accounting for the
number of oscillations will reduce the asymmetry for the case of
normal hierarchy by a factor of $20$).

\begin{figure}[tb]

\centerline{%
\epsfysize=4.5cm\epsfbox{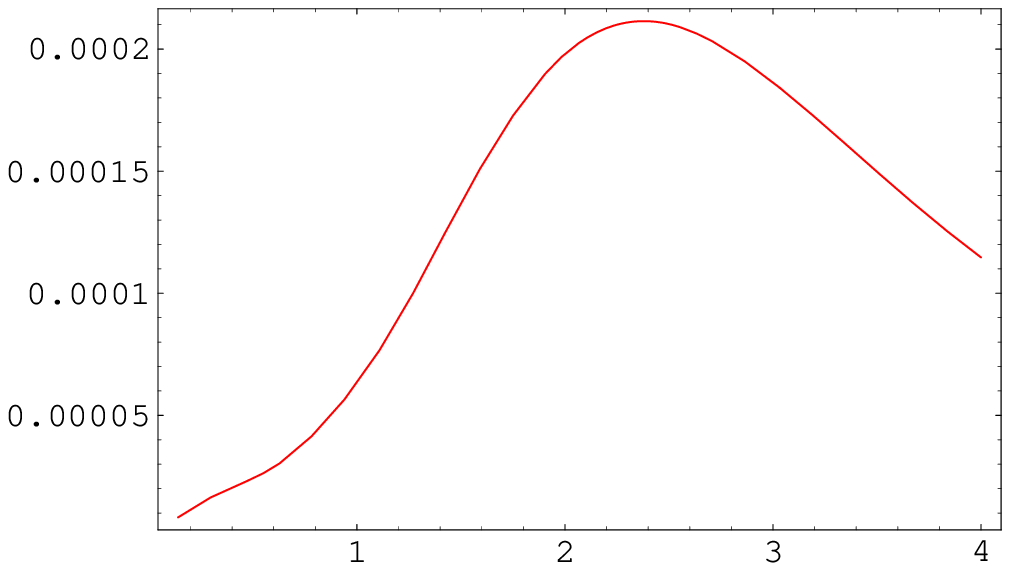}%
~~~~\epsfysize=4.5cm\epsfbox{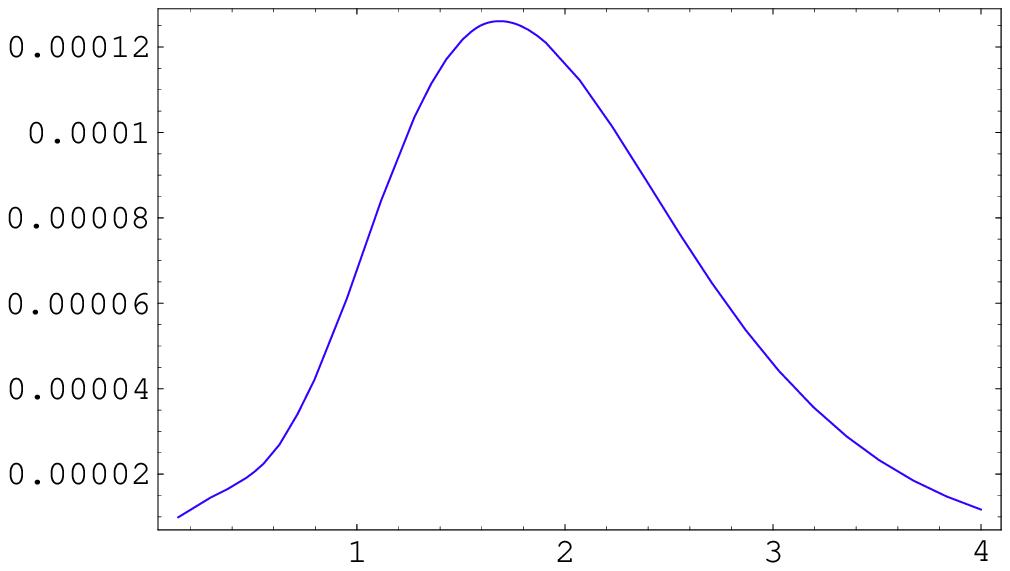}
}
\caption[a]{\small Maximal possible lepton asymmetry (vertical axis)
generated at $T_+$ and survived till $T_-$ for the Scenario I  for
normal (left) and inverted (right) hierarchies
as a function of singlet lepton mass (in GeV). We took $\epsilon
=1$.}
\la{fig:bauNI}
\end{figure}
 
In the {\bf Scenario II} for generic choice of parameters the results
for $Q_5$ stay the same as in the previous case.
In other words, no sufficient asymmetry in $Q_5$
can be produced at $T_+$ for this case. Potentially, in the {\bf
Scenario IIa} the leptonic charge $Q_4$ can survive. However, this can
only happen if $\epsilon < 2.4\times 10^{-2}$. For $\epsilon$ so small
the maximal asymmetry in $Q_4$ cannot exceed $\Delta_{max} \epsilon^2
\simeq 10^{-4}$, too small to have any effect on dark matter
production. Now, if the {\bf Scenario III} for singlet fermion mass
difference is realized, the asymmetry gets reduced by a factor
$m_{atm}/\delta M \ll 1$ in comparison with {\bf Scenario I}. Since
no large asymmetry can be produced in the {\bf Scenario I}, 
{\bf Scenario III} can be discarded as well. 

To summarise, no generation of large lepton asymmetry at $T\simeq
T_+$, which can survive till small temperatures, is possible.

Consider now a possibility of large lepton asymmetry generation at
lower temperatures, $T \simeq T_-$. The oscillations of singlet
fermions re-enter into coherence regime at $T \simeq T_-$,
corresponding to $t_-$. Then, one can simply change the region of
integration in (\ref{simpler}):
\be
\int_0^t dt'\int_0^{t'} dt''\to  \int_{t_-}^t dt'\int_{t_-}^{t'} dt''
\ee
accounting for the fact that at $t<t_-$ the oscillations were
exponentially damped. Correspondingly, the limits of integration
in the phase factor $\Phi(T)$ defined by (\ref{Phit}) must be changed.
We get:
\ba
\Phi(t)&=& {\rm const} \times {\rm Im}
\int_{x(T)}^\infty dz_1
e^{iz_1}z_1^{(n-1)/3}\int_{x(T)}^{z_1}dz_2e^{-iz_2}z_2^{(n-1)/3}
\nonumber\\
&=& \left(\frac{R(T)}{3H}\right)^2 F_-(x(T))~,
\label{smallxm}
\ea
where the plot of the function $F_-(x)$ for $n=-5$ is shown in Fig.
\ref{fig:Fplus}. It reaches the maximal value $F_-^{max}=0.167$ at
$x=0.47$.

In limiting cases the function $F_-(x)$ is given by
\be
F_-(x)=
\left\{
\begin{array}{c c}
-\frac{27}{(n+2)(n+5)(2n+7)} x &,~~~ x \ll 1 \\
-\frac{3}{2n+1} \frac{1}{x}&,~~~ x \gg 1 
\end{array}
\right.
\ee
which is valid for $n <-5$. The case of $n=-5$ requires a special
treatment, leading to the asymptotic value $F_-(x) = - x \log(e\gamma x)$ for
$x\ll 1$, where $e=2.718...$ and $\gamma =0.577...$ is the Euler
constant.

To estimate the leptonic asymmetry generated at this time one can
write
\be
\Delta \simeq \Delta_{max} 
\frac{\epsilon F_-(x(T_-))}{F_-^{max}}\frac{\delta n}{n_\nu}~,
\label{after}
\ee
where the factor $\delta n/n_\nu$ accounts for deviation of the
sterile neutrino concentration from the equilibrium one at
temperatures close to but below $T_-$. If $N_{2,3}$ decouple from the
plasma being relativistic, $M \lsim T_-$, the deviation of their
concentration from equilibrium is suppressed by the factor  
\be
\delta n/n_\nu \simeq
|n_{eq}(T,M)/n_{eq}(T,0) -1| \simeq 0.2 M^2/T_-^2~. 
\ee
If the decoupling occurs when $T_- < M$, the corresponding factor is 
\be
\delta n/n_\nu \simeq 0.7\left(\frac{M}{T_-}\right)^{3/2}
\exp\left(-\frac{M}{T_-}\right)~.
\label{ab}
\ee
Since all reactions which change different leptonic numbers are out
of equilibrium at temperatures below $T_-$, the asymmetries
(\ref{after}) stay intact.

Let us estimate the number of oscillations at $t\sim t_-$. Suppose
first that $T_- \gg M$ so that the high temperature approximation can be
used. Then with the use of eq. (\ref{apph}) the temperature $T_-$ is
given by
\be
T_- \simeq \left(\frac{\epsilon M}{\kappa B G_F^2 m_{atm} M_0}\right)^{1/3}
\label{Tmin}
\ee
leading to 
\be
x(t_-) \simeq \frac{0.15 \kappa B}{\epsilon}(G_F M_0)^2  m_{atm} \delta M~.
\ee
The asymmetry is maximal if the number of oscillations is minimal.
So, to get the maximal asymmetry we should take $\epsilon =1$ and 
the minimal $\delta M$. For the {\bf Scenario I} this corresponds to
the inverted hierarchy of neutrino masses and to  $x(T_-) \simeq 
3.6\times 10^3$.
So, the asymmetry cannot exceed $10^{-4}$, with the actual number
being smaller as one has to account for the factor $\delta n/n_\nu <1$
and for extra suppression
from CP-breaking phases. For the normal hierarchy of neutrino masses
the number of oscillations is larger by a factor of $\sim 50$, and
for the {\bf Scenario III} for the singlet fermion mass difference it is
even higher. We conclude, therefore, that large lepton asymmetry,
interesting for dark matter production,  cannot be generated at $T
\simeq T_-$ for {\bf Scenarios I} and {\bf III}, at least if $M \ll
T_-$.

Let us find the critical singlet fermion mass where the relativistic
approximation used above is not valid. Since the typical momentum of
a fermion in the plasma is $\langle p\rangle \sim 3 T$, we require $
3T \simeq M$ and find that the singlet fermions decouple being
non-relativistic if
\be
M > M_{crit} \simeq \left(\frac{27\epsilon}{\kappa B G_F^2 m_{atm}
M_0}\right)^{1/2}~,
\ee
giving $M_{crit} \simeq 30$ GeV for $\epsilon =1$. We will
demonstrate now that the lepton asymmetry is also very small if the
singlet leptons decouple in the non-relativistic regime (again
{\bf Scenarios I} and {\bf III} are considered).

At large singlet fermion masses one can neglect the influence of the
medium and consider the processes involving $N_{2,3}$ as if they were
in the vacuum. The fastest reactions at temperatures $T < M_W$  are
the decays $Z\rightarrow \nu N$ and $W \rightarrow l N$
(with the rate $\Gamma_V$), and decays or inverse decays of $N$ to all
possible leptonic or semi-leptonic channels (rate $\Gamma_N$). The
rates of inverse $W$ and $Z$ decays, responsible for thermalisation,
can be approximated as (at $M<M_W$)
\be
\Gamma_V \simeq \frac{1}{3} \theta_0^2 n_V\left[\Gamma_{W\to l\nu}
\exp\left(-\frac{M_W}{T}\right)\left(1-\frac{M^2}{M_W^2}\right)^{3/2}
+
2\Gamma_{Z\to \bar{\nu}\nu}\exp\left(-\frac{M_Z}{T_L}\right)
\left(1-\frac{M^2}{M_Z^2}\right)^{3/2}\right]~,
\ee
where $\Gamma_{W\to l\nu}\simeq 0.7$ GeV and $\Gamma_{Z\to
\bar{\nu}\nu}\simeq 0.5$ GeV are the widths of the intermediate vector
bosons, $n_V=3$, and $\theta_0^2\simeq \kappa m_{atm}/(2\epsilon M)$. 

The rate of inverse decays of $N$ is of the order
\be
\Gamma_N = A \frac{G_F^2 M^5 \theta_0^2}{192
\pi^3}\left(1-\frac{M^2}{M_W^2}\right)^{-2}
\exp\left(-\frac{M}{T}\right)~,
\ee
where $A$ is proportional to the number of open channels for $N_2$
decays,  $A \sim 10$ if $M>10$ GeV \cite{Gorbunov:2007ak}. The
temperature at which the oscillations of $N$ start to be coherent can
be determined from the condition $H=\Gamma_N +\Gamma_V$, and the
lepton asymmetry from the relations (\ref{after},\ref{ab}). The
results for the temperature $T_-$, the number of oscillations and the
lepton asymmetry are shown in Fig. \ref{fig:tminusNI}. Note that for
the non-relativistic case the number of oscillations is given by
\be
x \simeq \frac{M_0\delta M}{T^2}~.
\ee

\begin{figure}[tb]

\centerline{%
\epsfysize=4.5cm\epsfbox{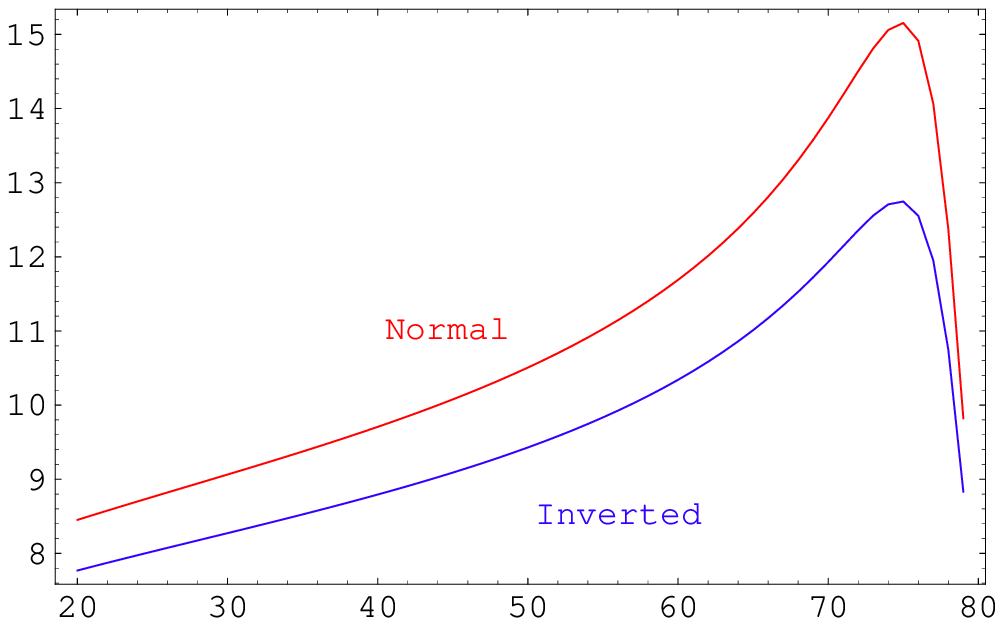}%
~~~\epsfysize=4.5cm\epsfbox{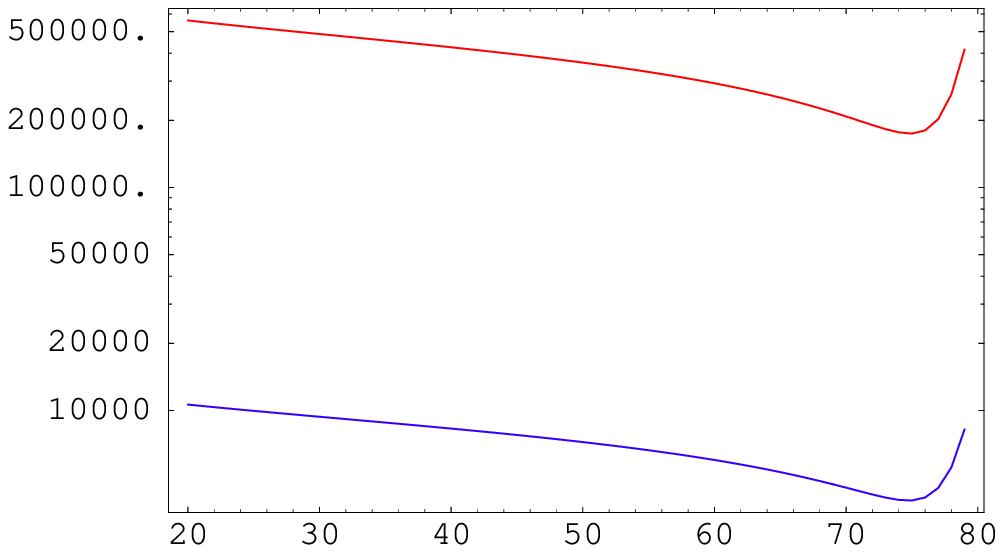}
}

\vspace{0.2cm}

\centerline{%
~~~\epsfysize=4.5cm\epsfbox{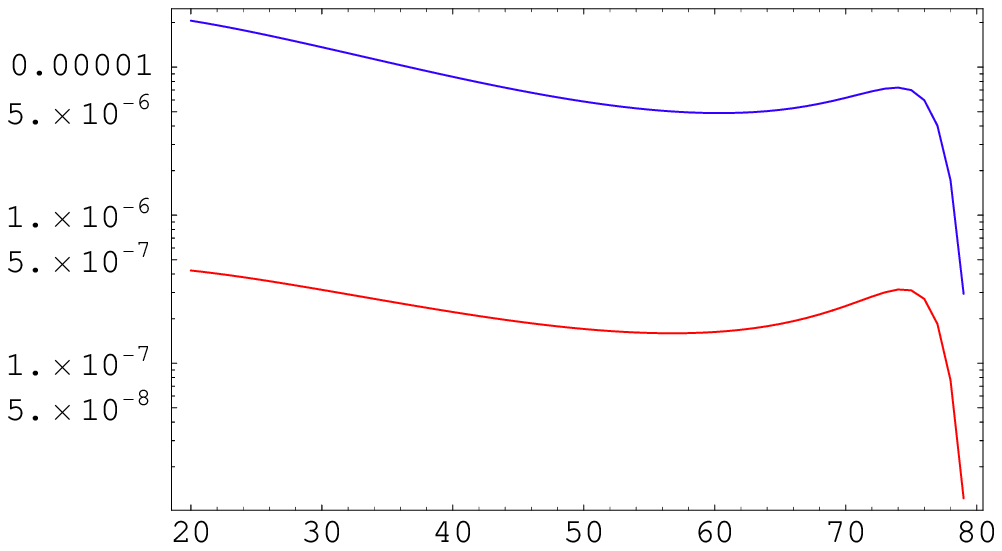}
}
\caption[a]{\small The temperature $T_-$ in GeV (upper left),  the
number of oscillations (larger for normal hierarchy) of singlet
fermions  at temperature $T_-$ (upper right) and  the maximal lepton
asymmetry (smaller for normal hierarchy) generated at $T_-$   as a
function of the fermion mass (in GeV) for normal (red curve) and
inverted (blue curve) hierarchies. We took $\epsilon =1$ and the {\bf
Scenario I}.
}
\la{fig:tminusNI}
\end{figure}

One can see that the asymmetry never exceeds $2\times 10^{-5}$
(inverted hierarchy) and $4\times 10^{-7}$ (normal hierarchy), which
is well below the threshold for the resonant production of dark
matter. We conclude, therefore, that no substantial asymmetry
generation can occur after singlet fermions decouple in {\bf
Scenarios I} and {\bf III}. The same conclusion is valid for the {\bf
Scenario II} for a generic choice of parameters. 

On the other hand, for a
special case of {\bf Scenario IIa}, when the sterile fermion  mass
difference is much smaller than the active neutrino mass difference,
the asymmetry production enters into resonance and the generation of
large lepton asymmetries $\Delta L/L\gsim 2\times10^{-3}$ at $T_-$
becomes possible for a variety of masses  and
couplings of singlet fermions. With the use of eqns.
(\ref{mssdeff},\ref{x},\ref{Tmin}) one finds that if $\delta M(0)=0$,
the number of oscillations at $T=T_-$ does not depend on $M$ and
$\epsilon$ (for $\epsilon \ll 1$) and is given by
\be
x(T_-)\simeq 0.15
\left(\frac{p}{T}\right)^2\frac{8}{\pi\alpha_W}\frac{7\pi^2}{360B}
(2 +\cos^2\theta_W) \simeq 10 ~.
\label{a2}
\ee
In other words, we are close to the resonance and a large  asymmetry
can be produced. In Fig.~\ref{fig:regiontm} we present the part of
the parameter-space where the asymmetry may exceed the critical
value.

\begin{figure}[tb]

\centerline{%
\epsfysize=4.5cm\epsfbox{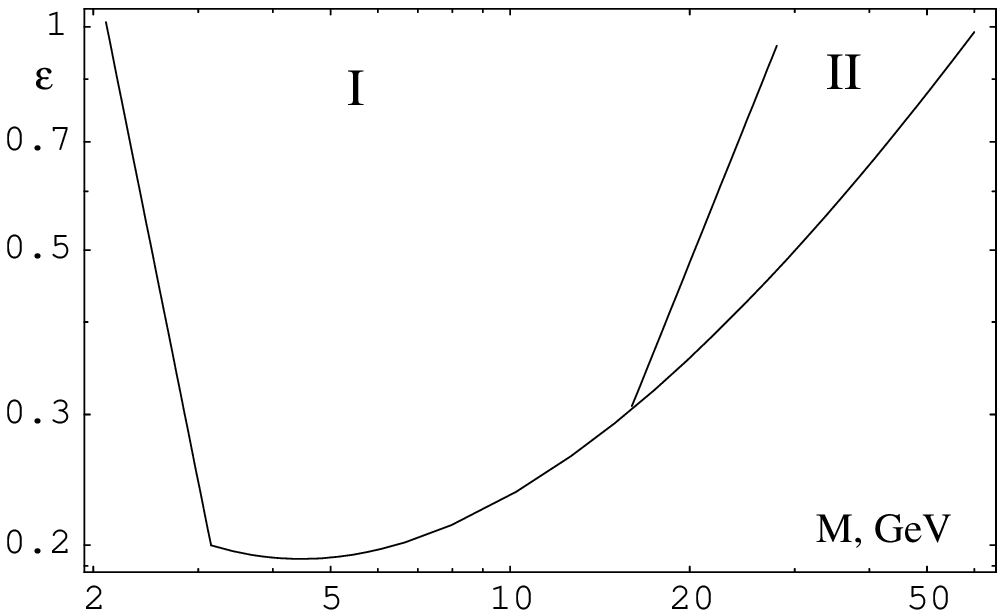}%
~~~~\epsfysize=4.5cm\epsfbox{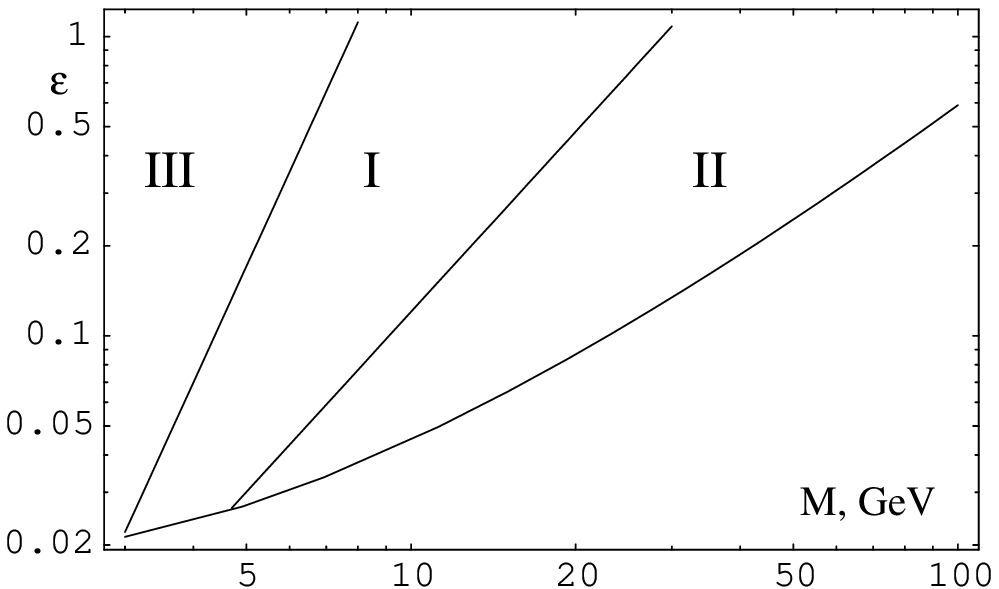}%
}
\caption[a]{\small Left panel: the parameter-space (I and II) which
can lead to  the lepton asymmetry, produced at $T=T_-$ and exceeding
$2\times 10^{-3}$.  Right panel: the parameter-space (I and II), which
can lead to  the lepton asymmetry, produced in decays of $N_{2,3}$ and
exceeding $2\times 10^{-3}$. In the region III $N_{2,3}$ decay below the
temperature 100 MeV and thus do not contribute to resonant production
of dark matter. In the region I (II) $N_{2,3}$ decouple being
relativistic (non-relativistic).
}
\la{fig:regiontm}
\end{figure}

In fact, yet another mechanism for late leptogenesis is possible in
the {\bf Scenario IIa} with the ``tuned'' mass difference. If $3T_- >
M$, the singlet fermions decouple from the plasma being relativistic.
Later, they decay with lepton number non-conservation and
CP-violation and, if they are degenerate enough, they will produce
large lepton asymmetries\footnote{The fact that CP-violation is
greatly enhanced in the decays of degenerate particles is well known
from $K^0$ physics. It was first suggested for baryogenesis in
\cite{Kuzmin:1970nx}, discussed in \cite{Ignatiev:1979tz} and studied
in detail for TeV scale Majorana fermions in
\cite{Pilaftsis:2003gt,Pilaftsis:2004xx,Pilaftsis:2005rv}.}. 

Let us estimate the value of the lepton asymmetry which can be created
in decays of $N_{2,3}$. Since the Yukawa couplings are very small, the
main contribution to CP asymmetry comes from the mixing between $N_2$
and $N_3$, as shown in Fig. \ref{fig:assdecay}.
\begin{figure}[tb]
\centerline{%
\epsfysize=4.0cm\epsfbox{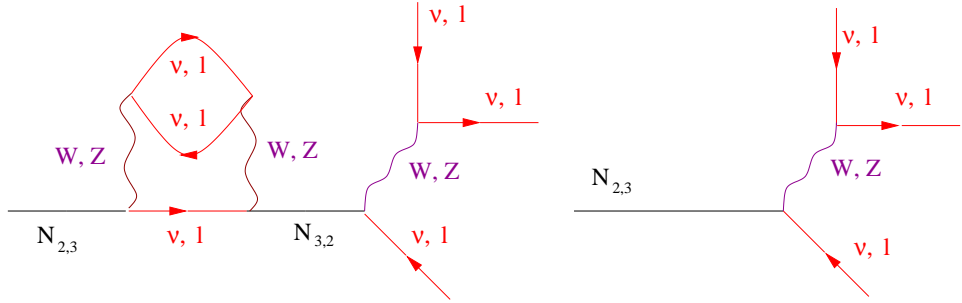}%
}
\caption[a]{\small Diagrams for $N_{2,3}$ decay which can lead to large
lepton asymmetry below the electroweak scale.
} 
\la{fig:assdecay}
\end{figure}
An estimate for asymmetry reads
\be
\Delta  \simeq \Delta_{max} \left(\frac{\epsilon \Gamma_N}{\delta
M}\right)
\left[\frac{M^2}{\Gamma_N M_0}\right]~,
\ee
where the the first term ($\epsilon$) comes from CP-violation, the
second term describes the resonance and is valid for $\delta M \gsim
\Gamma_N$ (it should be replaced by 1 in the opposite limit), the
third term accounts for equilibration of the asymmetry due to inverse
$N_2$ decays. It should be replaced by $1$ if $M^2 >\Gamma_N M_0$,
i.e. for
\be
\frac{M}{\rm GeV} < 19 \left(\frac{\epsilon}{2\times 10^{-3}}\right)^{1/2}
\left(\frac{10}{\kappa A}\right)^{1/2}~.
\ee
For the case $\delta M \sim\Gamma_N$ the asymmetry can be large and
lead to the resonant production of dark matter sterile neutrino,
provided $N_{2,3}$ decay above the temperature $\sim 100$ MeV, at
which $N_1$ are created most effectively. The latter requirement leads
to the constraint
\be
\frac{M}{\rm GeV} > 1.4  \left(\frac{\epsilon}{2\times 10^{-3}}\right)^{1/4}
\left(\frac{10}{\kappa A}\right)^{1/4}~.
\ee
In Fig. \ref{fig:regiontm} we present the part of the parameter-space
where the asymmetry created in $N_{2,3}$ decays may exceed the
critical value. 
  
To constrain further the parameter-space of the model one should take
into account the requirement that not only the low temperature lepton
asymmetry must be large enough, but also that the baryon asymmetry is
small. Given the number of CP-phases and other parameters we expect
that anywhere in the regions shown in Fig. \ref{fig:regiontm} the
required hierarchy can be achieved by some choice of Yukawa couplings.
However, for a generic case, in which no cancellation between
different CP-violating phases takes place, the region of small singlet
fermion masses and large $\epsilon$ is singled out. 

Indeed, in the {\bf Scenario IIa} the baryon asymmetry generation 
occurs in the resonant regime at $T_+>T_{EW}$, leading generally  to
large baryon asymmetries. The observed small baryon asymmetry can be
derived moving out of the resonance, i.e. for $T_+ < T_{EW}$. The
number of oscillations at the electroweak temperature $T_{EW}\simeq
175$ GeV is (for $\epsilon \ll 1$):
\be
x(T_{EW}) \simeq 
0.15 \frac{\kappa M_0 M m_{\rm atm}v^2(T_{EW})}{4\epsilon T_{EW}^3v^2}
\simeq \frac{0.12\kappa}{\epsilon} \left(\frac{M}{\rm GeV}\right)~
\ee
and smaller than one if $\epsilon$ is large and $M$ is small. In this
regime the baryon asymmetry is suppressed by a factor 
\be
(R/3H)^3 \frac{x(T_{EW})}{30F_+^{max}}\simeq 
2\times \left[\frac{0.02\kappa}{\epsilon} \left(\frac{M}{\rm
GeV}\right)\right]^4 ~,
\ee 
which is about $5 \times 10^{-6}$ for $M\simeq 2$ GeV, $\epsilon
\simeq 1~, \kappa =1$ (upper left corner in Fig. \ref{fig:regiontm}), 
producing roughly a correct hierarchy between high temperature baryon
asymmetry and low temperature lepton asymmetry.

Finally, let us discuss the possibility that large lepton asymmetries
$\Delta_0 \gsim 2\times 10^{-3}$  were generated well above the
electroweak temperature. Is it possible that they were not
transferred to baryon asymmetry but survived till low temperatures?

As we have already found, the only leptonic numbers that can survive
till low temperatures are related to the currents $J_\mu^4$ and
$J_\mu^5$, defined in (\ref{j4},\ref{j5}). Moreover,  the only
flavour structure of primordial asymmetry which is consistent with
small baryon asymmetry is the one in which $L + \Delta N_2 =0$, where $L$
is a lepton number of active fermions, and  $\Delta N_2$ is the asymmetry
in a more strongly interacting singlet fermion. Indeed, if  $L$ is
large it will lead to large baryon asymmetry due to sphalerons. If
$\Delta N_2$ is large, a part of it will be transferred to $L$ and then to
baryon asymmetry. The amount of $N_2$ going to $L$ is at least 
\be
\Delta_0 S_+(T_{EW}) > 2\times 10^{-4} \Delta_0 \gg \Delta B~,
\ee
where we used the minimal possible rate $R(T,M)$ corresponding to
$M\simeq m_\pi$ and $\epsilon =1$. In other words, the only
possibility is to have large asymmetry in $N_3$, $\Delta N_3 =
\Delta_0$  and assume that $\epsilon \ll 1$, suppressing the
transitions $N_3 \to L$.

Now, four different possibilities can be realised. If the reactions
changing $Q_4$ and $Q_5$ were both in thermal equilibrium, no
primordial asymmetry will survive. If, on the contrary,  none of
the reactions changing   $Q_4$ and $Q_5$ were in thermal equilibrium,
a large asymmetry in $N_3$ will not be transferred to an asymmetry in
active leptons, and, therefore, no resonant production of dark matter
sterile neutrinos is possible. So, to get large lepton asymmetry at
low temperatures one must require that one of charges out of $Q_4$ and
$Q_5$ must be conserved and the other equilibrate. In Fig. \ref{fig:surv}
we present the parameter-space in which the primordial asymmetry in
$N_3$ induces a baryon asymmetry smaller than the observed one but
leads to large low temperature lepton asymmetry. It requires rather
small values of the parameter $\epsilon$. 

\begin{figure}[tb]

\centerline{%
\epsfysize=4.5cm\epsfbox{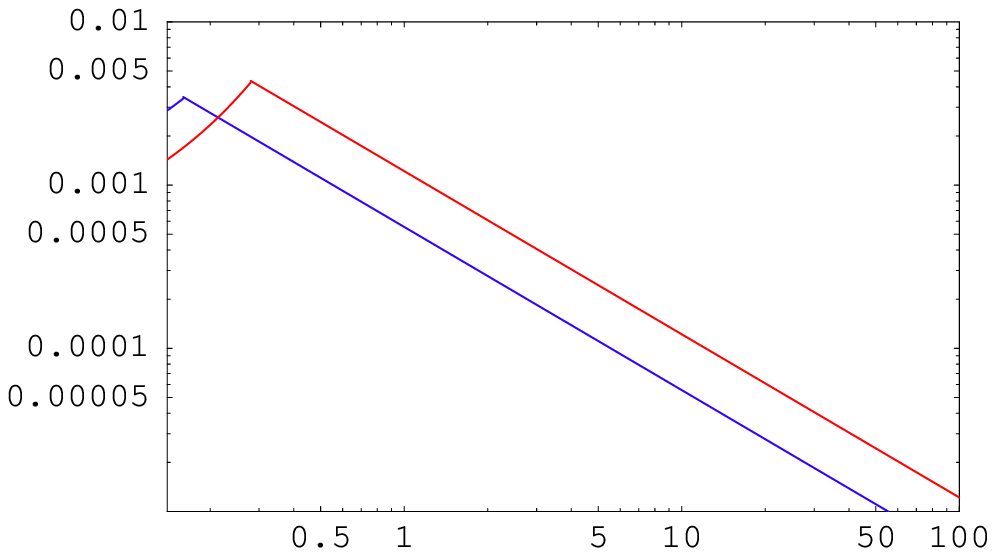}%
~~~~\epsfysize=4.5cm\epsfbox{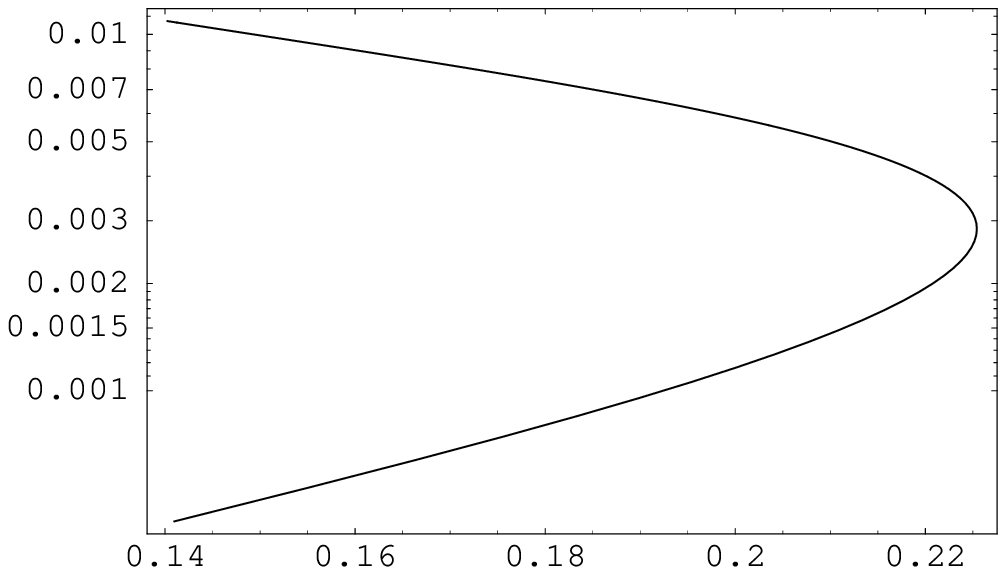}%
}
\caption[a]{\small Left panel: Part of the parameter space
corresponding to conservation of $Q_4$ and non-conservation of $Q_5$.
This can only be realised in {\bf Scenario IIa} in which $T_\beta <
T_-$. Vertical axis: $\epsilon$, horizontal axis: mass in GeV. The
admitted regions are below the curves. Upper red line - normal
hierarchy, lower blue line - inverted hierarchy. Right panel: Part of
the parameter space corresponding to conservation of $Q_5$ and
non-conservation of $Q_4$. It is required that  $T_\beta > T_-$.
Vertical axis: $\epsilon$, horizontal axis: mass in GeV. The admitted
region is to the left of the curve. No parameter space is allowed for
the inverted hierarchy case.
}
\la{fig:surv}
\end{figure}

\section{Fine tunings or new symmetries?}
\la{se:symmetry}
The requirement that the $\nu$MSM produces both baryon asymmetry and
dark matter in amounts required by observations puts very stringent
constraints on the parameters of the model. In this section we will
discuss whether these constraints, appearing as different fine-tunings
in the Lagrangian of the $\nu$MSM, can indicate the  existence of some
hidden approximate symmetries. These symmetries, if exist, cannot be
explained in the framework of the $\nu$MSM itself, as this model is
based on  a renormalizable field theory which may be valid all the way
up to the Planck scale \cite{Shaposhnikov:2007nj}. At the same time,
their presence can give some hints on the properties of more
fundamental theory, replacing the $\nu$MSM at high energies.  

We start from the relative strength of Yukawa interactions of singlet
fermions $N_2$ and $N_3$. A non-trivial constraint on the $\nu$MSM
parameters is coming from the requirement that the baryon asymmetry at
the electroweak scale must be much smaller than the lepton asymmetry
at small temperatures. It tells  that the parameter $\epsilon$ should
be close to its maximal value, $\epsilon \sim 1$.  For  $\epsilon$
that large the lepton number U(1) symmetry, introduced in
\cite{Shaposhnikov:2006nn}, is strongly broken in the singlet fermion 
Yukawa sector, but is respected by the Majorana masses of the singlet
fermions and by charged lepton Yukawas.  Therefore, one may wonder if
some other global symmetry, respected both by the Yukawa couplings and
by Majorana masses, may exist for the extreme case $\epsilon =1$. 

As was discussed in \cite{Shaposhnikov:2006nn}, such a symmetry does
not exist if {\em both charged and singlet} lepton Yukawa couplings
are taken into account.  If, however, charged lepton Yukawas are
disregarded, quite a symmetric singlet lepton interaction can be found
in the {\em inverted} hierarchy case. Indeed, for the case $m_1=m_2$,
$\theta_{23}=\pi/4$, and  $\theta_{13}=0$  the fields $L_2$ and $L_3$
defined in (\ref{L23inv}) are the {\em orthogonal} mixtures of
different leptonic flavours. Thus, for $\epsilon=1$ the Yukawa part of
Lagrangian  (\ref{pmm}) is symmetric with respect to the non-Abelian
flavour group SU(2) (broken, of course, by the charged lepton Yukawa
couplings). This group is broken down to U(1) by the Majorana mass
term $M\bar N_2^c N_3$. This U(1) group is then only slightly broken
by the diagonal mass terms $\sim \Delta M_M \ll M$ and by corrections
in the Yukawa sector, which can be as small as $\delta_{\rm inv}\sim
0.01$ defined in eq. (\ref{deltainv}). So, if the existence of
slightly broken approximate symmetry indeed matters, then the 
inverted hierarchy of neutrino masses with small $\theta_{13}\sim
\delta_{\rm inv}$ and small deviation of the angle 
$\delta\theta_{23}\sim \delta_{\rm inv}$ from the maximal value is
preferred. 

Interestingly, for $|\epsilon - 1| \sim \delta_{\rm inv}$ and  
$\delta_{\rm inv}\ll 1$ the interactions of the heavy neutral lepton 
mass eigenstates with intermediate weak vector bosons are universal
and characterized by the same mixing angle 
\be
\theta_M^2 = \frac{F^2 v^2}{M^2}=\frac{m_{atm}}{M}~. 
\ee
Also, both the high temperature
baryogenesis and low temperature leptogenesis can take place.

Let us now try to guess what kind of couplings of the dark matter
sterile neutrino with leptons may lead to some non-trivial symmetries.
The phenomenology of DM sterile neutrino requires its mass be much
smaller than the mass of the singlet fermions responsible for baryon
asymmetry and that its Yukawa constants are much smaller than those
for heavier neutral leptons. This leads to a conjecture that the
singlet fermion Majorana masses could be proportional to their Yukawa
couplings, satisfied already for $N_{2,3}$ in the construction
presented above. If true, then the mixing angles of all three sterile
leptons with neutrinos are the same, and the interaction of them with
$W$ and $Z$ bosons exhibits the global SU(3) symmetry, which exists
for charged leptons. If this hypothesis happens to be correct, the
mixing angle of DM sterile neutrino is predicted to be 
\be
\theta^2_{DM} =\frac{\sum_\alpha |h_{\alpha1}|^2v^2}{M_1^2}=
\theta_M^2= \frac{m_{atm}}{M}\simeq 2.5\times 10^{-11}~,
\ee
corresponding to $M\simeq 2$ GeV, a preferred value leading to  the
required hierarchy between baryon asymmetry and low temperature lepton
asymmetry. For this value of the mixing angle the mass of DM sterile
neutrino is bounded from above by $M_1 \lsim 8$ keV by X-ray
observations (see the plots presented in ref. \cite{Laine:2008pg}). If
the Lyman-$\alpha$ bounds of refs. \cite{Seljak:2006qw,Viel:2006kd}
are correct, then $M_1 \gsim 4$ keV (see the discussion in ref.
\cite{Laine:2008pg}). To exclude or verify this prediction, the current
X-ray constraints must be improved by a factor of 10.

Yet another fine-tuning which is necessary for creation of large
low-temperature lepton asymmetry, is eq. (\ref{finetun}), leading to 
{\bf Scenario IIa} for singlet fermion mass difference.  Though the
origin of different parameters even in the Standard Model remains a
mystery, it is tempting to speculate how this relation, equivalent
to  
\be
2(h^\dagger h)_{23} v^2 + M(\Delta
M_{22}^*+\Delta M_{33})\simeq 0~,
\label{fff}
\ee
may come from some more fundamental theory. In the $\nu$MSM described
by Lagrangian (\ref{lagr}) the first term in this condition is due to
the Higgs condensate while the second is due to Majorana masses of
singlet fermions and, therefore,  they have completely different
nature. Clearly, a correlation between two independent dimensionfull
parameters would be a miracle if the $\nu$MSM were the final
fundamental theory. This is not so if the mass parameters in the
$\nu$MSM have the common source, as in the model of
\cite{Shaposhnikov:2006xi}, where the Higgs boson and the neutral
fermion masses come from the vacuum expectation value of the nearly
conformally coupled scalar field $\chi$, singlet with respect to the
SM gauge group. In this case the relation (\ref{fff}) turns into a
connection between the Yukawa coupling constants in the sterile
neutrino sector of the $\nu$MSM. In fact, all phenomenological and
cosmological requirements to the parameters of the $\nu$MSM with extra
scalar field $\chi$ can be encoded in a simple Lagrangian, kind of
Effective Theory of Everything (ETOE), containing just few
dimensionless parameters, their powers, and one mass scale.  It has
the form:
\begin{eqnarray}
  {\cal L}_{\nu \rm{MSM}}\rightarrow
   {\cal L}_{\nu \rm{MSM}[M\rightarrow 0]} + 
  \frac{1}{2 f_0 f_1}(\partial_\mu\chi)^2
  - \frac{\chi}{2} \; \bar {N_I}^c m_{IJ} N_J  + \rm{h.c.} -
  V(\Phi,\chi)+{\cal L}_G
   \,,
  \label{lagr1}
\end{eqnarray}
where the first term is the $\nu$MSM Lagrangian without Higgs
potential and  with all dimensionfull parameters (Higgs and Majorana
masses) put to zero, the constants $f_0$ and $f_1$ will be specified
below. The scalar potential is given by
\be
V(\Phi,\chi) =
\lambda\left(\Phi^\dagger\Phi-\chi^2\right)^2 +
\beta(\chi^2-v^2)^2~,
\label{pot}
\ee
where $\lambda\sim \beta \sim 1/10$  are the Higgs and $\chi$
self-couplings correspondingly, $v$ is the Higgs vev. 
The gravity part is
\be
{\cal L}_G = -\left(\frac{1}{f_0^6}\chi^2 
+\frac{\lambda}{f_0}\Phi^\dagger\Phi\right)\frac{R}{2}~,
\label{GG}
\ee 
where $R$ is the scalar curvature. This is a Lagrangian of ``induced
gravity'' going back to refs. \cite{Zee:1978wi,Smolin:1979uz} (see
also \cite{Shaposhnikov:2006xi} in the $\nu$MSM context).   The
Yukawa couplings  $h_{\alpha I}$ in eq.(\ref{lagr}) are written as
\be
h_{\alpha I} = f_0 f_{\alpha J} m_{JI}~,
\ee
where $f_{\alpha J}$ is an arbitrary complex matrix with elements 
$f_{\alpha J} \sim 1$ and 
\be
m_{IJ} = f_1\left[ {\cal M}_0 -\frac{1}{2} f_0^2\left(f^\dagger f {\cal M}_0 +
{\rm transposed}\right)\right]
\label{mmm}
\ee
with
\be
{\cal M}_0 =  
\left(
\begin{array}{c c c}
0 & 0 & 0 \\
0 & 0 & 1 \\
0 & 1 & 0
\end{array}
\right)
+ 
f_0 \left(
\begin{array}{c c c}
1 & 0 & 0 \\
0 & a & 0 \\
0 & 0 & -a
\end{array}
\right)~,
\ee
where $a\sim 1$ is a real number. The second term in (\ref{mmm}) is
chosen in such a way that eq. (\ref{fff}) is automatically satisfied
for any choice of $f_0,~f_1$ and $f_{\alpha J}$.

The parameter $f_0 = (v/M_P)^{\frac{1}{3}} \simeq 4 \times 10^{-6}$,
where $M_P=(8\pi G_N)^{-\frac{1}{2}}=2.4\times 10^{18}$ GeV is the
reduced Planck scale,  appears in several places in Lagrangian
(\ref{lagr1}). It is fixed from the requirement to provide the known
Newton constant and the correct phenomenology of singlet fermions, as
we describe in what follows.  

The parameter $f_1 \sim \delta_{\rm inv} \sim 10^{-2}$ sets the mass
of the singlet fermions (giving baryon and lepton asymmetries) in the
GeV region, $M_{2,3}\sim f_1 v$. The mass of dark matter neutrino $
M_1 \sim f_0 M_{2,3}$ is then in ${\cal O}(10)$ keV region, masses of
active neutrinos in the fraction of eV region, $m_\nu \sim f_0^2
M_{2,3}$. In addition, $f_0$ makes the Yukawa coupling of sterile
neutrino to be small compared with the Yukawa couplings of $N_{2,3}$
by a factor $f_0$, exactly what is needed  to produce them in the
early universe in right amounts to play the role of dark matter.
Moreover, $\lambda/(2 f_0) \sim 2 \times 10^{4}$, appearing in the
conformal coupling of the Higgs field  to Ricci scalar $R$, leads to
inflation producing correct amplitude of primordial fluctuations
\cite{Bezrukov:2007ep}. The field $\chi$ is very light due to its
conformal coupling to gravity  ($m_\chi \sim \sqrt{\beta} f_0^3 v \sim
10^{-5}$ eV), but practically decouples from the fields of the
$\nu$MSM  \cite{CervantesCota:1995tz,vanderBij:1993hx} (see also
\cite{Bezrukov:2007ep}). In contrast with \cite{Shaposhnikov:2006xi},
where the terms (\ref{GG}) were not introduced,  it plays no role in
inflation and in production of dark matter sterile neutrinos.

The author has no idea from were the structures discussed above can
be coming from but is amazed by some numerical coincidences they
uncover.

\section{Conclusions}
\la{se:conclusion}
In this work we scrutinized the mechanism of leptogenesis via
oscillations of light singlet fermions and determined the parameter
space of the $\nu$MSM which can lead to successful baryogenesis. The
kinetic processes in the model are quite complicated as they are
characterised by a number of different time scales and by fluctuations
(deviations from thermal equilibrium) of different nature, interacting
with each other. 

The first sector includes {\em CP-even deviations} from thermal
equilibrium in the system of almost degenerate singlet fermions. These
fluctuations give a ``source'' term for baryogenesis; creation of
lepton asymmetry switches off when these deviations are damped away.
The kinetic evolution of these fluctuations is governed by four
different time scales: two equilibration rates for $N_{2,3}$, the rate
of losing of quantum coherence in oscillations of singlet fermions, 
and the rate of oscillations, related to the  mass difference between
singlet fermions.  In the paper we estimated all these time scales 
(\se\ref{se:CPeven}). We found, in particular,  the temperature
dependence of the oscillation time, essential for {\bf Scenarios I}
and {\bf II} for singlet fermion mass difference. 

The second sector includes {\em CP-odd deviations} from thermal
equilibrium in the system  of singlet fermions and active leptons.
There are 7 different essential kinetic time scales there. The first 4
are similar to those described above, three others govern the damping
of asymmetries in different active leptonic flavours.

We established that the oscillations of $N_{2,3}$ must be {\em
coherent} for effective leptogenesis. This is only true if {\em both}
$N_2$ and $N_3$ are out of thermal equilibrium. In other words, lepton
asymmetry increases in time till one of the singlet fermions, which
interacts more strongly with the plasma ($N_2$ in our notations)
enters in thermal equilibrium. After this moment the coherence in
singlet fermion oscillations is lost, and asymmetries in different
quantum numbers (which we identified) are damped with the rates,
which we determined  in \se\ref{se:CPodd}.

We found that baryogenesis may occur in a wide range of singlet lepton
masses ranging from $140$ MeV, allowed by experimental and BBN
constraints, to the masses exceeding the electroweak scale. An
essential requirement is a near degeneracy of a pair of the heavy
neutral leptons. In addition, the parameter $\epsilon$, characterising
the breaking of the U(1) leptonic symmetry, cannot be smaller than
$7\times 10^{-5}$.

We determined explicitly the CP-violating phase which drives
baryogenesis in the model and demonstrated that it cannot be expressed
only in terms of CP-violating phases of the active neutrino mixing
matrix. Moreover, we found that the baryon asymmetry is non-zero in
the limit of small $\theta_{13}$. 

We showed, furthermore, that the $\nu$MSM interactions of  singlet
fermions may  produce a significant low temperature lepton asymmetry,
being consistent with neutrino oscillation experiments and leading to
the observed baryon asymmetry of the universe. In a companion paper
\cite{Laine:2008pg}, we show that this lepton asymmetry can account
for all the dark matter in the universe. Thus, the $\nu$MSM {\em
without introduction of any new physics or fields} such as the
inflaton may happen to be a correct effective field theory all the way
up the Planck scale \cite{Shaposhnikov:2007nj}  explaining a variety
of phenomena that the SM  fails to deal with. It is intriguing that
the production of the baryon asymmetry of the universe and of the dark
matter is due to essentially the same mechanism, making a step towards
understanding why the abundances of dark and baryonic matters are
roughly the same. 

We also found that large lepton asymmetries in singlet fermions $N_3$,
which could have been generated above the electroweak scale, may not
be in conflict with the observed baryon asymmetry and can survive till
low temperatures in a specific part of the $\nu$MSM parameter space.
It corresponds  to masses above $140$ MeV and small $\epsilon <
5\times 10^{-3}$ and also require {\bf Scenario IIa} for the singlet
fermion mass difference. Another possibility is to have singlet
fermion masses near the pion mass and $\epsilon$ in the range $5\times
10^{-4} < \epsilon < 0.01$. These regions can be  explored in kaon
experiments and in  searches for singlet fermion decays 
\cite{Gorbunov:2007ak}.

The requirement that the $\nu$MSM produces a lepton asymmetry large
enough to speed up the dark matter production allows to constrain
considerably the parameters of the $\nu$MSM. The most non-trivial
requirement  is (\ref{finetun}), telling that  the  {\em
zero-temperature difference} between masses of the {\em physical}
singlet fermions must be much smaller than the active neutrino mass
differences. For this choice of parameters the baryon asymmetry is
generated at temperatures close to the sphaleron freeze-out, $T \sim
130-175$ GeV, and a large lepton asymmetry at relatively small
temperatures, $ T = T_- \sim 0.1~-~10$ GeV, corresponding to the
decoupling of singlet fermions from the plasma or to their decays.
Later the lepton asymmetry is transferred to the dark matter
population of sterile neutrinos. The asymmetry generation mechanism
works for all singlet lepton masses admitted by experimental and BBN
constraints discussed in \cite{Gorbunov:2007ak} and for both types of
neutrino mass hierarchies; to produce the low temperature lepton
asymmetry required for resonant dark matter production the parameter
$\epsilon$ should be large enough,  $\epsilon \gsim 2\times 10^{-2}$.
Moreover, the requirement of  having a much smaller baryon asymmetry
favours large $\epsilon\sim 1$ and singlet fermion masses in the
${\cal O}({\rm GeV})$ range.  Particles with these  properties can be
searched for at existing accelerators \cite{Gorbunov:2007ak}, which is
however very challenging due to the large value of $\epsilon$, leading
to a suppression of their production and to a decrease of their decay
rates. At the same time, the CP-asymmetry in their decays must be at
least  on the level of few \%.

We speculated on the origin of the necessary fine-tunings in the
$\nu$MSM and proposed a Lagrangian, containing two dimensionless
parameters and their powers, which encodes different relations
required for the phenomenological success of the model. We found, in
particular, that the theory with $\epsilon =1$ and inverted hierarchy
of neutrino masses exhibits a SU(2) flavour symmetry in the singlet
fermion Yukawa sector, broken to U(1) by the Majorana mass term. The
magnitude of the breaking of this U(1) group is small and is of the
order $\frac{\Delta m^2_{\rm sol}}{4\Delta m^2_{\rm atm}}\simeq
8\times10^{-3}$. In a search of a ``maximally symmetric'' version of
the $\nu$MSM we found that it is phenomenologically acceptable to
think that the strength of the weak interactions of all types of
singlet fermions is universal. This conjecture leads to a specific
prediction for the mixing angle of dark matter sterile neutrino $N_1$,
potentially testable with the help of {\em existing} X-ray satellites.

Finally, a word of warning. All the constraints discussed above are
applicable only in the case when at temperatures well above the
electroweak scale concentrations of all singlet leptons are zero. In
particular, if the dark matter sterile neutrinos are generated above
the electroweak scale in right amounts, no generation of large lepton
asymmetry is needed below the electroweak scale. 

%
\section*{Acknowledgements}

The work of M.S was supported in part  by the Swiss National Science
Foundation. I thank Takehiko Asaka for collaboration at the initial
stages of this work and to Mikko Laine for writing Appendix A and
preparing Fig. 3. Extremely helpful discussions with them are
greatly acknowledged.

\appendix
\renewcommand{\thesection}{Appendix~\Alph{section}}
\renewcommand{\thesubsection}{\Alph{section}.\arabic{subsection}}
\renewcommand{\theequation}{\Alph{section}.\arabic{equation}}

\section{The rates of singlet fermions production}

We specify in this Appendix the ingredients that 
went into producing Fig.~\ref{fig:Y}.
  
The basic formalism we follow is that of 
refs.~\cite{Asaka:2006rw,Asaka:2006nq}. More precisely, 
the quantity $Y$ in \fig\ref{fig:Y} is given by \eq(4.8) of 
ref.~\cite{Asaka:2006nq}, while $Y_M$ 
contains the additional weight $({q_0-q})/({q_0+q})$, 
cf.\ \eq\nr{eq:RM}.

The main difference with respect to the analysis of ref.~\cite{Asaka:2006nq}
is that we now consider the heavy sterile neutrinos, and that 
the temperatures are correspondingly higher. 
This implies that the exponentially suppressed 1-loop corrections
(\eq(3.1) of ref.~\cite{Asaka:2006nq}) start to dominate over the 2-loop
terms (Sec.~3.2 of ref.~\cite{Asaka:2006nq}). More precisely, the main
changes to the numerical code are as follows: 
\begin{itemize}

\item
Because of the higher temperatures, 
the contribution of the bottom quark has been added to the 2-loop
processes listed in Table~1 of ref.~\cite{Asaka:2006nq}.

\item
Once the temperature increases above 20 GeV or so, the treatment
of 2-loop effects through the Fermi model is no longer justified. 
Therefore we smoothly switch off the 2-loop contributions within the
range $T = (15 ... 30)$~GeV. 

\item
Concerning the 1-loop effects, the graphs to be considered are 
given in \fig\ref{fig:deltam} of the present paper (except that we use here 
a basis where the Majorana mass matrix is flavour-diagonal). 
In the top graph, the particle in the loop can either be a Higgs 
or a Goldstone. In addition, the self-energy of the active 
neutrino, appearing in the bottom graph, depends
on the gauge choice (because the active neutrino is off-shell). 

Now, the simplest gauge choice in this context is that of Feynman. 
Then the real part of the active neutrino self-energy (the function $b$)
can be taken directly from ref.~\cite{Quimbay:1995jn} 
and the imaginary part from
\eq(3.1) of ref.~\cite{Asaka:2006nq}.
At the same time, the top graph of \fig\ref{fig:deltam} amounts to 
\be
 \delta R(T,\vec{q})
 =  \frac{2 \nF{}(q^0)}{(2\pi)^3 2 q^0}
   \sum_{\alpha = 1}^{3}  
   \frac{|M_D|^2_{\alpha I}}{m_W^2}
   \tr\Bigl[\bsl{Q} a_L \im \bsl{\Sigma}_\rmi{Higgs}\, a_R\Bigr]
   \;,
\ee
where $\im \bsl{\Sigma}_\rmi{Higgs}$ has exactly the form
in \eq(3.1) of ref.~\cite{Asaka:2006nq}, with three channels
characterized by
$p_C = 1; m_C= m_H; m_{l_C} = m_{\nu_\alpha}$; 
$p_C = 1; m_C= m_Z; m_{l_C} = m_{\nu_\alpha}$; and
$p_C = 2; m_C= m_W; m_{l_C} = m_{l_\alpha}$.

Another possible choice is the unitary gauge. Then the Goldstone
contributions can be dropped from the top graph, but the active 
neutrino self-energy needs to be modified. We have checked that
after the appropriate changes, 
the numerical results in the two gauges differ by an amount 
which is insignificant on our resolution. 

\item
Once the temperature increases to several tens of GeV, the evolution
of the Higgs vacuum expectation value needs to be taken into account. 
We do this by scaling 
$\sqrt{2} v(T) = \mbox{246~GeV}\sqrt{1-T^2/T_0^2}$, 
where $T_0$ is fixed through
the knowledge that the sphaleron freeze-out temperature $T_{EW}$, where
we start our evolution, is characterized by 
$\sqrt{2} v(T_{EW}) \simeq T_{EW}$. We choose $m_H \simeq 200$~GeV and then, 
according to ref.~\cite{Burnier:2005hp}, $T_{EW} \simeq 175$~GeV.
All physical particle masses are rescaled by $v(T)/v(0)$.

\end{itemize}

Apart from these changes, the numerical techniques used are identical
to those in ref.~\cite{Asaka:2006nq}.

\section{Lower bounds on Yukawa couplings}

In this Appendix we present a lower bound on the following
combinations of Yukawa couplings which will appear in the analysis of
equilibration in the early universe,
\be
|f_{\alpha\alpha}|^2 \equiv
\left(|h_{\alpha 2}|^2 + |h_{\alpha3}|^2\right)~.
\label{combin}
\ee
With the use of (\ref{mass}) one can see that the minimal value of 
$|f_{\alpha\alpha}|^2$ is simply $|[M_\nu]_{\alpha\alpha}|M/v^2$. 
The smallest Yukawa couplings correspond to the smallest value of the
Majorana neutrino mass, which we take to be $M\simeq m_\pi\simeq 140$
MeV (the mass of the pion is introduced as a useful parametrisation)
(smaller values would be in conflict with predictions of BBN
\cite{Dolgov:2000pj,Dolgov:2000jw} and experiments devoted to the
search of singlet fermions 
\cite{Bernardi:1987ek,Shaposhnikov:2006nn,Gorbunov:2007ak}). 
Inserting the central
values for neutrino masses and mixing angles from
\cite{Strumia:2006db}: $\Delta m_{sol}^2 = 8.0\times 10^{-5}$
eV$^2$,    $\Delta m_{atm}^2 = 2.5\times 10^{-3}$ eV$^2$,
$\theta_{23} = \pi/4$, $\tan^2(\theta_{12}) =0.45$, $\theta_{13}=0$,
and choosing the unknown CP-violating phases in  a way to minimize
the Yukawa couplings, we get for the normal hierarchy:
\be
|f_{ee}|^2 > 1.3 \times 10^{-17}~,~~|f_{\mu\mu}|^2 >10^{-16},~~
|f_{\tau\tau}|^2 > 10^{-16}
\label{fnorm}
\ee
and for the inverted hierarchy
\be
|f_{ee}|^2 > 8.8 \times 10^{-17}~,~~|f_{\mu\mu}|^2 >4.4 \times 10^{-17},~~
|f_{\tau\tau}|^2 > 4.4 \times 10^{-17}~.
\label{finv}
\ee
These numbers change somewhat if the neutrino mixing parameters are
varied in the experimentally admitted ranges. To get a minimal
possible value of, say, $|f_{ee}|^2$ one should take the maximal possible
atmospheric mass difference  ($2.7\times 10^{-3}$ eV), minimal solar
mass difference ($7.7\times 10^{-5}$ eV), minimal $\theta_{12} \simeq
0.56 $ and maximal $\theta_{13} \simeq 0.11$, leading to 
\be
|f_{ee}|^2 > 8.4 \times 10^{-18}~.
\label{minimal}
\ee
If $M > m_\pi$ then the lower bounds are stronger by a factor
$M/m_\pi$.


\end{document}